\documentclass[aps,twocolumn,9pt,preprintnumbers,nofootinbib,superscriptaddress]{revtex4-1}
\usepackage{eurosym}
\usepackage{amsmath}
\usepackage{graphicx}
\usepackage{amsfonts}
\usepackage{array}
\usepackage{amsthm}
\usepackage{bm}
\usepackage{palatino}
\usepackage{mathpazo}
\usepackage{supertabular}
\usepackage{subfig}
\usepackage[breaklinks]{hyperref}
\usepackage[font=small,labelfont=bf,
justification=justified,
format=plain]{caption}
\usepackage[font=small,labelfont=bf,justification=justified]{caption}
\usepackage{color}
\usepackage{booktabs}
\usepackage{multirow}
\usepackage[labelfont=bf,labelsep=quad,justification=justified]{caption}
\newcommand{\be}{\begin{equation}}
\newcommand{\ee}{\end{equation}}
\newcommand{\ba}{\begin{eqnarray}}
\newcommand{\ea}{\end{eqnarray}}
\newcommand{\bal}{\begin{align}}
\newcommand{\eal}{\end{align}}
\newcommand{\lb}{\label}

\newcommand{\ta}{\theta}

\newcommand{\bw}{\begin{widetext}}
	\newcommand{\ew}{\end{widetext}}

\newcommand{\bq}{\begin{equation}}
\newcommand{\eq}{\end{equation}}
\newcommand{\bqn}{\begin{eqnarray}}
\newcommand{\eqn}{\end{eqnarray}}
\newcommand{\nb}{\nonumber}
\begin{document}
	
	\title{Quasinormal modes, quasiperiodic oscillations and shadow of rotating regular black holes in non-minimally coupled Einstein-Yang-Mills theory}
	\author{Kimet Jusufi}
	\email{kimet.jusufi@unite.edu.mk}
	\affiliation{Physics Department, State University of Tetovo, Ilinden Street nn, 1200,
		Tetovo, North Macedonia}
	\affiliation{Institute of Physics, Faculty of Natural Sciences and Mathematics, Ss. Cyril
		and Methodius University, Arhimedova 3, 1000 Skopje, North Macedonia}
	\author{Mustapha Azreg-A\"{i}nou}
	\email{azreg@baskent.edu.tr}
	\affiliation{Ba\c{s}kent University, Engineering Faculty, Ba\u{g}l{\i}ca Campus, 06790-Ankara, Turkey}
	\author{Mubasher Jamil}
	\email{mjamil@zjut.edu.cn (corresponding author)}
	\affiliation{Institute for Theoretical Physics and Cosmology
		Zhejiang University of Technology
		Hangzhou, 310023 China}
	\affiliation{United Center for Gravitational Wave Physics,
		Zhejiang University of Technology
		Hangzhou, 310023 China}
	\affiliation{Department of Mathematics, School of Natural
		Sciences (SNS), National University of Sciences and Technology
		(NUST), H-12, Islamabad, Pakistan}
	\affiliation{Canadian Quantum Research Center 204-3002 32 Ave Vernon, BC V1T 2L7, Canada}	
	\author{Shao-Wen Wei}
	\email{weishw@lzu.edu.cn}
	\affiliation{
		Institute of Theoretical Physics \& Research Center of Gravitation,
		Lanzhou University, Lanzhou 730000, People's Republic of China}
	\author{Qiang Wu}
	\email{wuq@zjut.edu.cn}
	\affiliation{Institute for Theoretical Physics and Cosmology
		Zhejiang University of Technology
		Hangzhou, 310023 China}
	\affiliation{United Center for Gravitational Wave Physics,
		Zhejiang University of Technology
		Hangzhou, 310023 China}
	
	\author{Anzhong Wang}
	\email{anzhong$\_$wang@baylor.edu}
	
	\affiliation{GCAP-CASPER, Physics Department,  Baylor University, Waco, Texas 76798-7316, USA}
	
	\begin{abstract}
		 In this paper we obtain an effective metric describing a regular and rotating magnetic black hole (BH) solution with a Yang-Mills electromagnetic source in Einstein-Yang-Mills  (EYM)  theory using the Newman--Janis algorithm via the non-complexification radial coordinate procedure. We then study the BH shadow and the quasinormal modes (QNMs) for massless scalar and electromagnetic fields and the quasiperiodic oscillations (QPOs). To this end, we also study the embedding diagram for the rotating EYM BH. The energy conditions, shadow curvature radius, topology and the dynamical evolution of scalar and electromagnetic perturbations using the time domain integration method are investigated. We show that the shadow radius decreases by increasing the magnetic charge, while the real part of QNMs of scalar and electromagnetic fields increases by increasing the magnetic charge. This result is consistent with the inverse relation between the shadow radius and the real part of QNMs. In addition, we have studied observational constraints on the EYM parameter $\lambda$ via frequency analysis of QPOs and the EHT data of shadow cast by the M87 central black hole. We also find that the decaying rate of the EYM BH is slower than that of the neutral and ends up with a tail. We argue that the rotating EYM black hole can be distinguished from the Kerr-Newman black hole with a magnetic charge based on the difference between the angular diameters of their shadows.
	\end{abstract}
	\maketitle
	\section{Introduction}
	
	It is generally believed that most of the giant elliptical and spiral galaxies contain supermassive black holes (SMBHs) at their galactic centers. For instance, the masses of SMBHs at the centers of Milky Way spiral galaxy and M87 elliptical galaxy are four million and six billion solar masses, respectively. Besides having huge masses, these SMBHs also possess spins (or angular momenta).
	Depending on the spacetime geometry, a BH can capture light received from nearby stars or accretion disks into bound orbits. A large collection of light orbits constitutes a ``photon sphere'' around the BH. If the orbit of light is unstable, then photons (quanta of electromagnetic field) can either fall into the BH or escape to infinity (or a distant observer at a finite distance). The Event Horizon Telescope (EHT) collaboration has detected the first shadow images of the SMBH at the center of M87 galaxy \cite{Akiyama:2019eap,m87}.  With this image, it is observed that the diameter of the center BH shadow was approximately 52 micro-arc-second with a deviation of less than 10 \% from circularity which leads to a measurement of the central mass of 6.5 billion solar mass. Importantly, these precise observations could provide a {potential} window to explore, distinguish or constrain physically viable BH solutions that exhibit small deviations from the Kerr metric. The distortion in the size and magnification of the shadow images {provides information about the BH properties (such as its mass and spin) and the nearby geometry (the} Schwarzschild, Kerr or modified Kerr spacetime). Moreover, the shadow image is a manifestation of strong gravitational lensings which can be used to distinguish various forms of BH spacetimes and naked singularities. Some of {such} studies on BH shadows in various gravitational theories were given in \cite{Jusufi:2019nrn, Zhu:2019ura,Haroon:2018ryd,Haroon:2019new,Amir:2018pcu,Bambi:2019tjh,Vagnozzi:2019apd,Khodadi:2020jij,Allahyari:2019jqz,Paul:2019trt,Kumar:2020yem,Ghosh:2020ece}.
	
	In  the literature, numerous static and spherically symmetric BH solutions have been derived in the modified gravity theories (MGTs). However, the task of deriving the exact rotating black {hole} solutions analytically by solving the coupled field equations in any MGT has remained  {daunting due to the complexity of the non-linear partial differential equations of the underlying theory}. For instance, under reasonable assumptions of stationary, axial symmetry and asymptotic flatness, the governing equations in f(R) gravity are highly non-linear having the fourth order derivatives,  while in the general Horndeski theories the field equations are second order. Still one is able to generate the metrics of stationary and axis-symmetric BHs using the Newman--Janis algorithm (NJA)~\cite{NJA} and its modifications by starting with any seed static and spherically symmetric spacetime~\cite{Azreg-Ainou:2014pra}. Among the modifications to NJA there is the noncomplexification procedure of the radial coordinate~\cite{Azreg-Ainou:2014pra}. This method has been extensively used in the literature for obtaining rotating BH solutions~\cite{core}-\cite{Shaikh:2019fpu}. From the astrophysical and astronomical perspectives, almost all known candidates of BHs are rotating. The signature of rotation of a BH would be determined by the distortion of its shadow images or deviation from the spherical symmetry. The solution obtained by NJA method is acceptable only if the resulting solution is free from geometrical pathologies, and satisfies the energy conditions, causality and regularity everywhere except at some spacetime singularities,  while allowing the existence of a spatial hypersurface where a timelike Killing vector becomes null.
	
	A rigorous proof about the existence of an infinite number of BH solutions to the Einstein-Yang-Mills (EYM) equations with the gauge group $SU(2)$ for any event horizon was provided in \cite{smol}. In the literature, slowly rotating non-abelian BHs, numerical rotating BHs in the  {minimally coupled} EYM theory as well as nonstatic spherically symmetric EYM BHs were previously derived~\cite{1,2,5,55}, in addition to the static, spherically symmetric constant curvature BHs \cite{fR}. Recently, new BH solutions have been also derived by adding Lorentz group symmetry in the
	{minimally coupled} EYM theory \cite{3} and loop quantum corrections \cite{4}. In this paper,  we focus on the {non-minimally coupled} EYM theory where the curvature couples with the SU(2) gauge fields non-trivially \cite{Balakin:2015gpq,Balakin}. Our aim is to test the {non-minimally coupled EYM} theory via constructing rotating BHs, and then systematically {investigate the consistency of the theory with the} current and forthcoming observations, including the observations of M87 BH shadow. Furthermore, we {would like to relate the shadow size with the quasi-normal modes (QNMs)} of the BHs. Here gravitational waves will be treated as massless particles {propagating}  along null geodesics and slowly leaking to infinity.
	
	Among numerous astrophysical events, the quasiperiodic oscillations (QPOs) are very common phenomena in the X-ray power density spectra of stellar-mass BHs. The frequency of QPOs can be related to the matter orbiting in the vicinity of the innermost stable circular orbit (ISCO) of the BH. The appearance of two peaks at 300 Hz and 450 Hz in the X-ray power density spectra of Galactic microquasars, representing possible occurrence of a lower QPO and of an upper QPO in a ratio of 3 to 2, has stimulated a lot of theoretical works to explain the value of the $3/2$-ratio. Some theoretical models, including the parametric resonance, forced resonance and Keplerian resonance,  have been proposed. Therefore, the study of QPOs not only help us understand the physical processes in BH mechanics, but also provides a powerful approach to explore the nature of the BH spacetime in the {strong field regime}.
	
	The structure of our paper is laid out as follows: In section~\ref{secrot}, we review the {non-minimally coupled EYM theory} and the {static} BH solution. Henceforth, we apply the NJA modified by the noncomplexification procedure of the radial coordinate to generate the rotating counterpart of the static solution. In sections~\ref{secem} and~\ref{secec}, we study the embedding diagram and  energy conditions, respectively. In sections~\ref{secs} and~\ref{secoc1}, we study the geometrical and astronomical features of the BH shadows and constraints on the free parameters. In Section~\ref{seccr}, we investigate the curvature radius and its relation with the topology of the shadow. Section~\ref{secrq} is devoted to the investigation of QNMs of the static BH and their relationship with the radius of the shadow, as well as the dynamical evolution of scalar and electromagnetic perturbations. Section~\ref{secqpos} is devoted to QPOs and their resonances. First, we {derive the} generic expressions for the radial and vertical QPOs, and  then apply them to the rotating solution.
	{In particular,} we show how to obtain good and complete curve fits to the data of three microquasars. Finally, in Section~\ref{seccon}, we discuss our main results and provide some concluding remarks. There are also two appendices, in which we provide the exact expressions of the Einstein tensor and of some physical quantities pertaining to section~\ref{secqpos}, respectively.
	
	\section{Rotating regular Einstein-Yang-Mills BH\label{secrot}}
	
	Let us start by writing down the action of the  non-minimally coupled EYM  theory in  four-dimensional spacetimes is given by \cite{Balakin:2015gpq,Balakin}
	\begin{equation}
	S=\frac{1}{8 \pi}\int d^4 x \sqrt{-g} \left[ R+\frac{1}{2}F_{\mu \nu}^{(a)}F^{\mu \nu (a)}  +\frac{1}{2} \mathcal{R}^{\alpha \beta \mu \nu}F_{\alpha \beta}^{(a)}F^{(a)}_{\mu \nu} \right],
	\end{equation}
	in which $g$ is the determinant of the metric tensor and $R$ is the Ricci scalar. Furthermore,  the Greek indices run from $0$ to $3$,
	while the Latin indices run from $1$ to $3$. On the other hand, the Yang-Mills (YM) tensor $F^{(a)}_{\mu \nu}$ is connected to the YM potential $A_{\mu}^{(a)}$ by the following relation
	\begin{equation}
	F^{(a)}_{\mu \nu}=\nabla_{\mu}A_{\nu}^{(a)}-\nabla_{\nu}A_{\mu}^{(a)}+f^{(a)}_{(b)(c)}A_{\mu}^{(b)}A_{\nu}^{(c)}.
	\end{equation}
	
	In the last equation $\nabla_{\mu}$ represents the covariant derivative and $f^{(a)}_{(b)(c)}$ denote the real structure constants of the 3-parameters YM gauge group $SU(2)$. The
	{tensor} $ \mathcal{R}^{\alpha \beta \mu \nu}$ is given by \cite{Balakin}
	\begin{eqnarray}\notag
	\mathcal{R}^{\alpha \beta \mu \nu}&=&\frac{\xi_1}{2}\left(g^{\alpha \mu} g^{\beta \nu}-g^{\alpha \nu} g^{\beta \mu}\right)\\\notag
	&+& \frac{\xi_2}{2}\left(R^{\alpha \mu}g^{\beta \nu}-R^{\alpha \nu}g^{\beta \mu}+R^{\beta \nu}g^{\alpha\mu}-R^{\beta \mu}g^{\alpha \nu}\right)\\
	&+& \xi_3 R^{\alpha \beta \mu \nu},
	\end{eqnarray}
	in which $R^{\alpha \beta}$  and $R^{\alpha \beta \mu \nu}$ are the Ricci and Riemann tensors respectively. In addition,  {$\xi_i(i=1,2,3)$ are the non-minimally coupled parameters between the YM field and the gravitational field.} With the assumptions that the gauge field is characterized by the Wu-Yang ansatz and $\xi_1=-\xi$, $\xi_2=4 \xi$, $\xi_3=-6 \xi$ along with $\xi>0$,
	a regular, static and spherically symmetric BH was found  \cite{Balakin:2015gpq,Balakin,liu}
	\begin{equation}\label{static}
	ds^2=-f(r)dt^2+\frac{dr^2}{g(r)}+h(r)(d\theta^2+\sin^2\theta d\phi^2),
	\end{equation}
	with $f(r)=g(r)$, $h(r)=r^2$ and
	\begin{equation}
	\lb{eq5}
	g(r)=1+\left(\frac{r^4}{r^4+2 \lambda}\right)\left(-\frac{2GM}{c^2r}+\frac{GQ^2}{4\pi\epsilon_0 c^4r^2}\right).
	\end{equation}
	Note that $\lambda=\xi Q^2$, while $M$ is the BH mass and $Q$ is the magnetic charge.  {When} $\lambda=0$ and $Q=0$, the above metric {reduces} to the Schwarzschild BH. Furthermore the total effective energy-momentum tensor consists of the pure Yang-Mills field and the effect of the coupling between the gravity and the Yang-Mills field  \cite{Balakin:2015gpq,Balakin}. From the Einsteins field equation the energy  density, the radial and tangential pressures are derived as follows
	\begin{align}
&\rho=-p_r=\frac{ [Q^2 (r^4-6 \lambda )+16 M r \lambda ]}{(r^4+2 \lambda )^2},\nonumber\\
&p_\theta=p_\phi=\frac{Q^2(r^8-24\lambda r^4+12 \lambda^2)-8M\lambda r (6 \lambda-5r^4)}{(r^4+2\lambda)^3}.
\end{align}
	
	Now, we apply the modified NJ algorithm  recently proposed in \cite{Azreg-Ainou:2014pra} to the static metric~\eqref{static}.  The essence of the procedure
	is to drop the complexification of the $r$ coordinate normally done in the NJ algorithm~\cite{NJA}, as  there does not exist a unique way to carry out it ~\cite{Azreg-Ainou:2014pra}.
	Dropping the complexification of $r$ implies dropping the complexification of the metric functions $f(r)$, $g(r)$ and $h(r)$. {Taking this advantage, Azreg-A\"{i}nou replaced
		them by $F \equiv F(r, a, \theta)$, $ G \equiv G(r, a, \theta)$ and $H \equiv H(r, a, \theta)$, respectively,
		\begin{equation}
		f(r)\to F(r, a, \theta),\;g(r)\to G(r, a, \theta),\;h(r)\to H(r, a, \theta).
		\end{equation}
		This combined algorithm should be called NJAA algorithm or just NJAAA. Then, the remaining steps, as described  in detail in~\cite{Azreg-Ainou:2014pra}, lead to the explicit  expressions for $F/H$ and $GH$,
		\begin{align}
		\frac{F}{H}=\frac{g h+a^2 \cos^2\theta}{[k(r)+a^2 \cos^2\theta]^2},\quad GH=g h+a^2 \cos^2\theta,
		\end{align}
		where $k(r)\equiv\sqrt{g/f}~h$, for which  the rotating metric takes the form}
	\begin{align}
	&ds^2 = -\frac{(g(r) h(r)+a^2 \cos^2\theta) H}{(k(r)+a^2 \cos^2\theta)^2}dt^2+\frac{H dr^2}{g(r) h(r)+a^2}\nonumber\\
	&- 2 a \sin^2\theta \left[\frac{k(r)-g(r)h(r)}{(k(r)+a^2 \cos^2\theta)^2}\right]H dt d\phi+H d\theta^2\nonumber\\
	&+H \sin^2\theta \Bigg[a^2 \sin^2\theta \left(\frac{2k(r)-g(r)h(r)+a^2 \cos^2\theta}{(k(r)+a^2 \cos^2\theta)^2}\right)\nonumber\\
	& ~~~~~~~~~~~~~~~~~~~~~ + 1\Bigg] d\phi^2.
	\end{align}
	Now since $f(r)=g(r)$,  and $h(r)=r^2$, one finds $k(r) = h(r)=r^2$. Furthermore, the function $H(r, \theta, a)$ is still arbitrary and can be chosen so that the
	cross-term {$G_{r \theta}$ of the Einstein tensor vanishes,  i.e. $G_{r \theta}=0$, which} yields the differential equation
	\begin{eqnarray}
	(h(r)+a^2 y^2)^2(3 H_{,r}H_{,y^2}-2H H_{,r y^2})=3 a^2 h_{,r}H^2,
	\end{eqnarray}
	where {$y\equiv\cos \theta$. It can be shown  that the solution of the above equation takes the form \cite{Azreg-Ainou:2014pra},}
	\begin{eqnarray}
	H=\Sigma\equiv r^2+a^2 \cos^2\theta .
	\end{eqnarray}
	
	{Thus, summarizing all the above,   the metric of the rotating BH finally  reads}
	\begin{align}\label{metric}
	&ds^2=-c^2\left(1-\frac{2\Upsilon(r) r}{\Sigma}\right)dt^2 -2 ac \sin^2\theta \frac{2\Upsilon(r) r}{\Sigma}dt d\phi\nonumber\\
	&+\frac{\Sigma}{\Delta}dr^2+\Sigma d\theta^2 + \frac{[(r^2+a^2)^2-a^2\Delta \sin^2\theta] \sin^2\theta}{\Sigma} d\phi^2,\nonumber\\
	\end{align}
	where
	\bqn
	\lb{eq12}
	\Upsilon(r)&=&\frac{r (1-g(r))}{2},\\
	\lb{eq13}
	\Delta(r)&=&g(r)h(r)+a^2\nonumber\\
	&=&r^2-\frac{r^6}{(r^4+2 \lambda)}\left(\frac{2GM}{c^2r}-\frac{GQ^2}{4\pi\epsilon_0 c^4r^2}\right)+a^2. ~~~~~
	\eqn
	
	 The above metric is an effective metric describing a regular and rotating magnetic black hole (BH) solution with a Yang-Mills electromagnetic source in the non-minimal Einstein-Yang-Mills  theory. Metric~\eqref{metric} reduces to the Kerr--Newman BH with a magnetic charge instead of the electric charge if $\lambda=0$. Thus,  by continuity it is certainly an exact solution to the field equations (Eq.~(7) of~\cite{Balakin:2015gpq}) at least for small $\lambda$. As we shall see in the subsequent sections, it is also free from geometrical pathologies and satisfies the energy conditions outside the outer horizon. In addition, it   is  free of spacetime singularity, too, as its curvature and Kretschmann scalar invariants are all regular for $\lambda>0$,
	\begin{align}
	&R=\frac{8 \lambda  r^2 [\mathcal{Q}^2 (5 r^4-6 \lambda )+\mathcal{M} (-6 r^5+20 \lambda  r)]}{(r^2+a^2 \cos^2\theta ) (r^4+2 \lambda )^3},\\
	&R^{\alpha\beta\mu\nu}R_{\alpha\beta\mu\nu}=\frac{P(r,\cos^2\theta,\mathcal{M},\mathcal{Q}^2,a^2,\lambda)}{(r^2+a^2 \cos^2\theta )^6 (r^4+2 \lambda )^6},
	\end{align}
	where $P$ is a polynomial of its arguments and  {finite}, $\mathcal{M}=GM/c^2$ and $\mathcal{Q}^2=GQ^2/(4\pi\epsilon_0 c^4)$.
	
	\begin{figure*}[!htb]
		\includegraphics[width=8.2cm]{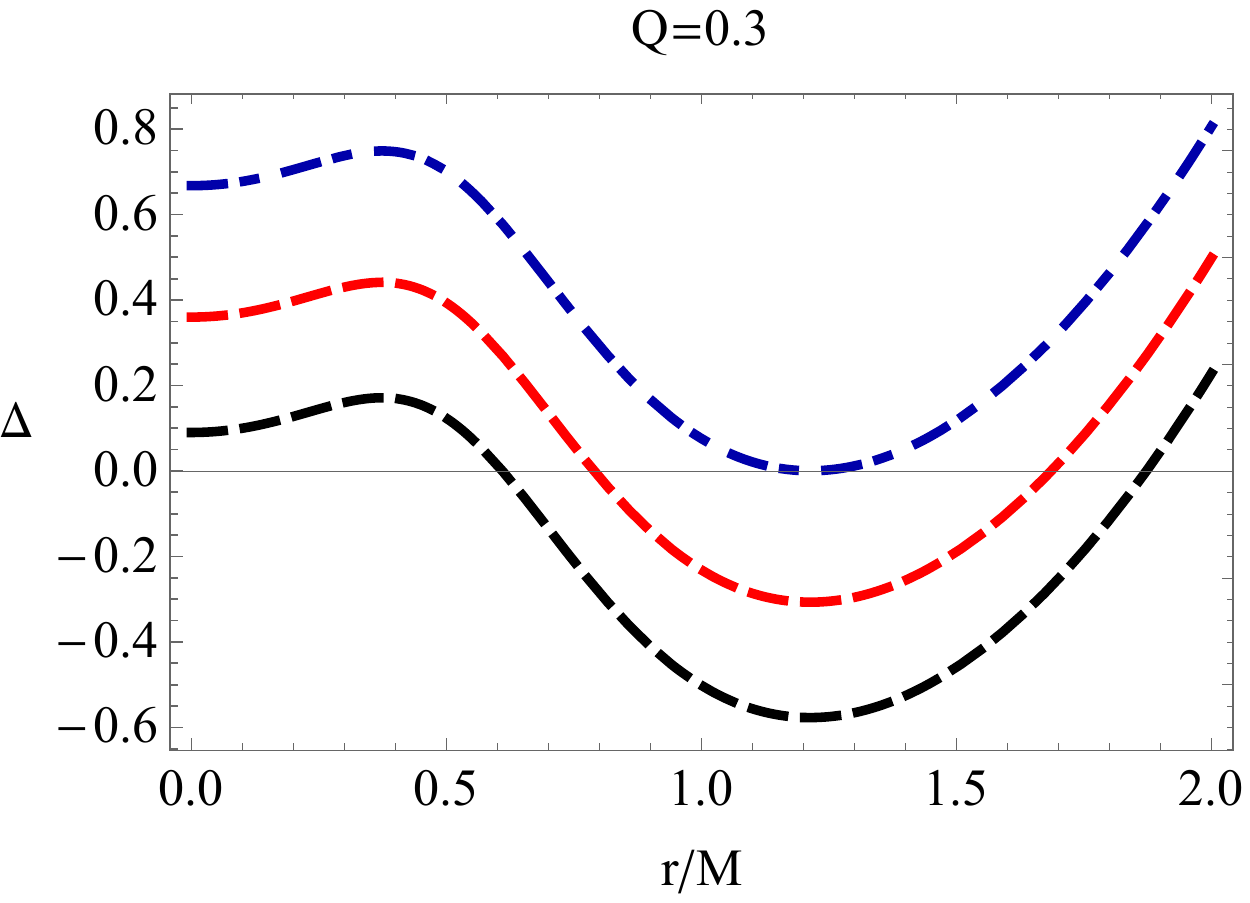}
		\includegraphics[width=8.2cm]{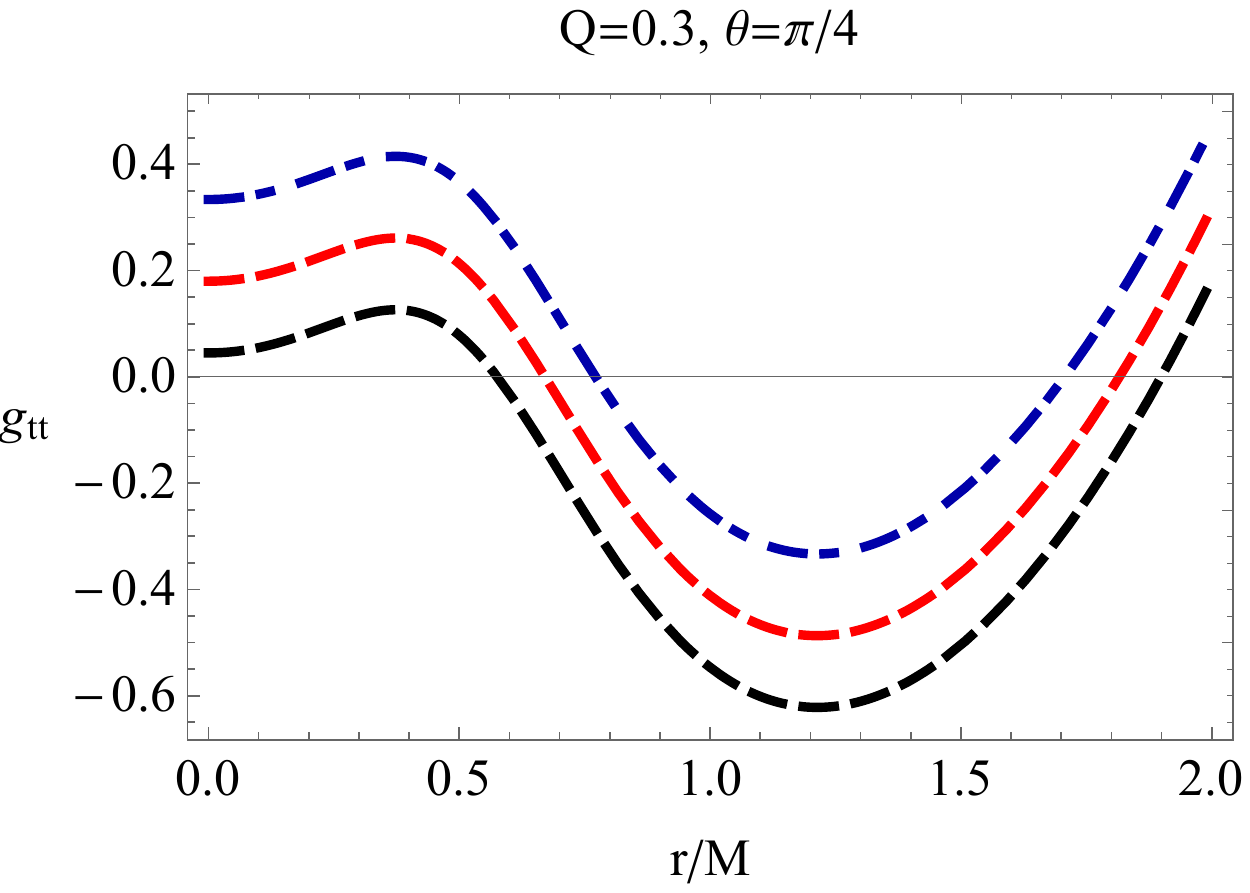}
		\caption{Right panel: Variation of $\Delta$ as a function of $r$. For $Q=0.3$, there is a critical value at $a=0.817$ such that the horizon disappears. Left panel: Variation of $g_{tt}$ as a function of $r$,
			with $\lambda=0.1$. We choose $a=0.3$ (black curve), $a=0.6$ (red curve),  $a=0.817$ (blue curve), in both plots. }
	\end{figure*}
	
	\begin{figure*}[!htb]
		\includegraphics[width=8.2cm]{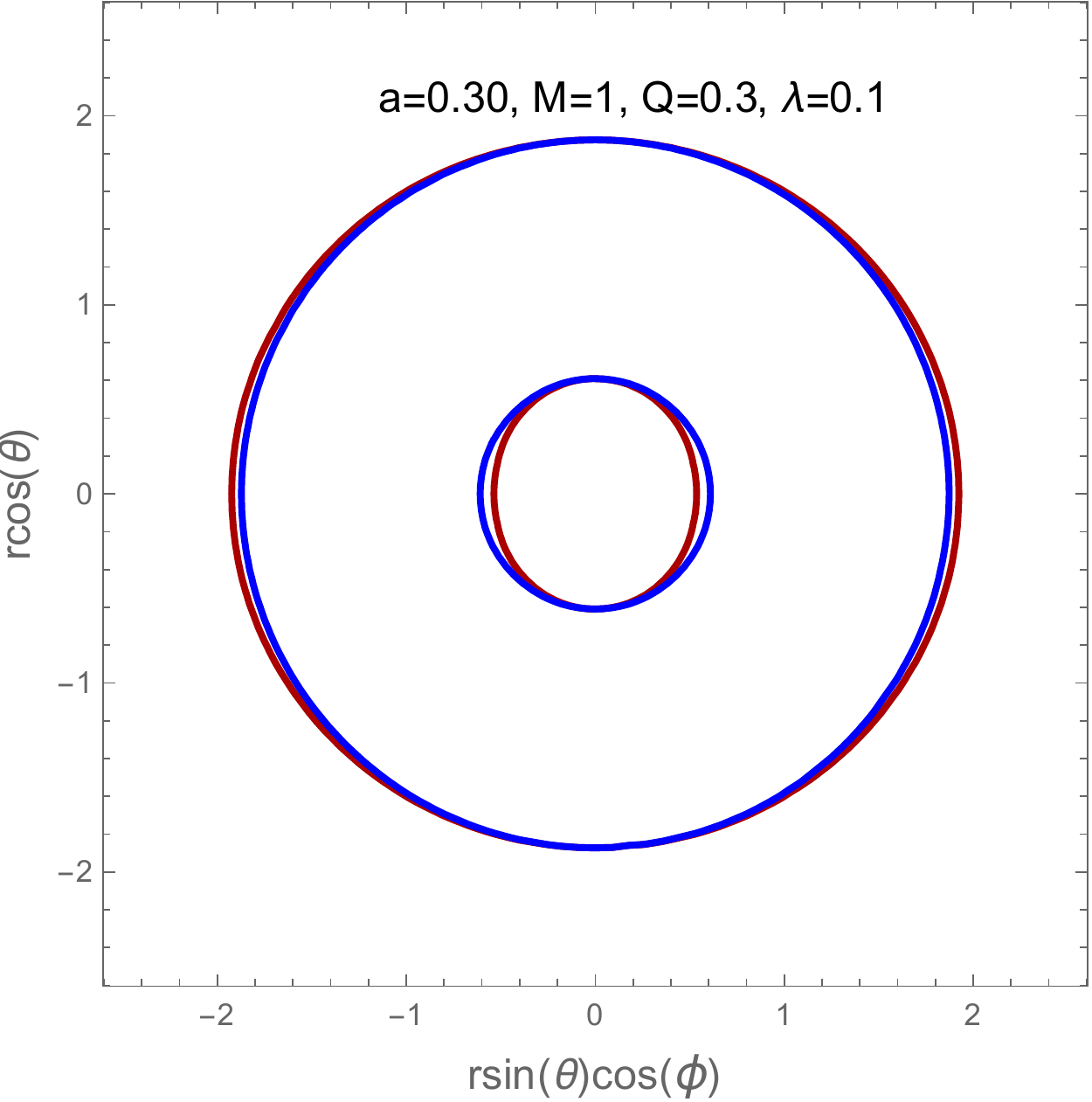}
		\includegraphics[width=8.2cm]{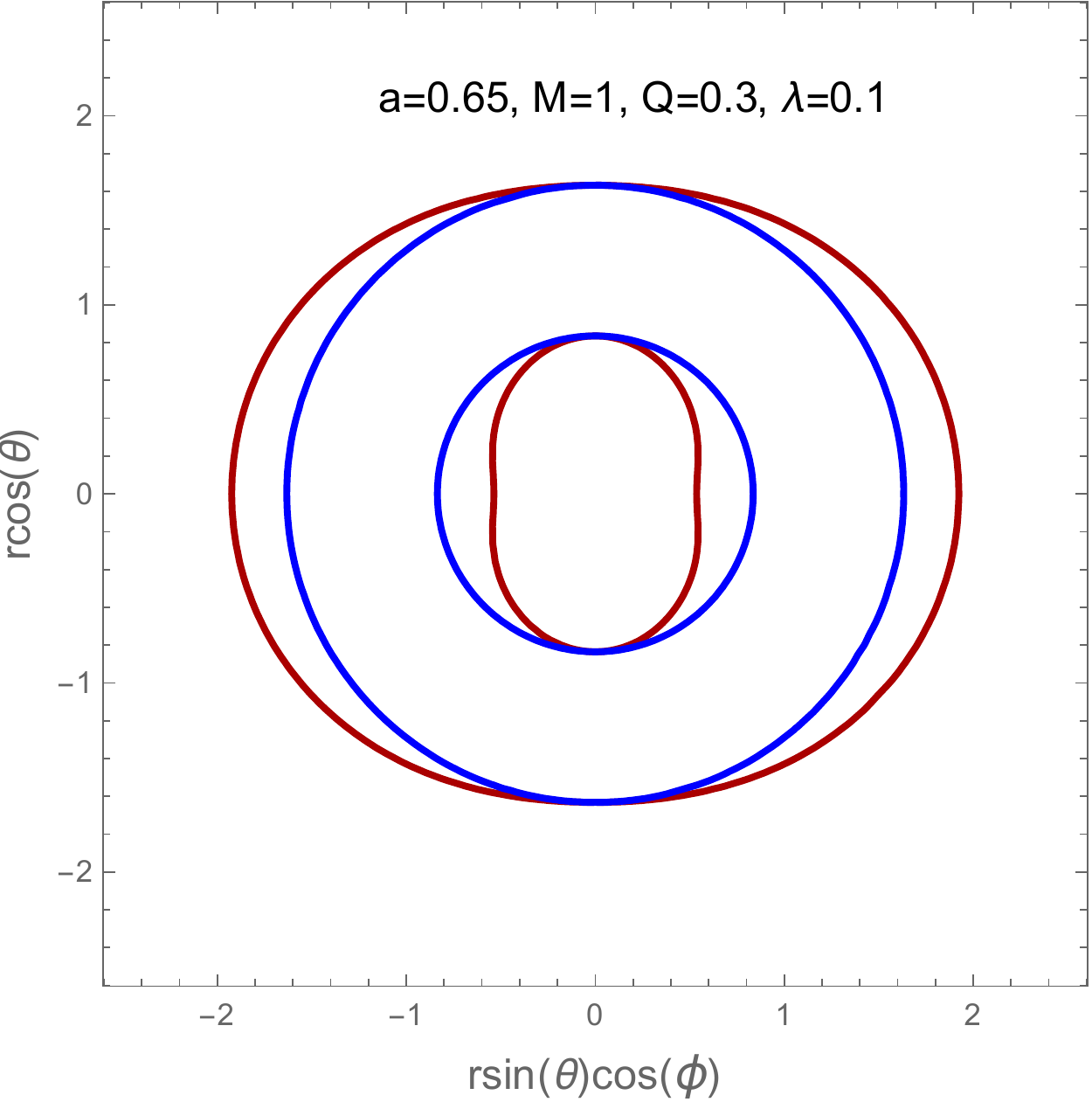}
		\includegraphics[width=8.2cm]{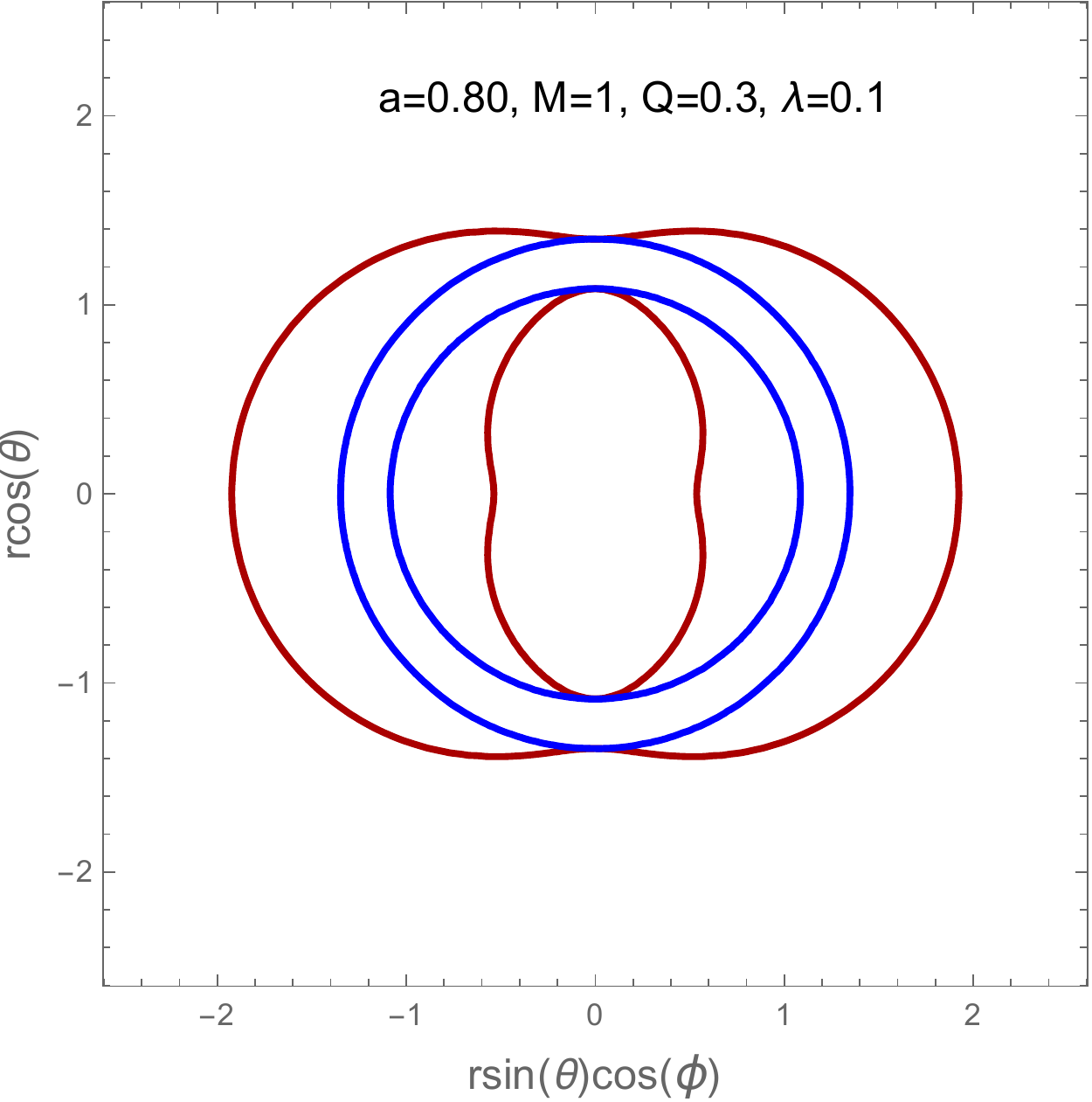}
		\includegraphics[width=8.2cm]{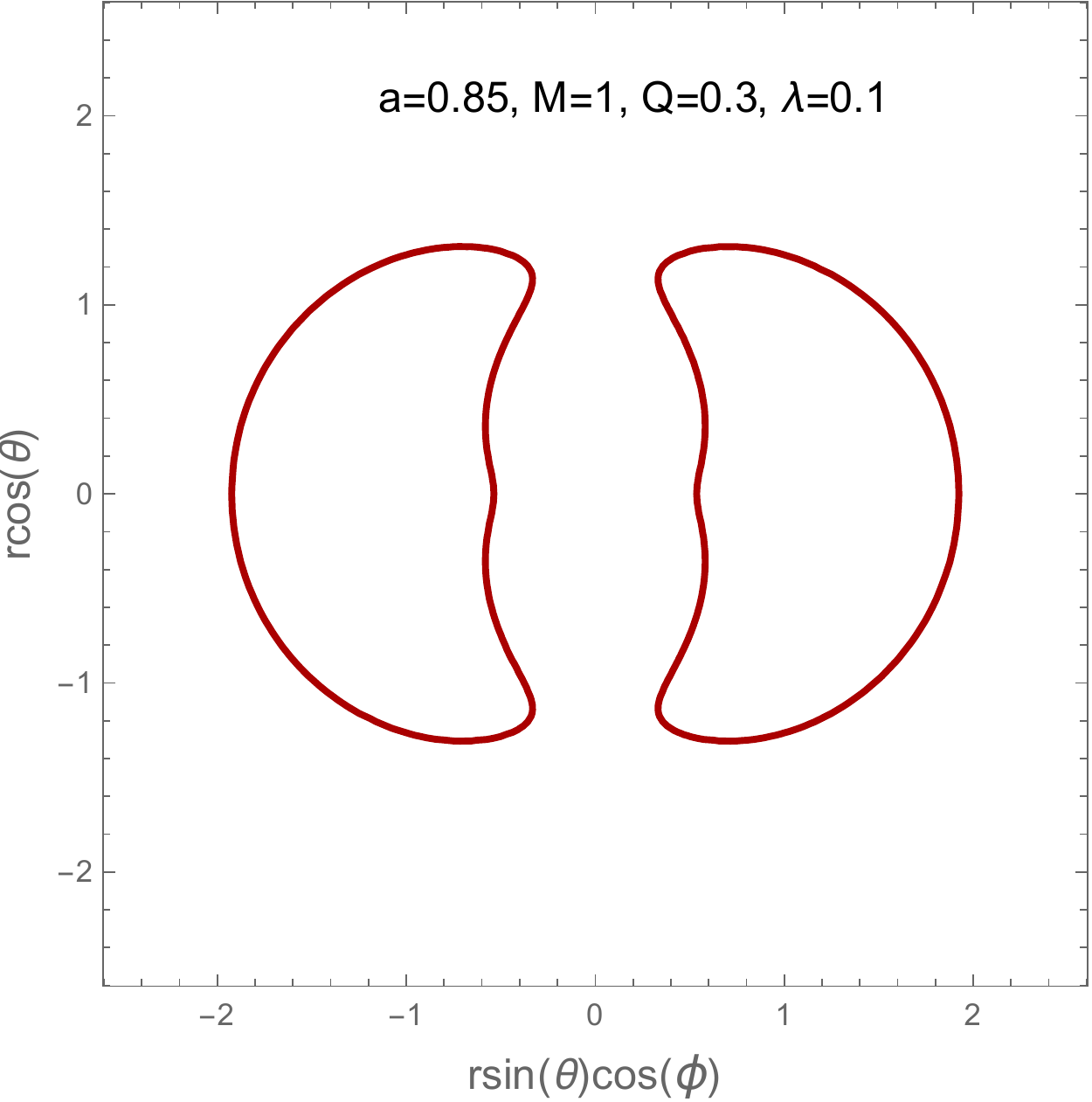}
		\caption{Surface horizon (blue color) and ergoregion (red color) of the BH for different values of $a$. For $Q=0.3$, and $a=0.817$ we find that the horizon disappears. On the other hand, the magnetic charge has strong effects on the ergoregion surface.} 
	\end{figure*}
	
	\begin{figure*}[!htb]
		\includegraphics[width=8.2cm]{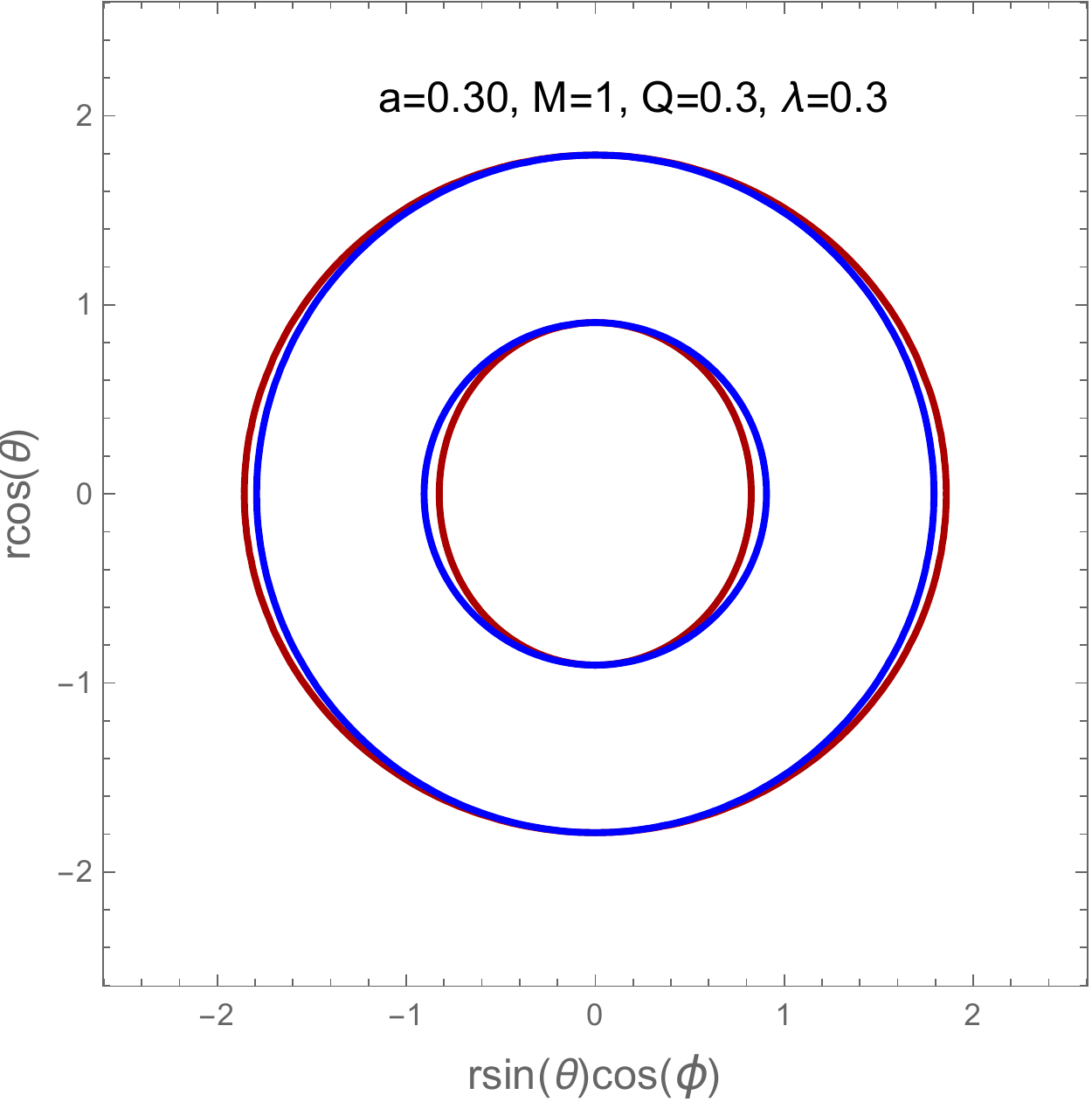}
		\includegraphics[width=8.2cm]{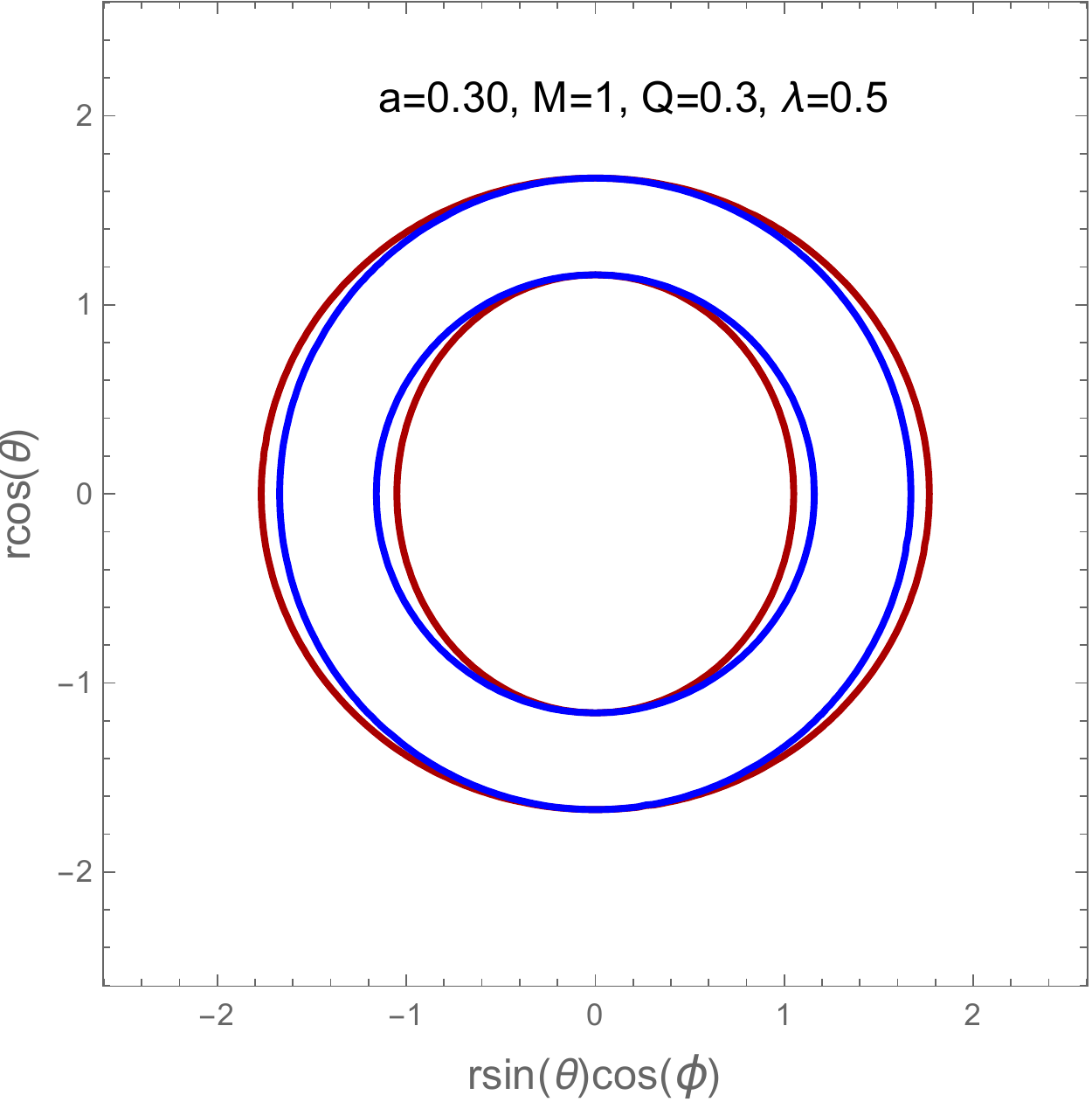}
		\includegraphics[width=8.2cm]{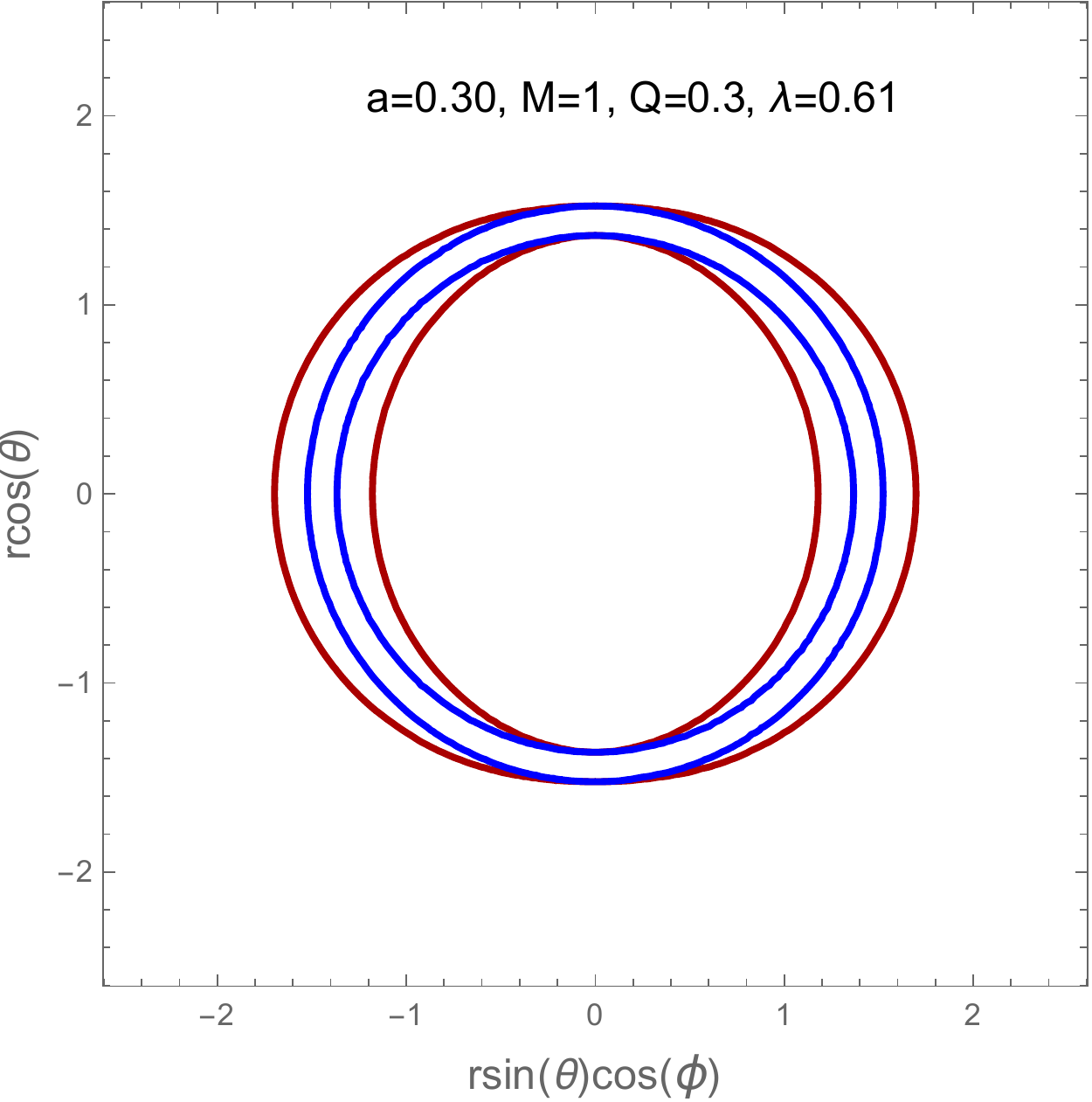}
		\includegraphics[width=8.2cm]{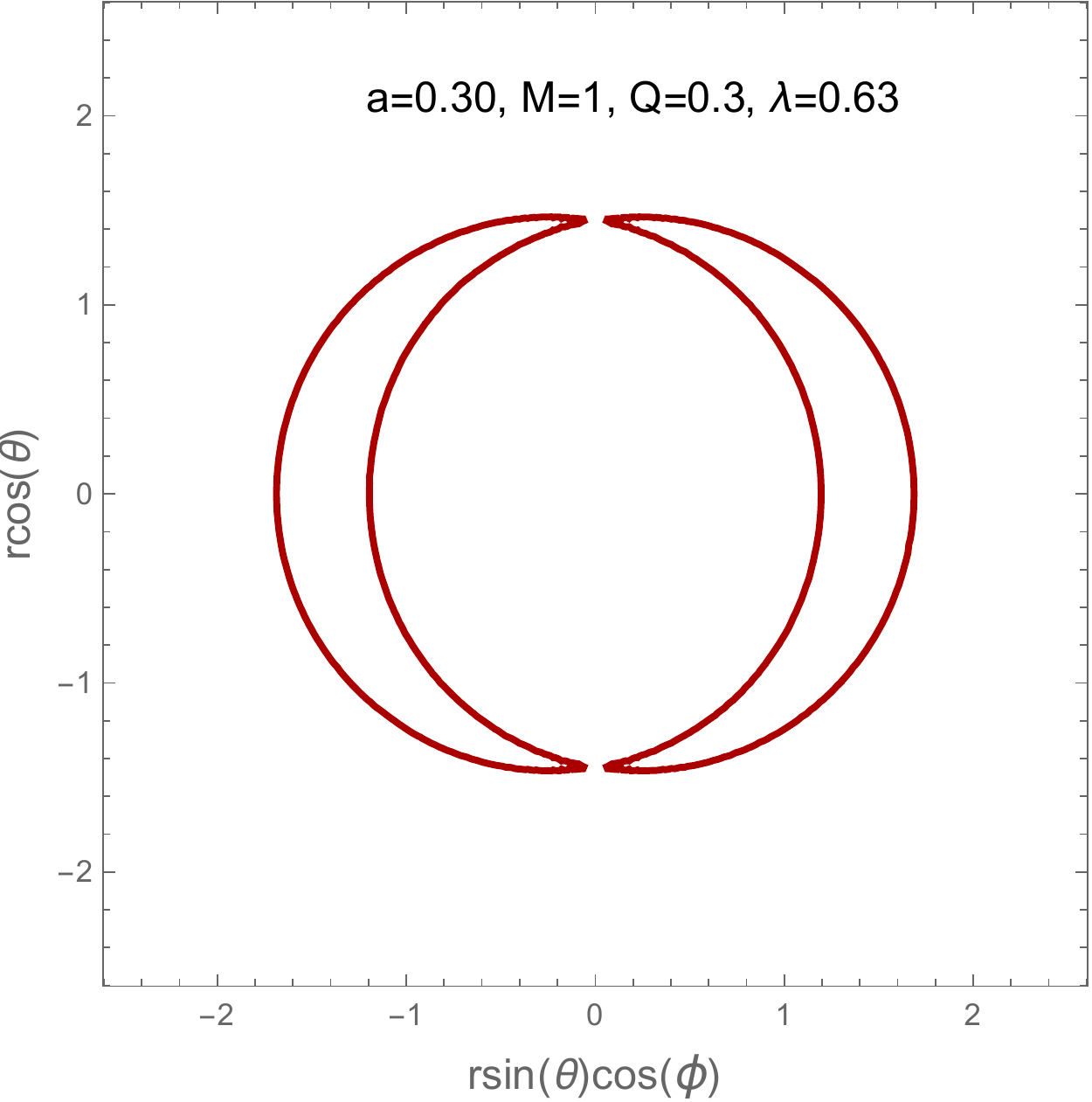}
		\caption{Surface horizon (blue color) and ergoregion (red color) of the BH for different values of  $\lambda$ and fixed $a$ and $Q$. For small values of $\lambda$ the black hole has two horizons but if we increase $\lambda$ an extremal black hole is obtained and the two horizons coincide. When $\lambda = 0.6225$ we find that the horizon disappears.  }
	\end{figure*}
	
%	\begin{figure*}[!htb]
%		\includegraphics[width=8.4cm]{embe1.pdf}
%		\includegraphics[width=8.4cm]{embe2.pdf}
%		\caption{The BH spacetime embedded
%			in a three-dimensional Euclidean space. Left panel: We  choose $M=1$, $a=0.1$, $Q=0.1$ and $\lambda=0.1$. Right panel: We choose $M=1$, $a=0.8$, $Q=0.3$ and $\lambda=0.1$  }
%	\end{figure*}
	
\begin{figure*}[!htb]
		\includegraphics[width=8.4cm]{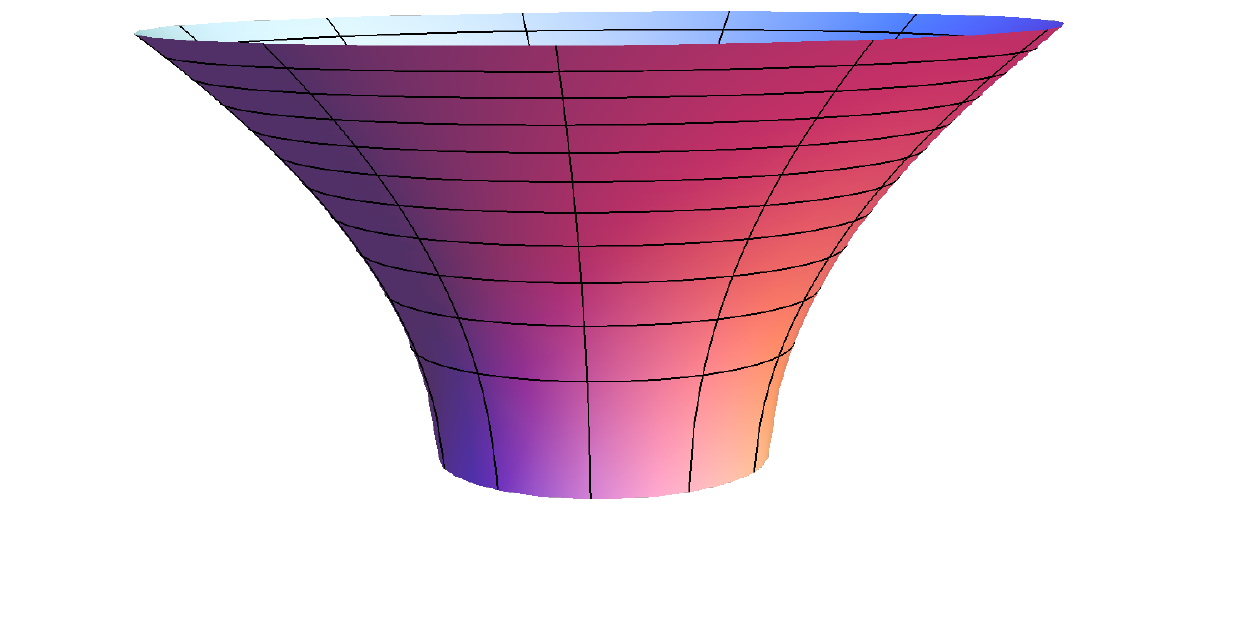}
		\includegraphics[width=8.4cm]{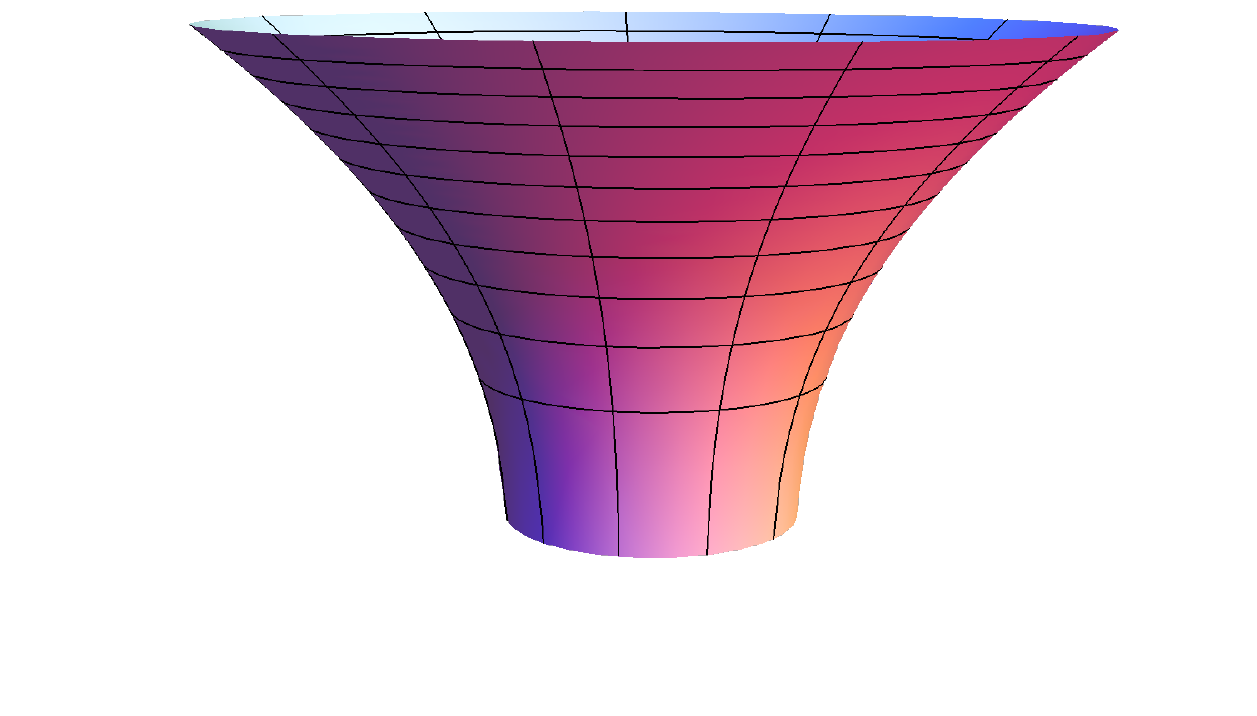}
		\caption{The BH spacetime embedded
			in a three-dimensional Euclidean space. Left panel: We  choose $a=0.3$ and $\lambda=0.1$. Right panel: We choose $a=0.3$ and $\lambda=0.5$. In both plots we have used $M=1$ along with $Q=0.3$.   }
\end{figure*}

\subsection{Effective energy-momentum tensor}

In the NJAAA the rotating solution is sourced by an anisotropic fluid the effective energy-momentum tensor $T_{\text{eff}}^{\mu\nu}$, which is diagonal in the orthonormal basis $(e_t,\,e_r,\,e_{\ta},\,e_{\phi})$ defined by~\cite{core,A1}
\begin{align}\label{eff1}
&e^{\mu}_t=\frac{(h+a^2,0,0,a)}{\sqrt{\Sigma \Delta}},& & e^{\mu}_r=\frac{\sqrt{\Delta}(0,1,0,0)}{\sqrt{\Sigma}},\nonumber\\
&e^{\mu}_{\theta}=\frac{(0,0,1,0)}{\sqrt{\Sigma}},& & e^{\mu}_{\phi}=\frac{(a\sin^2\theta,0,0,1)}{\sqrt{\Sigma}\sin\theta}.
\end{align}
Being given by the expression $T_{\text{eff}}^{\mu\nu}=\rho e^{\mu}_te^{\nu}_t+p_re^{\mu}_re^{\nu}_r+p_{\ta}e^{\mu}_{\ta}e^{\nu}_{\ta}+p_{\phi}e^{\mu}_{\phi}e^{\nu}_{\phi}$, where $\rho$ is the energy density and ($p_r,\,p_{\theta},\,p_{\phi}$) are the components of the pressure, it can be shown that the effective energy-momentum tensor has the following components (examples of detailed calculations can be found in~\cite{core,A1,s8,s9})
\begin{align}\label{eff2}
&\rho=-p_r=\frac{r^4 [Q^2 (r^4-6 \lambda )+16 M r \lambda ]}{(r^2+a^2 \cos ^2\theta )^2 (r^4+2 \lambda )^2},\nonumber\\
&p_\theta=p_\phi=\frac{r^2}{(r^2+a^2 \cos ^2\theta )^2 (r^4+2 \lambda )^3} \big\{8 M r \lambda  [r^2 (5 r^4-6 \lambda )\nonumber\\
&+(3
r^4-10 \lambda ) a^2 \cos ^2\theta ]+Q^2 [r^2 (r^8-24 \lambda  r^4+12 \lambda ^2)\nonumber\\
&-4 \lambda  (5 r^4-6 \lambda )
a^2 \cos ^2\theta ]\big\}. 
\end{align}
We can therefore say that rotating effective metric \eqref{metric} is a  solution to the field equation derived from the action given by Eq. (1) if all matter terms are replaced by the effective energy-momentum as defined in Eq. (17) and in Eq. (18). To have more elucidating expressions we seek their Taylor expansions about $\lambda =0$ in order to compare them with their counterparts of the Kerr--Newman black hole
\begin{align}\label{eff3}
&\rho=-p_r=\frac{Q^2}{(r^2+a^2 \cos ^2\theta )^2}-\frac{2 (5 Q^2-8 M r)}{r^4 (r^2+a^2 \cos ^2\theta )^2} \lambda +\cdots ,\nonumber\\
&p_\theta=p_\phi=\frac{Q^2}{(r^2+a^2 \cos ^2\theta )^2}\nonumber\\
&+\frac{1}{r^6
	(r^2+a^2 \cos ^2\theta )^2}~\big[8 M r (5 r^2+3 a^2 \cos ^2\theta )\nonumber\\
&-10 Q^2 (3 r^2+2 a^2 \cos ^2\theta )\big]\lambda +\cdots ,
\end{align}
where the first term in each expression corresponds to the Kerr–-Newman component of the energy-momentum tensor.  We see clearly that the corrections added to the Kerr--Newman counterparts can be neglected recalling that most observers are at large spatial distances from the sources. These corrections, proportional to $\lambda$, behave as the inverse of $r^7$ in the limit of large $r$ while the leading Kerr--Newman terms behave as the inverse of $r^4$.

We have discussed some relevant observable quantities and there remain some other observable quantities, mainly the usual electromagnetic fields and their Yang--Mills extensions. In the literature there are ansatze~\cite{5} for the general expressions of electromagnetic and Yang--Mills fields but no exact analytical solutions were found, see, for example, \cite{2,2b,2c} and references therein. The determination of the electromagnetic and Yang--Mills fields of the rotating black hole is more involved than the determination of the metric itself. This necessitates the resolution of coupled nonlinear differential equations and to the best of our knowledge only numerical solutions are available in the literature (see~\cite{2,2b,2c} and references therein). However,  for our current purpose, such solutions are not needed.

\subsection{Shape of the ergoregion}

	{After obtaining  the rotating BH solution \eqref{metric}, now let us turn to investigate its shape of the ergoregion.} Usually, one plots the shape of the ergoregion in the $xz$-plane. The corresponding horizons of  our BH can be found by solving the following equation $\Delta=0$,
	\begin{equation}
	\Delta=g(r)h(r)+a^2=0.
	\end{equation}
	{Meantime, the so-called static limit or ergo-surface, inner and outer,} is obtained via $g_{tt}=0$, i.e.,
	\begin{equation}
	r^2g(r)+a^2 \cos^2\theta=0.
	\end{equation}
	
	 From Fig. 1 we observe that in general for a given $Q$ and $\lambda$, one gets two horizons if $a<a_c$. However, when $a=a_c$ (the  blue line) the two horizons coincide, which means that we have an extremal BH with degenerate horizons.  It is interesting to note that going beyond this critical value, $a>a_c$, one can see that  event horizons no longer exist  and the solution represents a compact object without horizons and singularities at the center. Moreover, by varying the angular momentum parameter $a$ while having constant value of magnetic charge, say,  $Q=0.3$ and a constant value of the parameter $\lambda$, say, $\lambda=0.1$ one can see the effect of the magnetic charge on the surface horizon and ergoregion given in Fig. 2. For a given domain of parameters we find that at some for the angular momentum $a=0.817$, the horizon disappears. In Fig. 3 we depict  the effect of the magnetic charge on the black hole horizons and ergoregions by varying the parameter $\lambda$, wile having constant values of $a$ and $Q$. It is shown that there is a domain of parameters and a critical value of $\lambda_c$ such that the two horizons coincide, and for $\lambda > \lambda_c$ the horizons disappear.
	
	\section{Embedding Diagram\label{secem}}
	
	{In this section, we investigate the geometry of the BH spacetime, by embedding it into a higher-dimensional Euclidean space. To this purpose, let us consider
		the  equatorial plane $\theta=\pi/2$ at a fixed moment  $t=$ Constant, for which the metric can be written as }
	\begin{equation}
	ds^2=\frac{dr^2}{1-\frac{b(r)}{r}}+\mathcal{R}^2d\phi^2, \label{emb}
	\end{equation}
	where
	\bqn
	\lb{eq19}
	b(r)&=& \frac{ r^5}{(r^4+2 \lambda)}\left(\frac{2M}{r}-\frac{Q^2}{r^2}\right) -\frac{a^2}{r},\\
	\lb{eq20}
	\mathcal{R}(r)&=&\left[r^2+a^2+ \frac{a^2 r^4}{(r^4+2 \lambda)}\left(\frac{2M}{r}-\frac{Q^2}{r^2}\right) \right]^{1/2}. ~~~~
	\eqn
	{Let us embed this reduced  BH metric into three-dimensional Euclidean space  in the cylindrical coordinates,
		\begin{eqnarray}\notag
		ds^2&=&dz^2+d\mathcal{R}^2+\mathcal{R}^2d\phi^2\\
		&=& \left[ \left(\frac{d\mathcal{R}}{dr}\right)^2+\left(\frac{dz}{dr}\right)^2  \right]dr^2+\mathcal{R}^2d\phi^2.
		\lb{eq21}
		\end{eqnarray}
		From Eqs.(\ref{emb}) and (\ref{eq21}), } we find that
	\begin{equation}
	\frac{dz}{dr}=\pm \sqrt{\frac{r}{r-b(r)}-\left(\frac{d\mathcal{R}}{dr}\right)^2},
	\end{equation}
	where $b(r)$ is given by Eq.(\ref{eq19}). Note that the integration of the last expression cannot
	be accomplished analytically. Invoking numerical techniques allows us to illustrate the embedding diagrams given in Fig. 4. It is seen that by varying the parameter $\lambda$ the black hole geometry is significantly changed.

%	\begin{figure*}[!htb]
%		\includegraphics[width=7.6 cm]{ec1.pdf}
%		\includegraphics[width=8.7cm]{ec2.pdf}
%		\includegraphics[width=7.6 cm]{ec3.pdf}
%		\includegraphics[width=8.7cm]{ec4.pdf}
%		\caption{Top left panel: Plot of $\rho$ with $Q=0.3$ and $a=0.5$. Top right %panel: Plot of $\rho+2p_r+2p$ with $Q=0.3$ and $a=0.5$. Bottom left panel: Plot of $\rho$ with $Q=0.6$ and $a=0.5$. Bottom Right panel: Plot of $\rho+2p_r+2p$ with  $Q=0.6$ and $a=0.5$. Note that we have used $x=\cos\theta$ and set $M=1$ along with  $\lambda=0.1$ in all the plots. }
%	\end{figure*}
	
	\begin{figure*}[!htb]
		\includegraphics[width=7.6 cm]{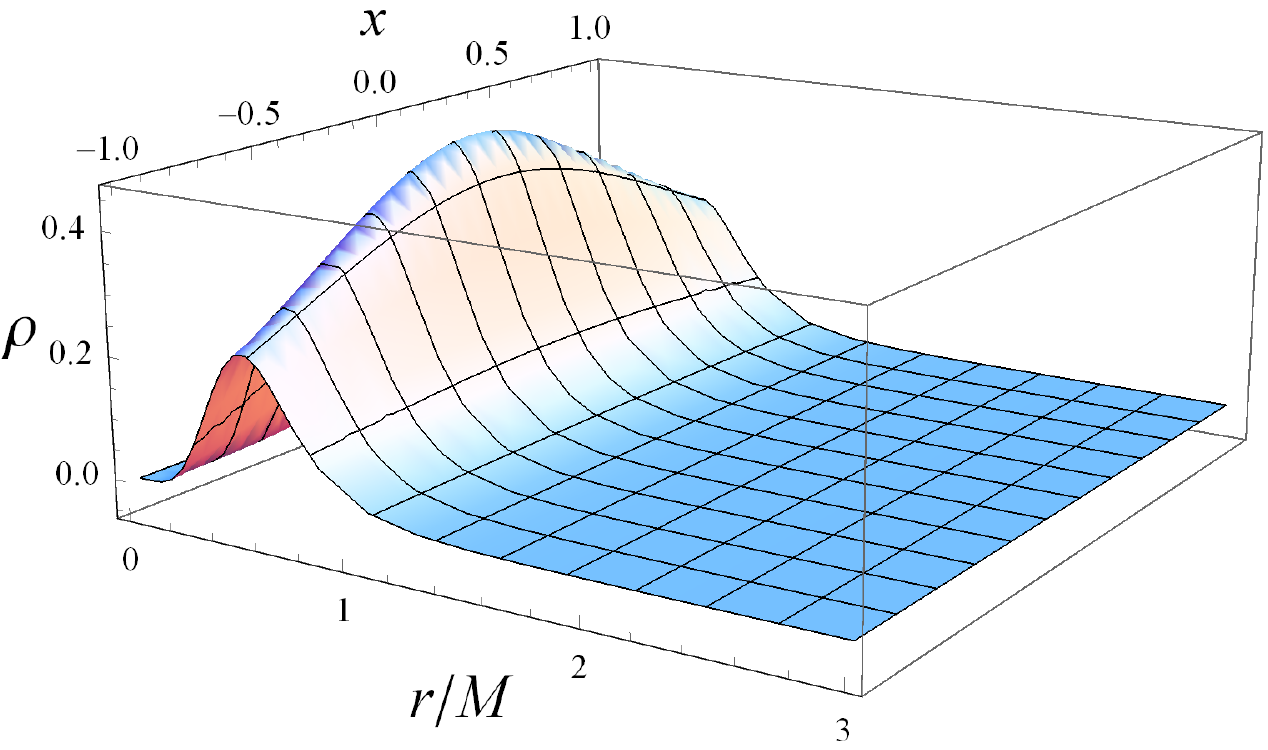}
		\includegraphics[width=8.7cm]{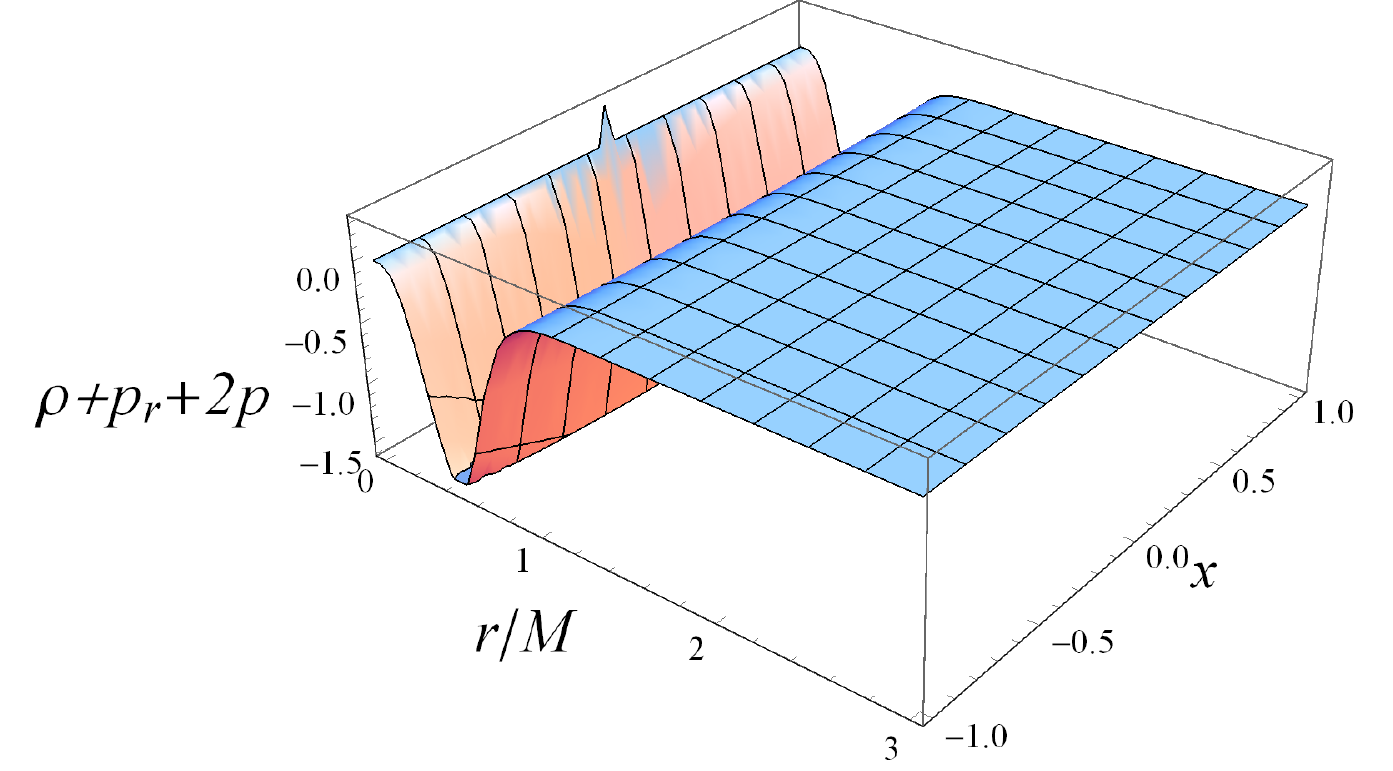}
		\includegraphics[width=7.6 cm]{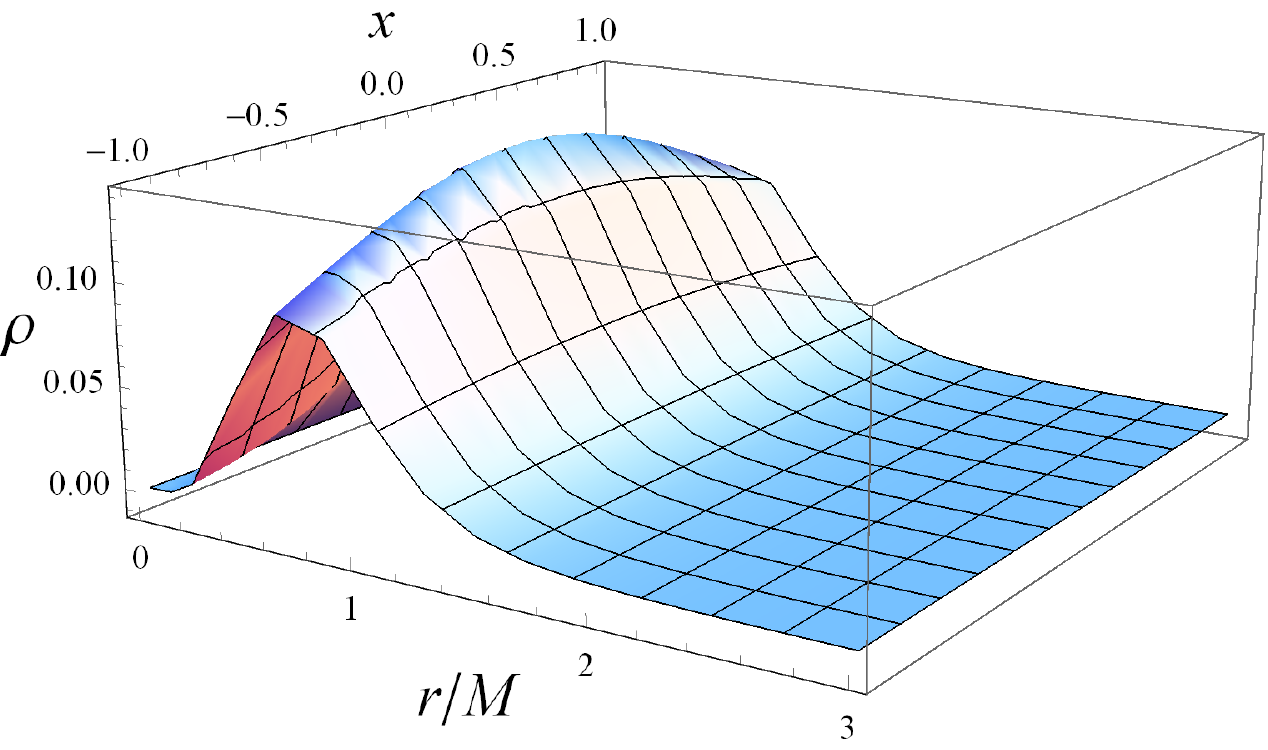}
		\includegraphics[width=8.7cm]{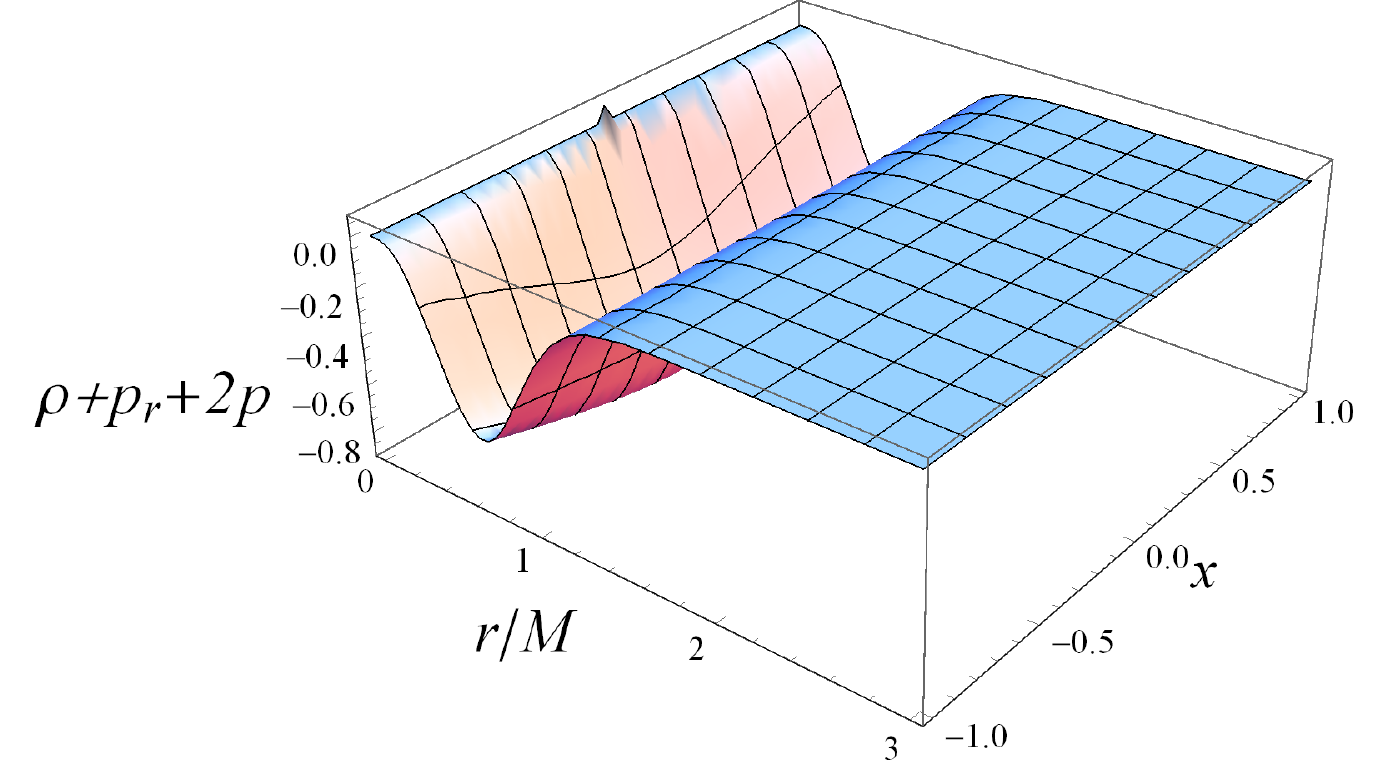}
		\caption{Top left panel: Plot of $\rho$ with $\lambda=0.1$ and $a=0.3$. Top right panel: Plot of $\rho+2p_r+2p$ with $\lambda=0.1$ and $a=0.3$. Bottom left panel: Plot of $\rho$ with $\lambda=0.5$ and $a=0.3$. Bottom Right panel: Plot of $\rho+2p_r+2p$ with  $\lambda=0.5$ and $a=0.3$. Note that we have used $x=\cos\theta$ and set $M=1$ along with  $Q=0.3$ in all the plots. }
	\end{figure*}

	\section{Energy Conditions\label{secec}}
	
	In this section we are going to explore the energy conditions for the rotating EYM BH. For this purpose we use the Einstein field equations $G_{\mu\nu}=8\pi T_{\mu\nu}$ along with the effective energy-momentum tensor represented by a properly chosen tetrad of vectors given by $T^{\mu \nu}=e^{\mu}_{a}e^{\nu}_{b}T^{ab}$, where $T^{ab}=(\rho,p_r,p_{\theta},p_{\phi})$. In terms of the orthogonal basis, the
	{non-vanishing components of the energy momentum tensor} are given as follows~\cite{core},
	\begin{eqnarray}\nonumber
	\rho &=&\frac{1}{8\pi}\emph{e}^\mu_t\,\emph{e}^\nu_t \,\emph{G}_{\mu\nu},\quad
	p_r =\frac{1}{8\pi}\emph{e}^\mu_r\,\emph{e}^\nu_r \,\emph{G}_{\mu\nu},\\\label{m1}
	p_\theta &=&\frac{1}{8\pi}\emph{e}^\mu_\theta\,\emph{e}^\nu_\theta \,\emph{G}_{\mu\nu},\quad
	p_\phi =\frac{1}{8\pi}\emph{e}^\mu_\phi\, \emph{e}^\nu_\phi\, \emph{G}_{\mu\nu}.
	\lb{eq23}
	\end{eqnarray}
	The Einstein tensor $\emph{G}_{\mu\nu}$ is given in Appendix A.
	%{It can be shown that the orthogonal bases for our rotating BH are  given by \cite{Azreg-Ainou:2014pra,A1,core},}
%	\begin{eqnarray}\label{basis}
%	{\emph{e}}^\mu_t&=&\frac{1}{\sqrt{\Sigma \Delta}}\left(r^2+a^2,0,0,a\right),\quad
%	\emph{e}^\mu_r=\frac{\sqrt{\Delta}}{\sqrt{\Sigma}}\left(0,1,0,0\right), \nonumber\\
%	\emph{e}^\mu_\theta&=&\frac{1}{\sqrt{\Sigma}}\left(0,0,1,0\right),\quad
%	\emph{e}^\mu_\phi=\frac{1}{\sqrt{\Sigma} \sin\theta}\left(a \sin^2\theta,0,0,1\right), \nonumber\\
%	\end{eqnarray}
	Using the orthogonal bases given by Eq. (16) the corresponding physical quantities defined in Eq.(\ref{eq23}) now read,
	\begin{eqnarray}\notag
	\rho &=& -p_r = \frac{2\Upsilon'(r)r^2}{8 \pi \Sigma^2}, \\
	p_{\theta}&=&p_{\phi}=p_r-\frac{\Upsilon''(r)r+2\Upsilon'(r)}{8 \pi \Sigma}.
	\end{eqnarray}
	
	 In Fig. 5 by varying the parameter $\lambda$  we plot out the quantities $\rho$ and $\rho+p_r+2p$, where $p=p_{\theta}=p_{\phi}$, for a given values of $(Q, a, M)$, from which we find that
		%\begin{equation}
		% \rho \geq 0,
		%\end{equation}
		%%and SEC energy conditions
		%\begin{equation}
		%\rho+p_r+2p \geq 0,
		%\end{equation}
		%hold.  From Fig. 4 we observe
		the strong energy condition (SEC) is not satisfied. In other words,  the matter supporting this configuration is exotic, although we note that the cosmological constant does not satisfy the
		SEC either.
	
	\section{Shadow of the Rotating BHs\label{secs}}
	
	In order to find the contour of a BH shadow,  we need to separate the null geodesic equations in the general rotating spacetime \eqref{metric}, by using the Hamilton-Jacobi equation given by
	\begin{equation}
	\frac{\partial \mathcal{S}}{\partial \sigma}=-\frac{1}{2}g^{\mu\nu}\frac{\partial \mathcal{S}}{\partial x^\mu}\frac{\partial \mathcal{S}}{\partial x^\nu},
	\label{eq:HJE}
	\end{equation}
	where $\sigma$ is the affine parameter, $\mathcal{S}$ is the Jacobi action.   In order to find a separable solution we  express the action in terms of the known constants of the motion as follows
	\begin{equation}
	\mathcal{S}=\frac{1}{2}\mu ^2 \sigma - E t + J \phi + \mathcal{S}_{r}(r)+\mathcal{S}_{\theta}(\theta),
	\label{eq:action_ansatz}
	\end{equation}
	where $\mu$ is the mass of the test particle,  $E=-p_t$ the conserved energy, and  $J=p_\phi$ the  conserved angular momentum (with respect to the symmetry axis). For a photon, we have $\mu=0$. From these equations it is straightforward to obtain the following equations of motion (see for instance~\cite{Azreg-Ainou:2014pra}),
	\begin{eqnarray}\notag
	\Sigma\frac{dt}{d\sigma}&=&\frac{r^2+a^2}{\Delta}[E(r^2+a^2)-aJ]-a(aE\sin^2\theta-J),\\\notag
	\Sigma\frac{d\phi}{d\sigma}&=&\frac{a}{\Delta}[E(r^2+a^2)-aJ]-\left(aE-\frac{J}{\sin^2\theta}\right),\\\notag
	\Sigma \frac{dr}{d\sigma}&=&\pm \sqrt{\mathfrak{R}(r)},\\
	\Sigma \frac{d\theta}{d\sigma}&=&\pm \sqrt{\Theta(\theta)},
	\end{eqnarray}
	{where
		\bqn
		\mathfrak{R}(r)&=&\left[X(r)E-aJ\right]^2-\Delta(r)\left[\mathcal{K}+\left(J-aE\right)^2\right],\\
		\Theta(\theta)&=& \mathcal{K}+a^2E^2\cos^2\theta-J^2\cot^2\theta.
		\eqn
		with  $X(r)=(r^2+a^2)$. The function} $\Delta(r)$ is defined by Eq.(\ref{eq13}), while $\mathcal{K}$ is the Carter separation constant. If we define the following two quantities  $\xi=J/E$ and $\eta=\mathcal{K}/E^2$, and make use the fact the the unstable circular photon orbits in the general rotating spacetime must satisfy $\mathfrak{R}(r)(r_{ph})=0$, $\mathfrak{R}(r)'(r_{ph})=0$ and $\mathfrak{R}(r)''\geq 0$, we obtain (see,
	for example \cite{Shaikh:2019fpu})
	\bqn
	\label{eq:Req0}
	&& \left[X(r_{ph})-a\xi\right]^2-\Delta(r_{ph})\left[\eta+\left(\xi-a\right)^2\right]=0,\\
	&& 2X'(r_{ph})\left[X(r_{ph})-a\xi\right]-\Delta'(r_{ph})\left[\eta+\left(\xi-a\right)^2\right] = 0,\nb\\
	\label{eq:Rpeq0}
	\eqn
	where  $r=r_{ph}$ is the radius of the unstable photon orbit. Furthermore, if we eliminate $\eta$ from the last two equations and then solve for $\xi$, we find  that \cite{Shaikh:2019fpu}
	\begin{equation}
	\xi=\frac{X_{ph}\Delta'_{ph}-2\Delta_{ph}X'_{ph}}{a\Delta'_{ph}},
	\label{eq:xi}
	\end{equation}
	\begin{equation}
	\eta=\frac{4a^2X'^2_{ph}\Delta_{ph}-\left[\left(X_{ph}-a^2\right)\Delta'_{ph}-2X'_{ph}\Delta_{ph} \right]^2}{a^2\Delta'^2_{ph}},
	\label{eq:eta}
	\end{equation}
	where we note that the subscript ``$ph$" indicates that the quantities are evaluated at $r=r_{ph}$. Equations (\ref{eq:xi}) and (\ref{eq:eta}) give the general expressions for the critical impact parameters $\xi$ and $\eta$ of the unstable photon orbits,  which describe the contour of the shadow.
	
	\begin{figure*}[!htb]
		\includegraphics[width=8.2cm]{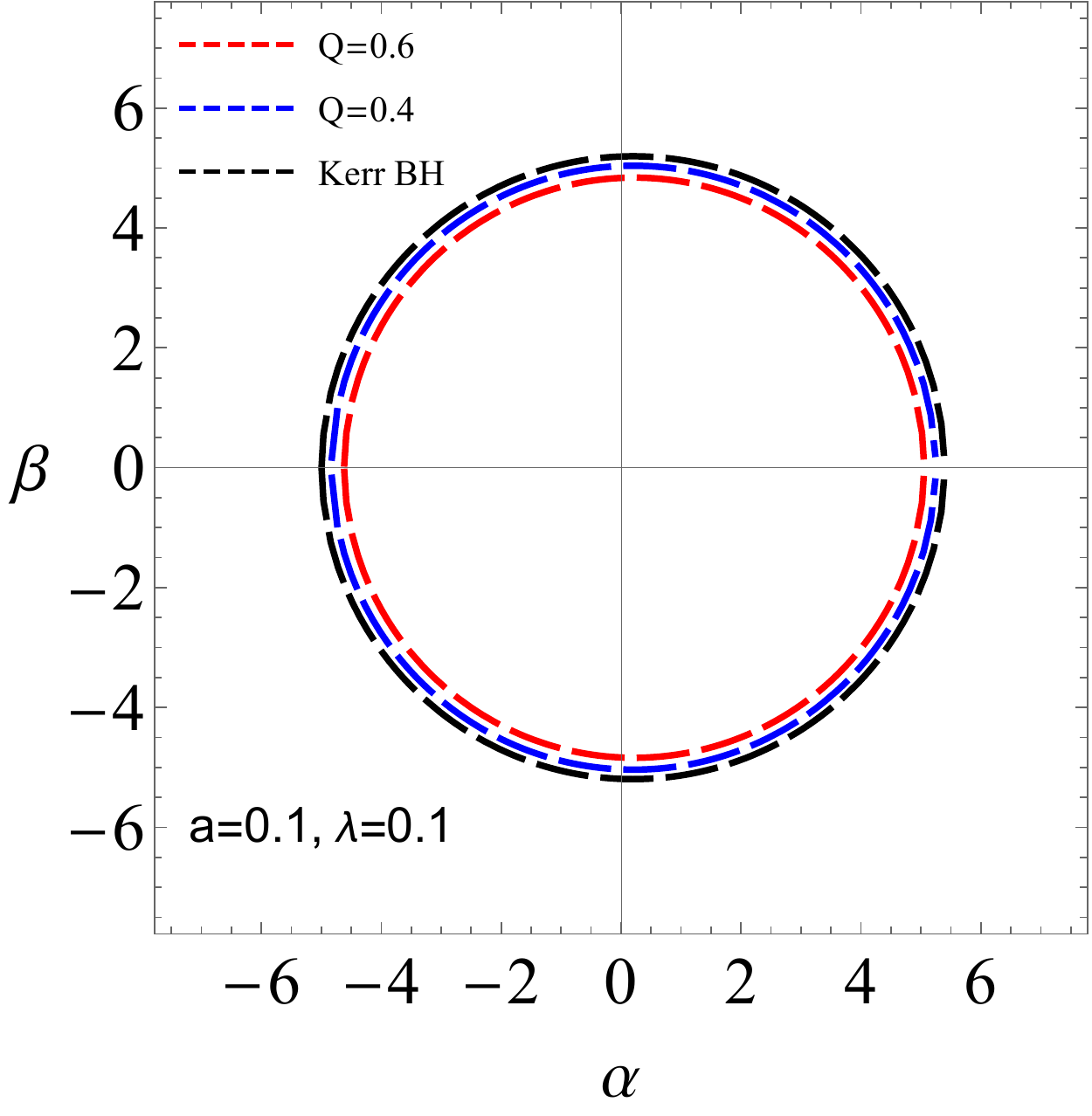}
		\includegraphics[width=8.2cm]{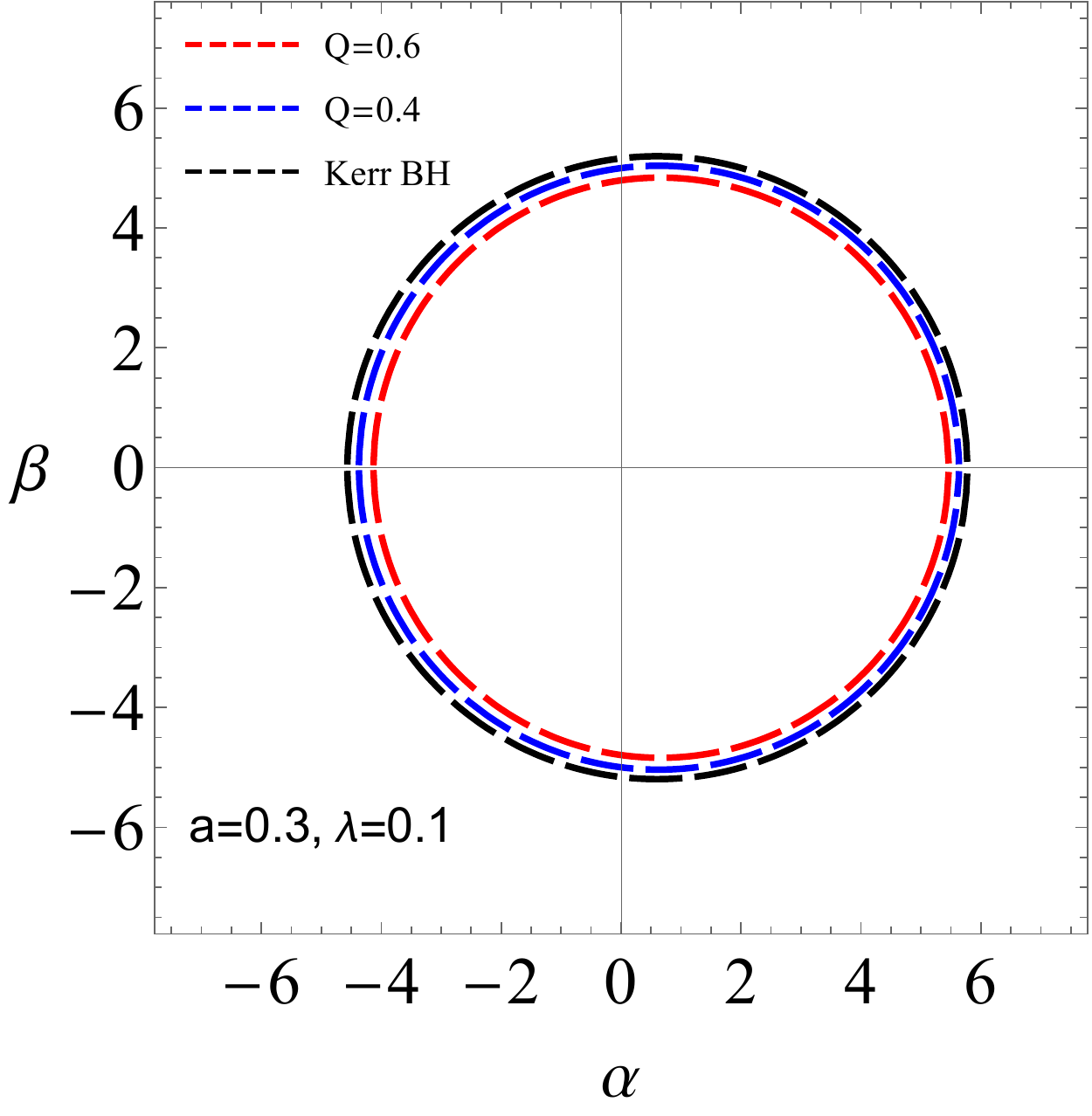}
		\includegraphics[width=8.2cm]{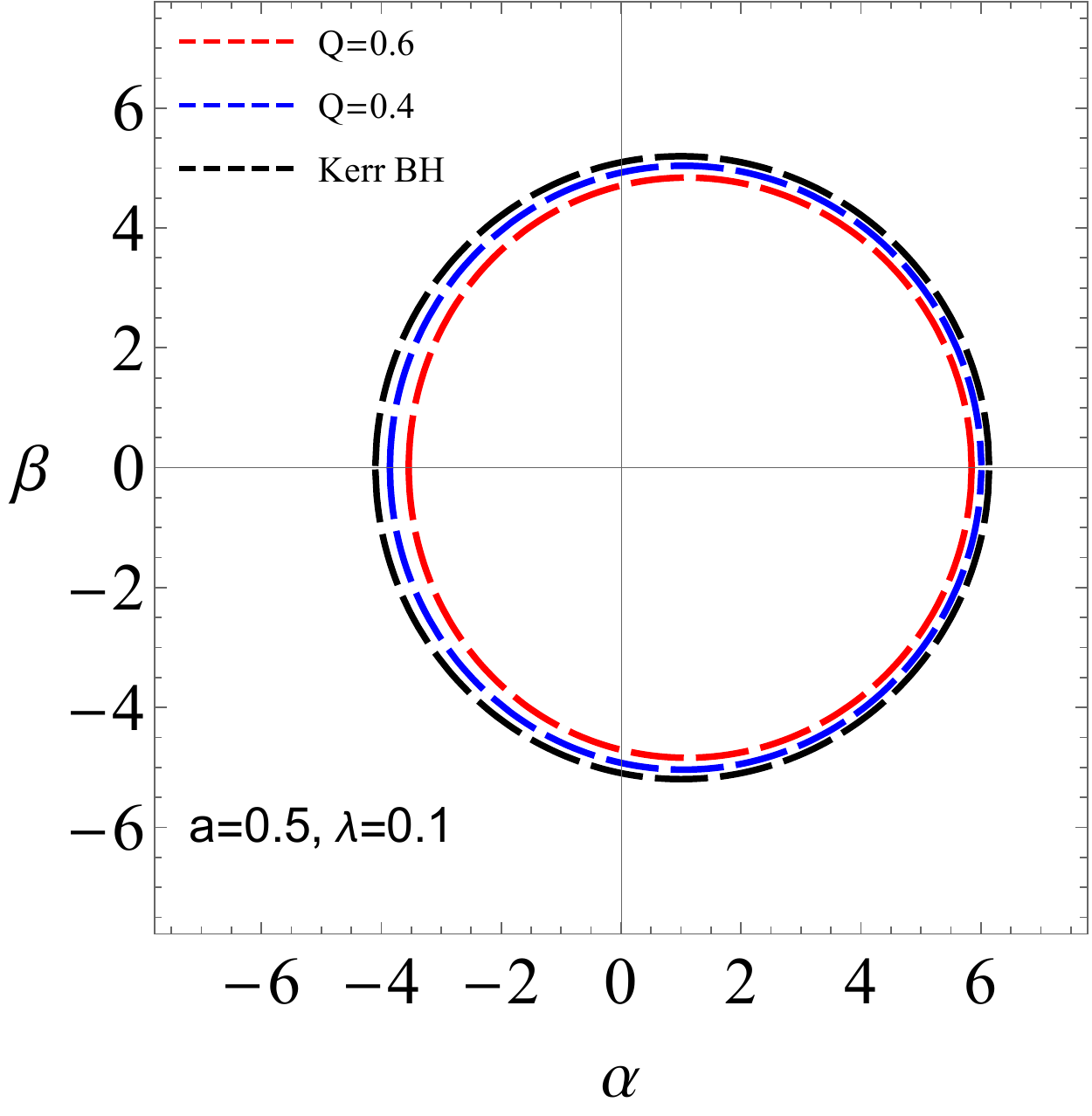}
		\includegraphics[width=8.2cm]{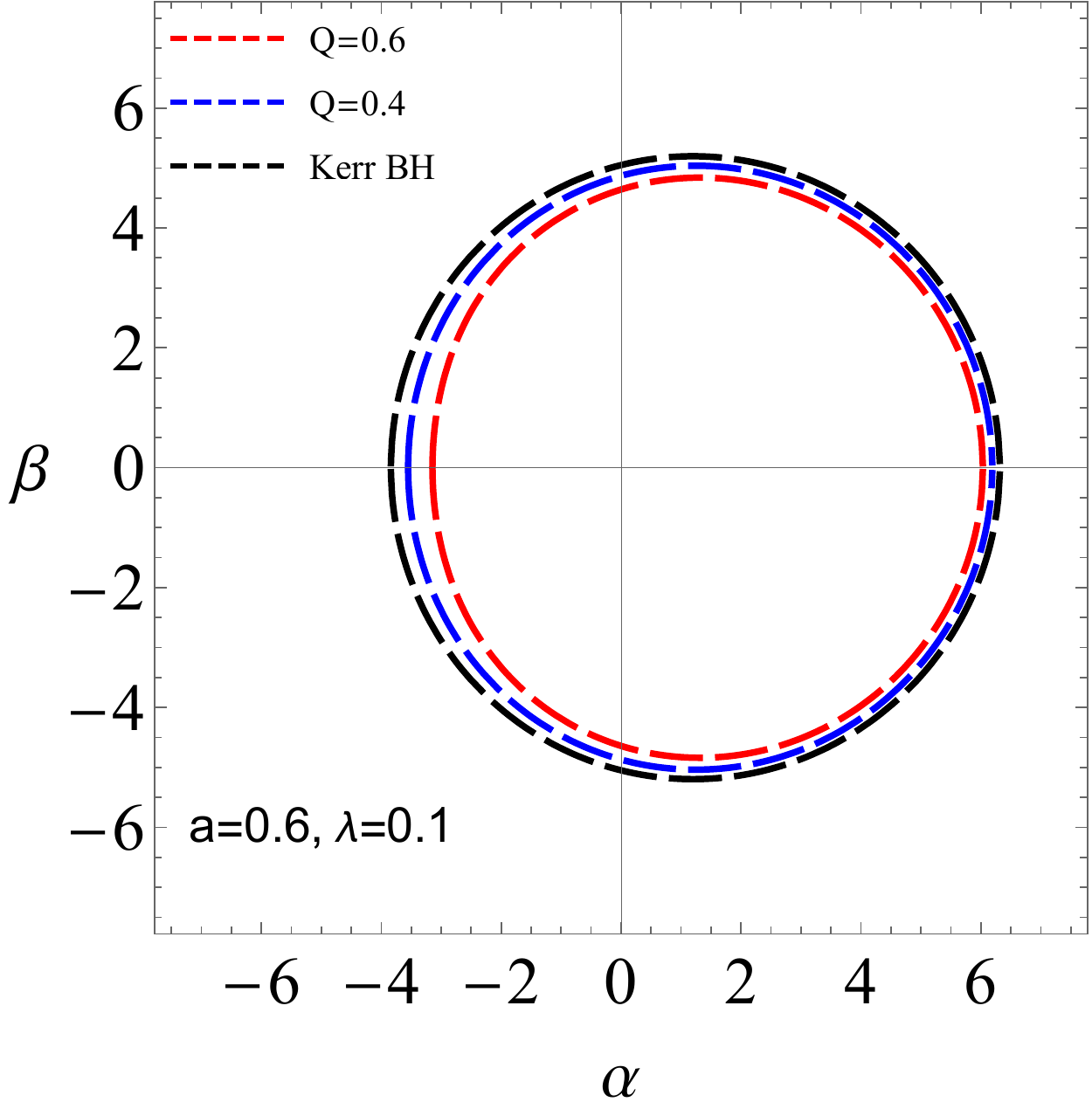}
		\caption{Variation in the shape of the shadow of the rotating BH described by the metric (\ref{metric}) for different values of $a$ and $Q$. }
	\end{figure*}
	
	\begin{figure*}[!htb]
		\includegraphics[width=8.2cm]{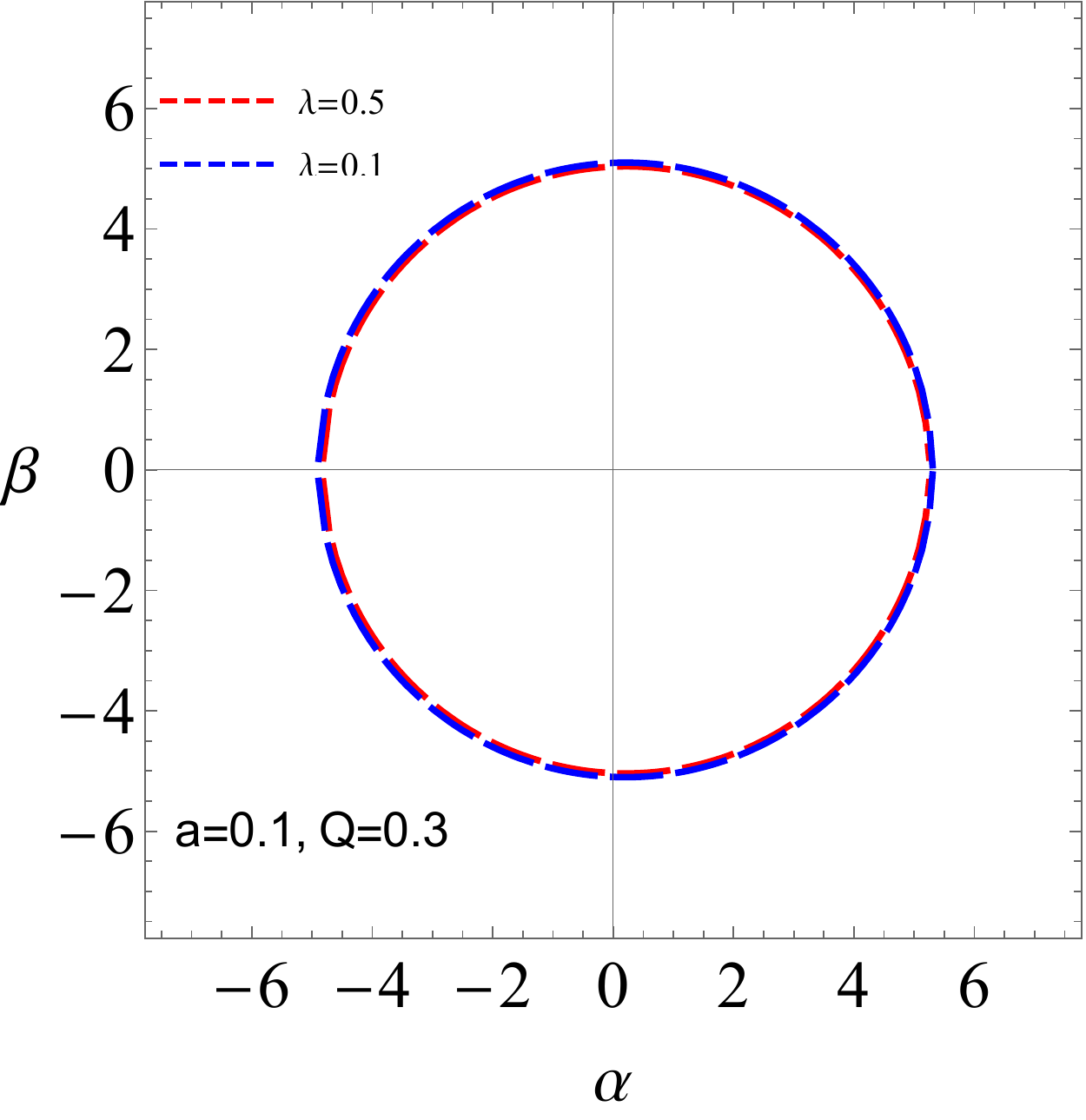}
		\includegraphics[width=8.2cm]{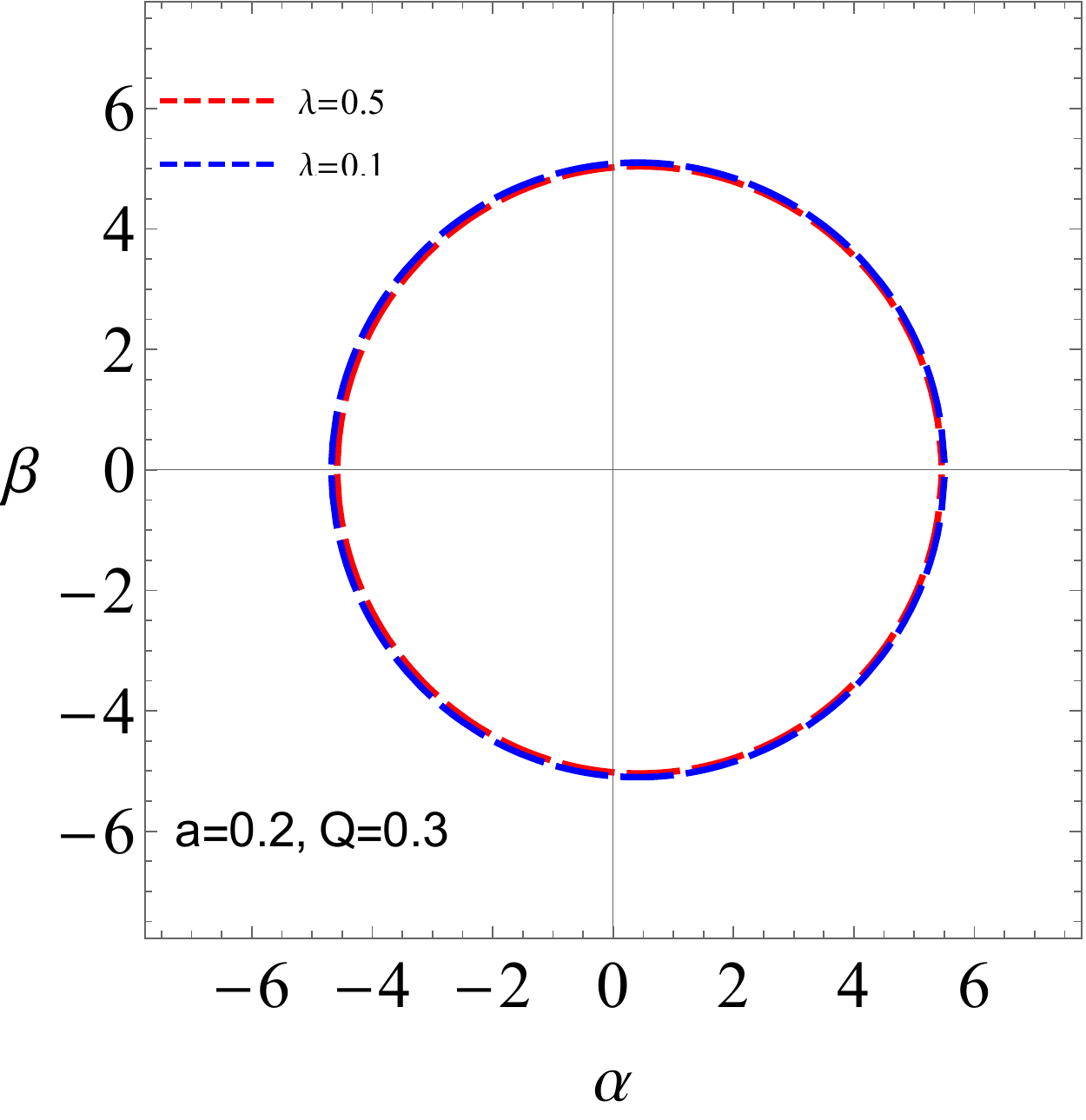}
		\includegraphics[width=8.2cm]{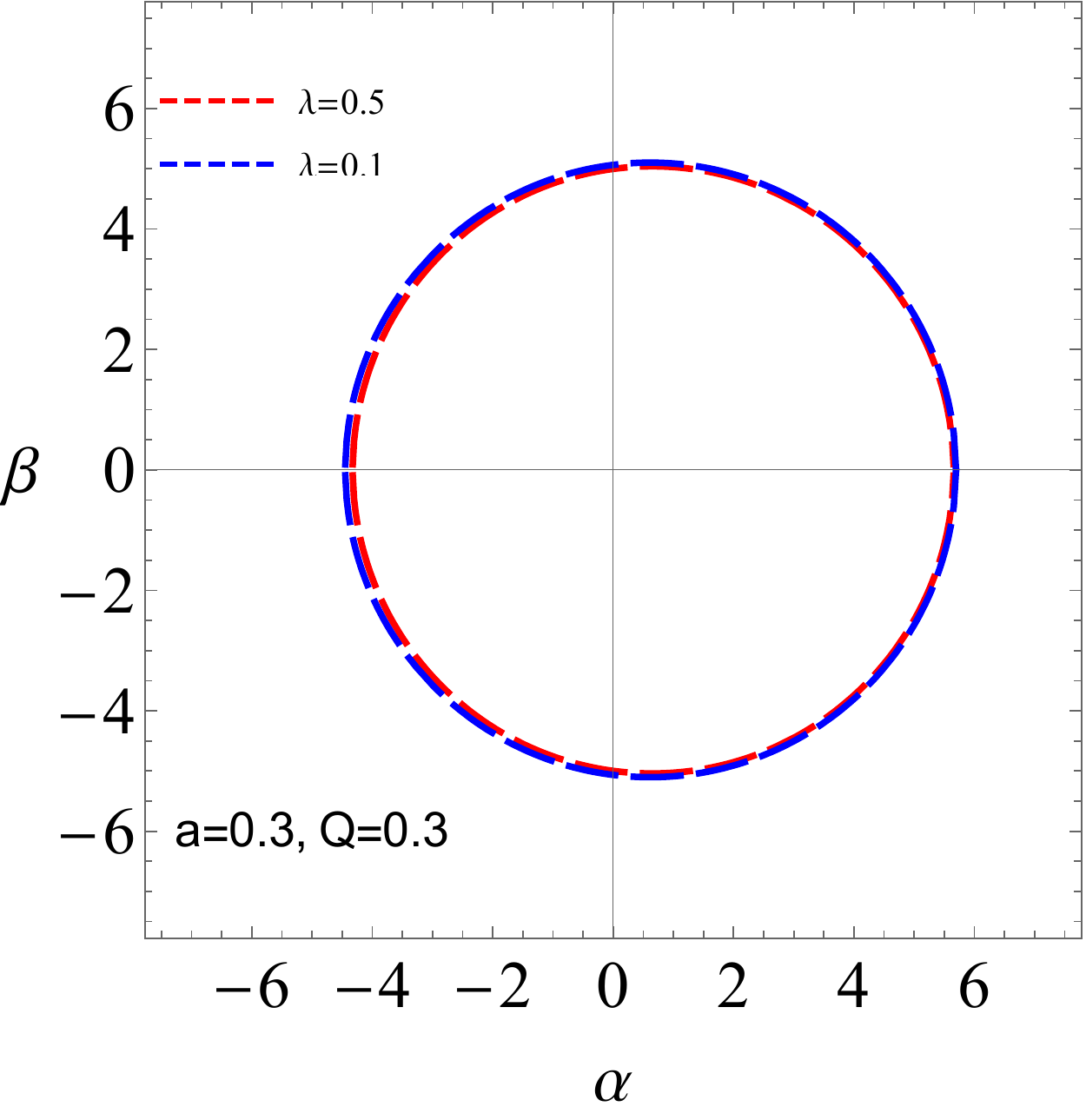}
		\includegraphics[width=8.2cm]{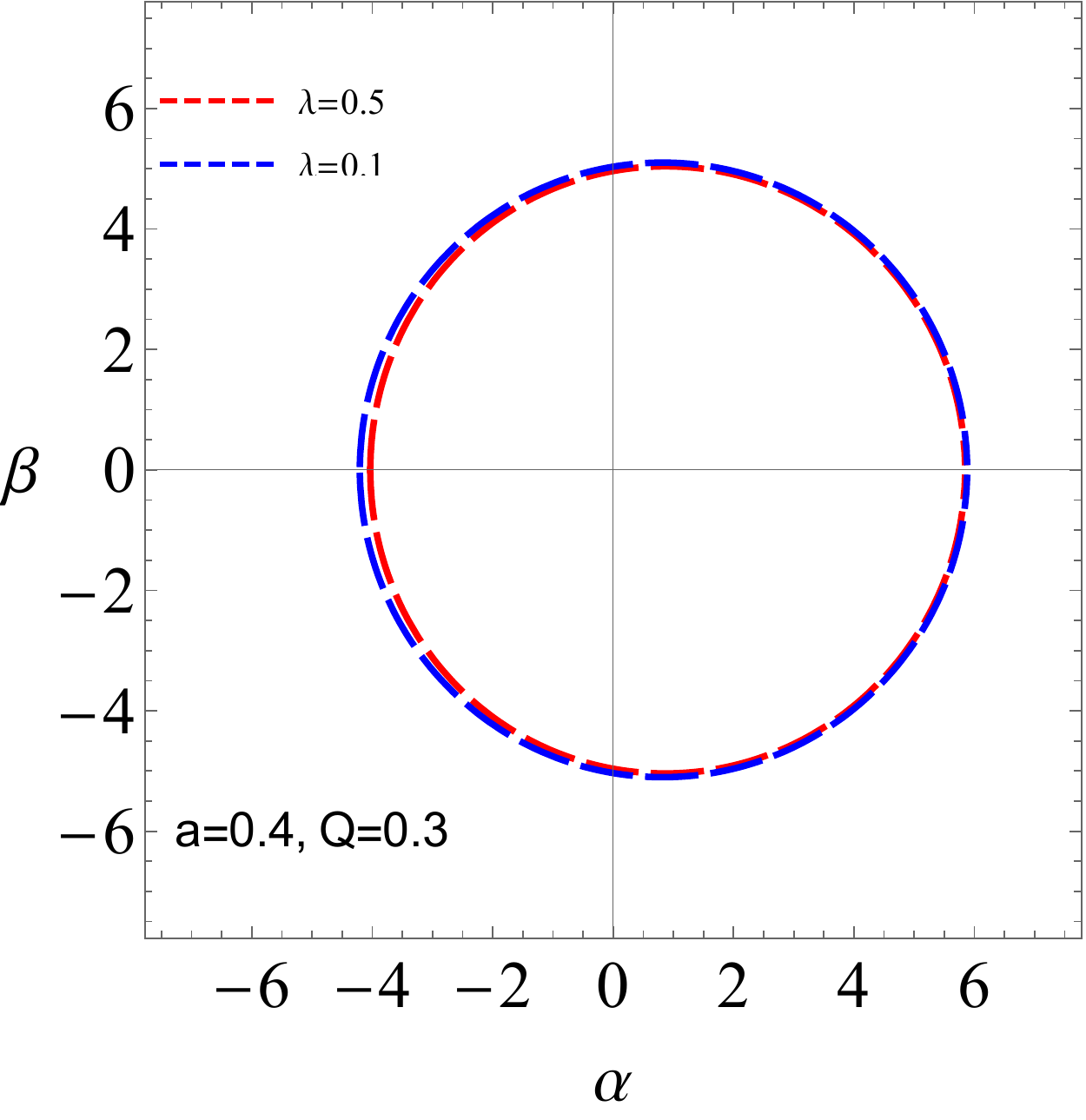}
		\caption{Variation in the shape of the shadow of the rotating BH described by the metric (\ref{metric}) for different values of $a$ and $\lambda$. }
	\end{figure*}

	The unstable photon orbits form the boundary of the shadow. The apparent shape of the shadow is obtained by using the celestial coordinates $\alpha$ and $\beta$,
	which lie in the celestial plane perpendicular to the line joining the observer and the center of the spacetime geometry. The coordinates $\alpha$ and $\beta$ are defined by
	\begin{equation}
	\alpha=\lim_{r_0\to\infty}\left(-r_0^2\sin\theta_0\frac{d\phi}{dr}\Big\vert_{(r_0,\theta_0)}\right),
	\end{equation}
	\begin{equation}
	\beta=\lim_{r_0\to\infty}\left(r_0^2\frac{d\theta}{dr}\Big\vert_{(r_0,\theta_0)}\right),
	\end{equation}
	where $(r_0,\theta_0)$ are the position coordinates of the observer.  After taking the limit, we obtain
	\begin{equation}
	\alpha=-\frac{\xi}{\sin\theta_0},
	\label{eq:alpha}
	\end{equation}
	\begin{equation}
	\beta=\pm \sqrt{\eta+a^2\cos^2\theta_0-\xi^2\cot^2\theta_0}.
	\label{eq:beta}
	\end{equation}
	\begin{table}
\begin{tabular}{|l|l|l|l|l|}
			\hline
			$\lambda$ [M]$^4$   &  $\bar{R}_s^{EYM}$ [M] & $ \theta_s^{EYM}$ [$\mu$as] & $ \Delta\theta_s$ [$\mu$as]    \\ \hline
			0.1 & 5.134496817 &  39.21861310 & 0.11175138 \\
			0.2 & 5.119287867  & 39.10244321 & 0.22792127  \\
			0.3 & 5.103426195 & 38.98128766   & 0.34907682  \\
			0.4 & 5.086818521  & 38.85443395  & 0.47593053 \\
			0.5 &  5.069344363 & 38.72096183 & 0.60940265 \\\hline
		\end{tabular}
		\caption{Shadow radius of EYM black hole for different values of $\lambda$ when viewed from the equatorial plane. In all these cases we have set $M=1$, $a=0.2$ and $Q=0.2$. Note that we have defined $\Delta\theta_s=\theta_s^{KN}-\theta_s^{EYM}$.  For the Kerr-Newman black hole in terms of the above parameters  we have the typical shadow radius $\bar{R}_s^{KN}=5.149127296$ and an angular diameter $\theta_s^{KN}=39.33036448 \mu$as which corresponds to the case of $\lambda=0$.}
	\end{table}
	
	The  shadow is constructed by using the unstable photon orbit radius $r_{ph}$ as a parameter and then plotting out $\alpha$ and $\beta$ using Eqs. (\ref{eq:xi}), (\ref{eq:eta}), (\ref{eq:alpha}) and (\ref{eq:beta}).  In Fig. (6) we show the effect of the magnetic charge by varying $Q$ and a given values of $(a,M,\lambda)$. It is observed that the black hole shadow radius decreases with the increase of $Q$. On the other hand, in Fig. (7) we show the effect of the magnetic charge by varying $\lambda$ and a given values of $(a,M,Q)$. It is observed that the black hole shadow radius monotonically decreases with the increase of $\lambda$, although the effect is very small compared to Fig. (6). Thus, for any $Q>0$ and $\lambda>0$, we see that the shadow radius is smaller compared to the Kerr--Newman black hole with a magnetic charge. As we see from Fig. 7, the effect of magnetic charge on the shadow radius
is very small when we increase $\lambda$ and, as a result, the EYM black hole is hard to be distinguishable from the Kerr-Newman black hole based on their shadows. The small effect of $\lambda$ can be understood from the fact that if we consider a Teylor expansion of $g(r)$ around $\lambda$, we obtain [working in natural units]
\begin{equation}
g(r)=1-\frac{2M}{r}+\frac{Q^2}{r^2}+\frac{2(2Mr-Q^2)\lambda}{r^6}+\hdots
\end{equation}
thus the leading correction term behave as the inverse of $r^6$.
Despite the fact that the effect of $\lambda$ is small, we are going to elaborate more on the possibility of distinguishing a rotating Kerr-Newman black hole with magnetic charge from a EYM black hole based on the physical observable such as the shadow radius $R_s$ and the angular diameter $\theta_s = 2R_s M/D$, where $M$ is the black hole mass and $D$ is the distance between the black hole and the observer. 
Our aim is to compute the shadow radius, however in general the shape of the shadow depends on the observer's viewing angle $\theta_0$. In the present work, we are going to use an expression for the typical shadow radius of rotating black holes obtained by Jusufi \cite{Jusufi:2020dhz}
\begin{equation}
\bar{R}_s=\frac{\sqrt{2}}{2}\left(\sqrt{\frac{ r_0^{+}}{g'(r)|_{r_0^{+}}}}+\sqrt{\frac{ r_0^{-}}{g'(r)|_{r_0^{-}}}}\right),
\end{equation}
provided the black hole shadow is viewed from the equatorial plane. In addition, the radius of circular null geodesics $r_0^{\pm}$ for the prograde/retrograde orbit must be chosen such that both are outside of the horizon and can be obtained by solving the equation \cite{Jusufi:2020dhz}
\begin{equation}
r_0^2-\frac{2 r_0}{g'(r)|_{r_0}}g(r_0)\mp 2 a \sqrt{\frac{2 r_0}{g'(r)|_{r_0}}}=0.
\end{equation} 
In particular we are going to use the M87 black hole with $M =6.5 \times  10^{9}$M\textsubscript{\(\odot\)} and $D =16.8$ Mpc. The angular diameter can be further expressed as $\theta_s =2\times 9.87098 \times 10^{-6} R_s(M/$M\textsubscript{\(\odot\)})$(1kpc/ D)$ $\mu$as. In Table I we show the numerical values obtained for the typical shadow radius of a rotating EYM black hole by varying the parameter $\lambda$. From these numerical results we can see that, as $\lambda$ increases,  the shadow radius and the corresponding angular diameter decreases while the numerical values for the angular diameter are in the range  $42\pm3 \mu as$ reported in \cite{m87,Akiyama:2019eap}. In other words as $\lambda$ increases, it is more easy to distinguish the EYM black hole from the Kerr-Newman black hole since the difference between their angular diameters given in terms of $\Delta \theta_s$, increases.

%\section{Black hole shadows and observables}
%\label{observa}

Now we would like to study the observables of the shadow, which is useful for us to fit the observed data and determine the values of the black hole parameters.

Let us first introduce several characteristic points, the right point ($\alpha_r$, 0), left point ($\alpha_l$, 0), top point ($\alpha_t$, $\beta_t$) and bottom point ($\alpha_b$, $\beta_b$) of the shape. According to the symmetry of the shadow, one easily gets $\alpha_t=\alpha_b$ and $\beta_t=-\beta_b$. Following Ref. \cite{Hioki}, we can construct the size and distortion of the shadow. The size of the shadow is described by the reference circle passing the top, bottom and right points of the shadow. The reference circle cuts the $\alpha$ axis at ($\tilde{\alpha}_l$, 0).

The radius of the reference circle can be calculated with these characteristic points
\begin{eqnarray}
 R_s=\frac{(\alpha_t-\alpha_r)^2+\beta_t^2}{2(\alpha_r-\alpha_t)}.
\end{eqnarray}
In the following, we will focus on two distortions $\delta_s$ and $k_s$, which can be defined as
\begin{eqnarray}
 \delta_s&=&\frac{\alpha_l-\tilde{\alpha}_l}{R_s},\\
 k_s&=&\frac{\beta_t-\beta_b}{\alpha_r-\alpha_l}.
\end{eqnarray}
For the nonrotating black hole, we can get $\delta_s$=0 and $k_s$=1, which means the shadow shape is a standard circle. However when the black hole spin is nonzero, both these distortions deviates from these values. 

\begin{figure*}
\includegraphics[width=8.2cm]{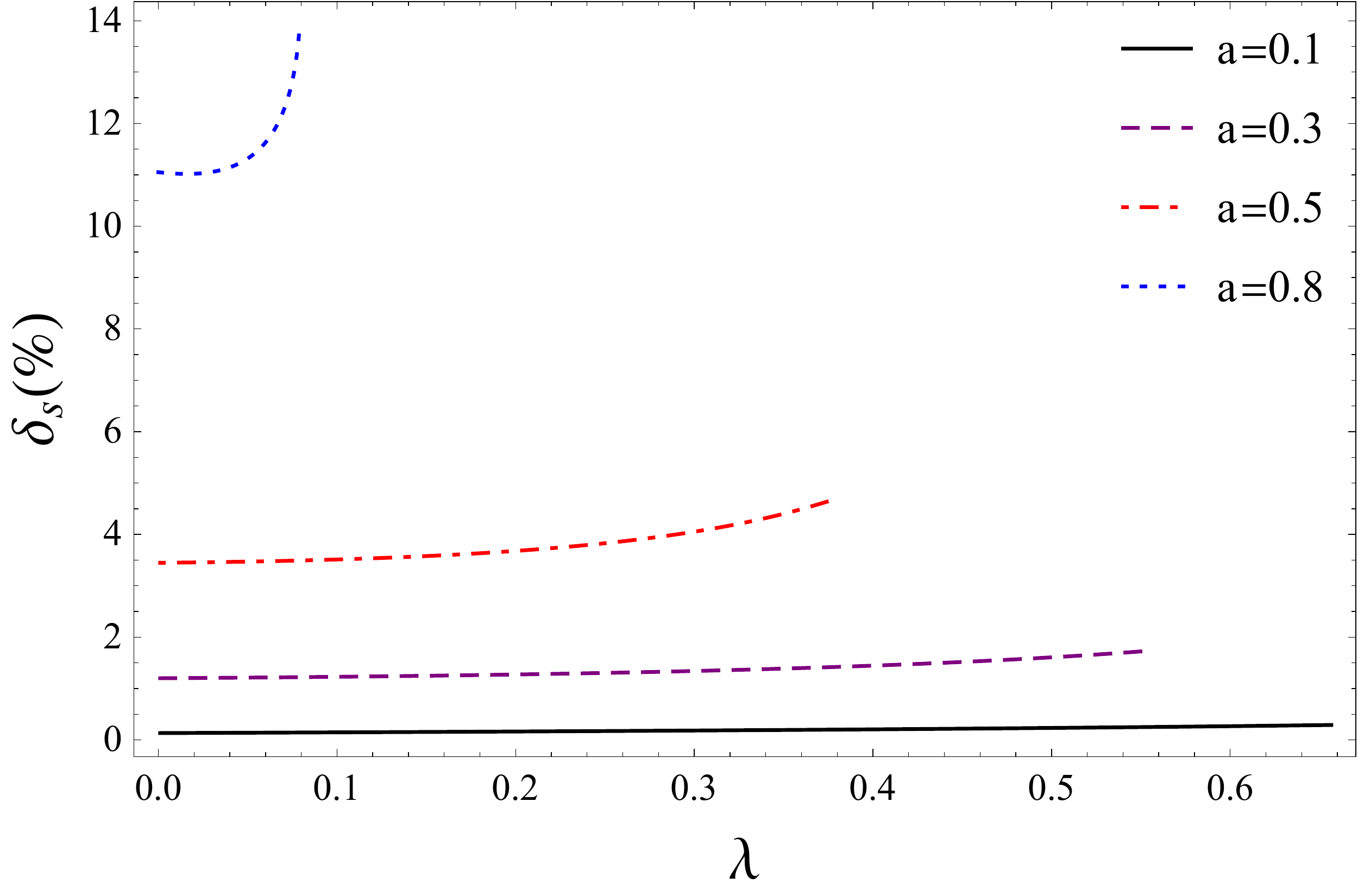}
		\includegraphics[width=8.2cm]{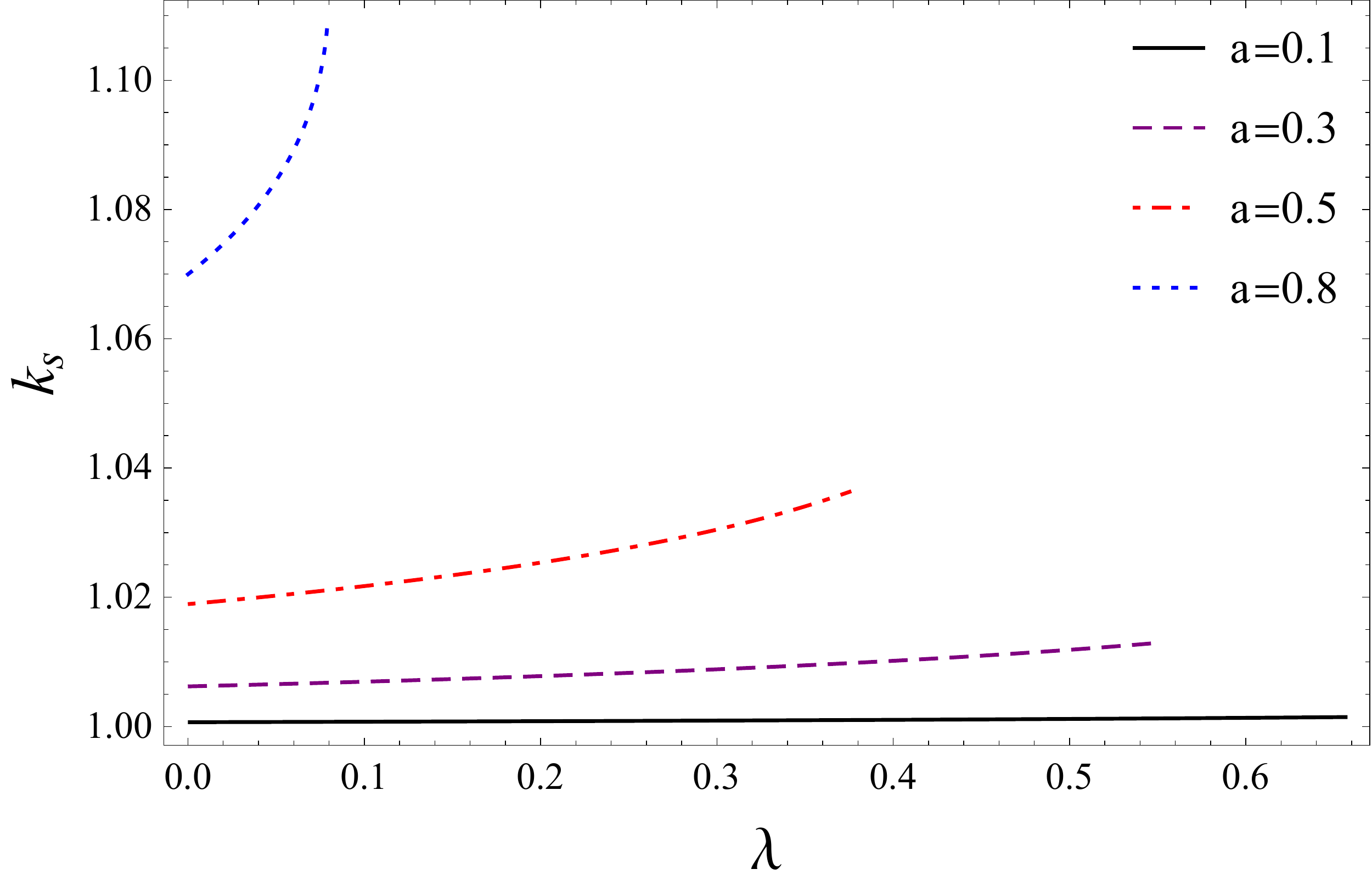}
		\includegraphics[width=8.2cm]{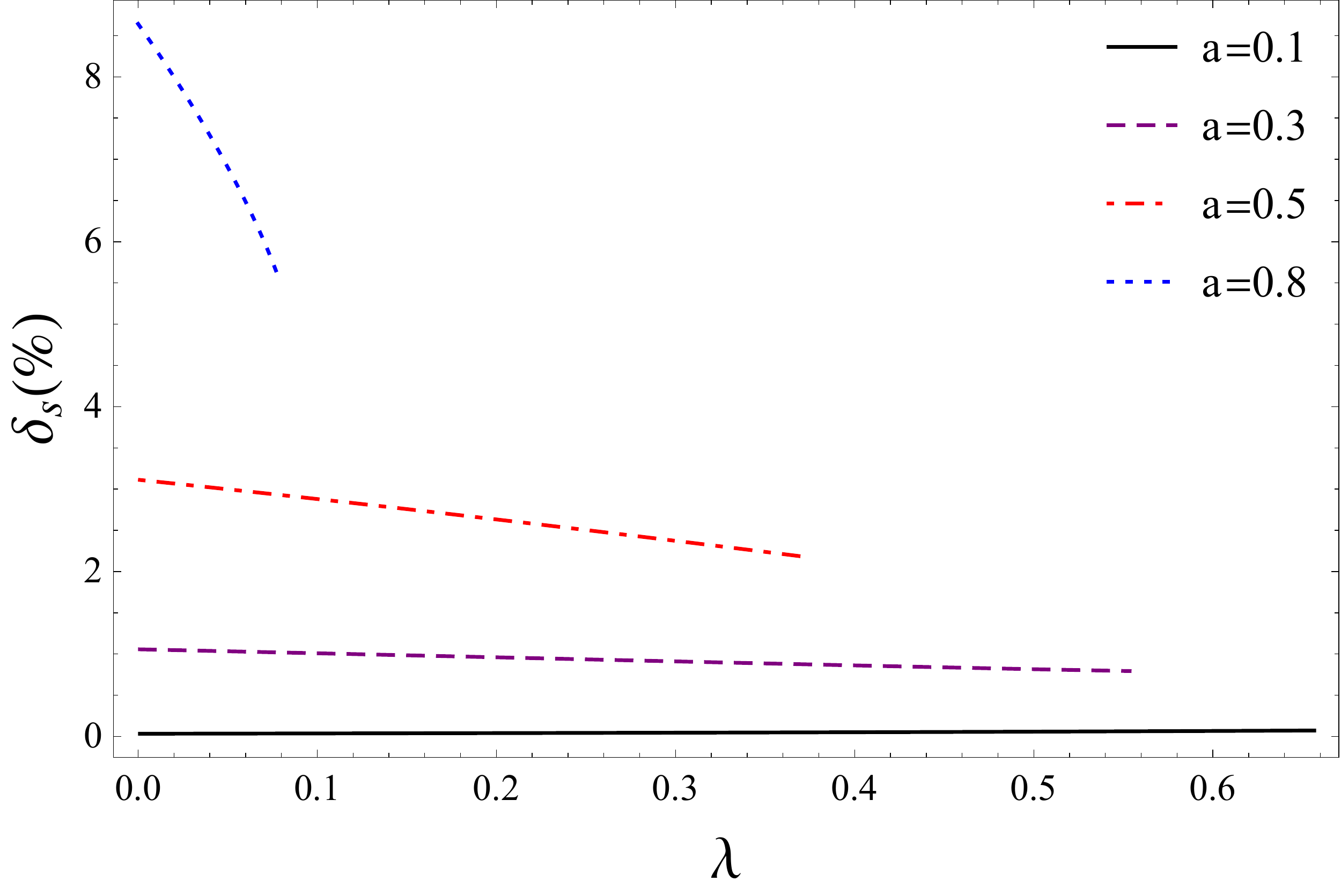}
		\includegraphics[width=8.2cm]{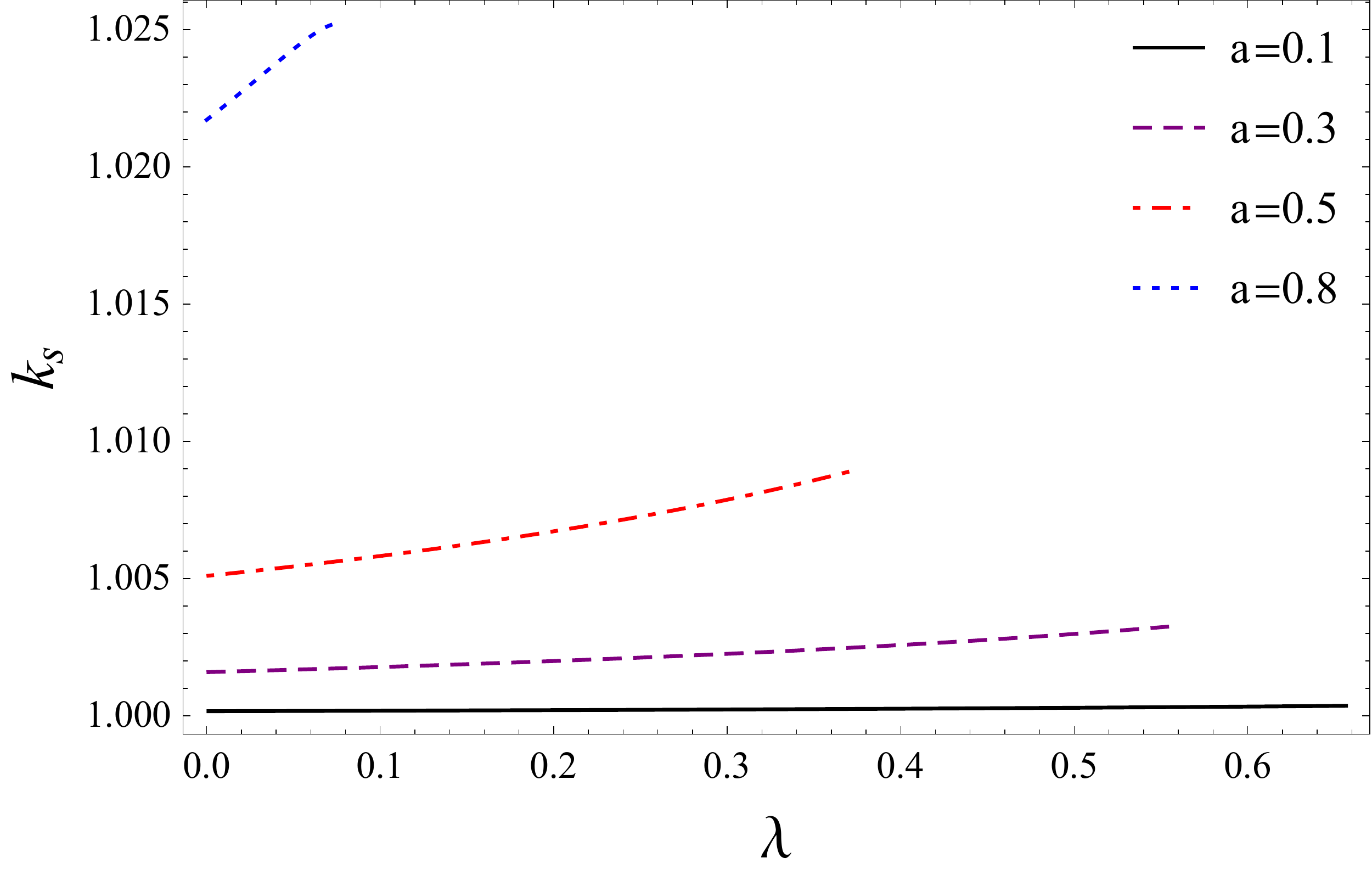}
\caption{Observables $\delta_s$ and $k_s$ for $Q$=0.4 and $M$= 1. The spin is set as $a$=0.1, 0.3, 0.5, and 0.8 from bottom to top. In the top left and top right plots, we fixed $\theta=\pi/2$, whereas in the bottom left and bottom right figures, we used $\theta=\pi/6$.}\label{ppOds90}
\end{figure*}

In order to show how these two distortions vary with $\lambda$, we plot them in Fig. \ref{ppOds90} for $Q$=0.4 and $M$= 1. From the figures, we can find that for low spin, the influence of $\lambda$ on distortions $\delta_s$ and $k_s$ is very tiny. $\delta_s$ and $k_s$ almost keep 0 and 1, respectively. These indicates the shadow shapes are very close to standard circle. For $\theta_0=\frac{\pi}{2}$, both $\delta_s$ and $k_s$ increases with $\lambda$, and approach to their maximal values for the extremal black holes. For example when $a$=0.8, the distortion $\delta_s$ takes 14\% and $k_s$ takes 1.12, indicating the shadows have a big deformation from a standard circle. For $\theta_0=\frac{\pi}{6}$, we find that the distortion $k_s$ still increases with $\lambda$ for different black hole spin. However $\delta_s$ decreases, which is resulted by the decrease of the shadow size. On the other sides, comparing with these figures, we can easily obtained the result that both the distortions get smaller with the decrease if $\theta_0$. So decreasing with $\theta_0$, the shadows get less deformation.

In summary, if the observer locates near the equatorial plane, both the observables $\delta_s$ and $k_s$ of the shadows increase with $\lambda$. If the observer is far off the equatorial plane, $\delta_s$ decreases while $k_s$ increase with $\lambda$. This provides us a possible way to test the magnetic black hole in EYM theory by making use the shadows.

\section{Observational constraints\label{secoc1}}
	
We can apply the numerical results of shadow size to the black hole of M87. The first M87 Event Horizon Telescope (EHT) results published the image of shadow of black hole with a ring diameter of $42\pm3 \mu as$ \cite{m87,Akiyama:2019eap}. Adopting this measurement value and distance $D=16.8\pm0.8 \text{Mpc}$, we performed the Monte-Carlo simulations for the parameters space $(M,Q,\lambda)$. The constraints on the parameter $\lambda$ and mass of M87 are  shown in Fig. \ref{mq} . In $95\%$ confidence level, the  parameter $\lambda$ is constrained as $\lambda=0.53^{+0.93}_{-0.53} $ where we have applied the prior $\lambda>0$. The mass of M87 is estimated as $M=(7.52^{+1.85} _{-1.56})\times 10^9 M_{\odot}$ which covers the range of value derived by EHT $M=(6.5 \pm 0.7)\times 10^9 M_{\odot}$ in Schwarzschild black hole. From above results, we found that there is a large parameter range to fit the EHT shadow size, so it is necessary to compare the above constraints with those obtained from other astrophysical observations.

\begin{figure*}
\includegraphics[width=6.6cm]{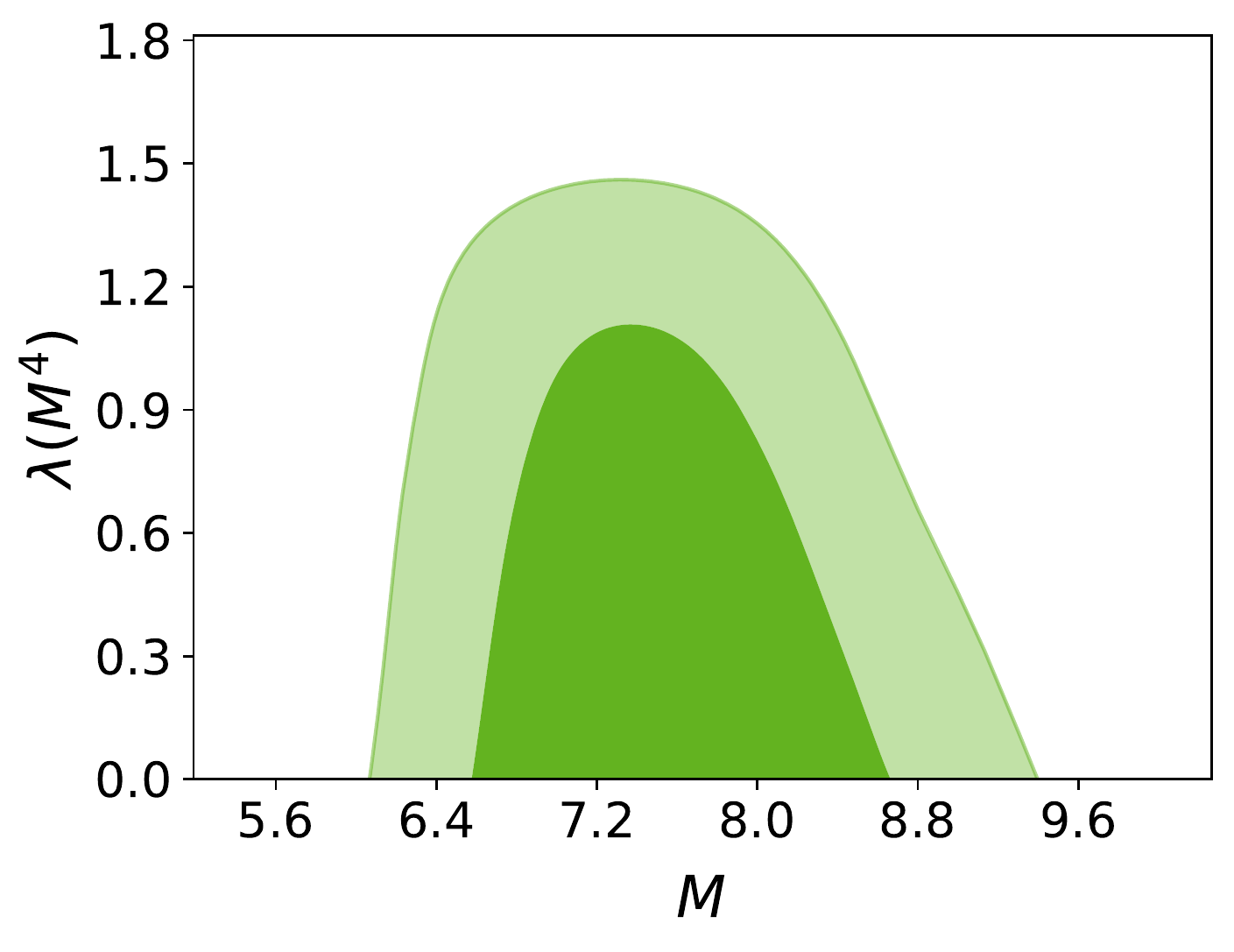}\label{mq}
\caption{Marginalized constrains for the parameter $\lambda$ and estimated M87 black hole mass $M(\times 10^9M_{\odot})$ using M87 shadow size in $68\%$ and $95\%$ confidence levels. } \label{mq}
\end{figure*}

	\section{Curvature radius and topology of shadow\label{seccr}}
	
	It is believed that the curvature radius has an important application in testing the BH shadow. Here we aim to study the curvature radius and then discuss its topology following \cite{Weiliu,WeiChuan}.
	
	Since the curvature is parameterized by the length parameter, we first show the perimeter of the shadow. For a given shadow, its perimeter can be calculated with the following formula
	\begin{equation}
	l_s=2\int \sqrt{(\partial_r\alpha)^2+(\partial_r\beta)^2}dr,
	\end{equation}
	where $\alpha$ and $\beta$ are the celestial coordinates describing the shadow~\cite{Weiliu}, {as mentioned above}. The factor 2 comes from the $\mathcal{Z}_2$ symmetry of the shadow. For $a=0.5$, $Q=0.4$ and $M=1$, the perimeter $l_s$ slightly decreases as $\lambda$ increases (from 31.4936 to 30.9322 as $\lambda$ varies from 0.01 to its maximum value 0.3752, corresponding to an extremal BH).
	
\begin{figure}[!htb]
		\includegraphics[width=8cm]{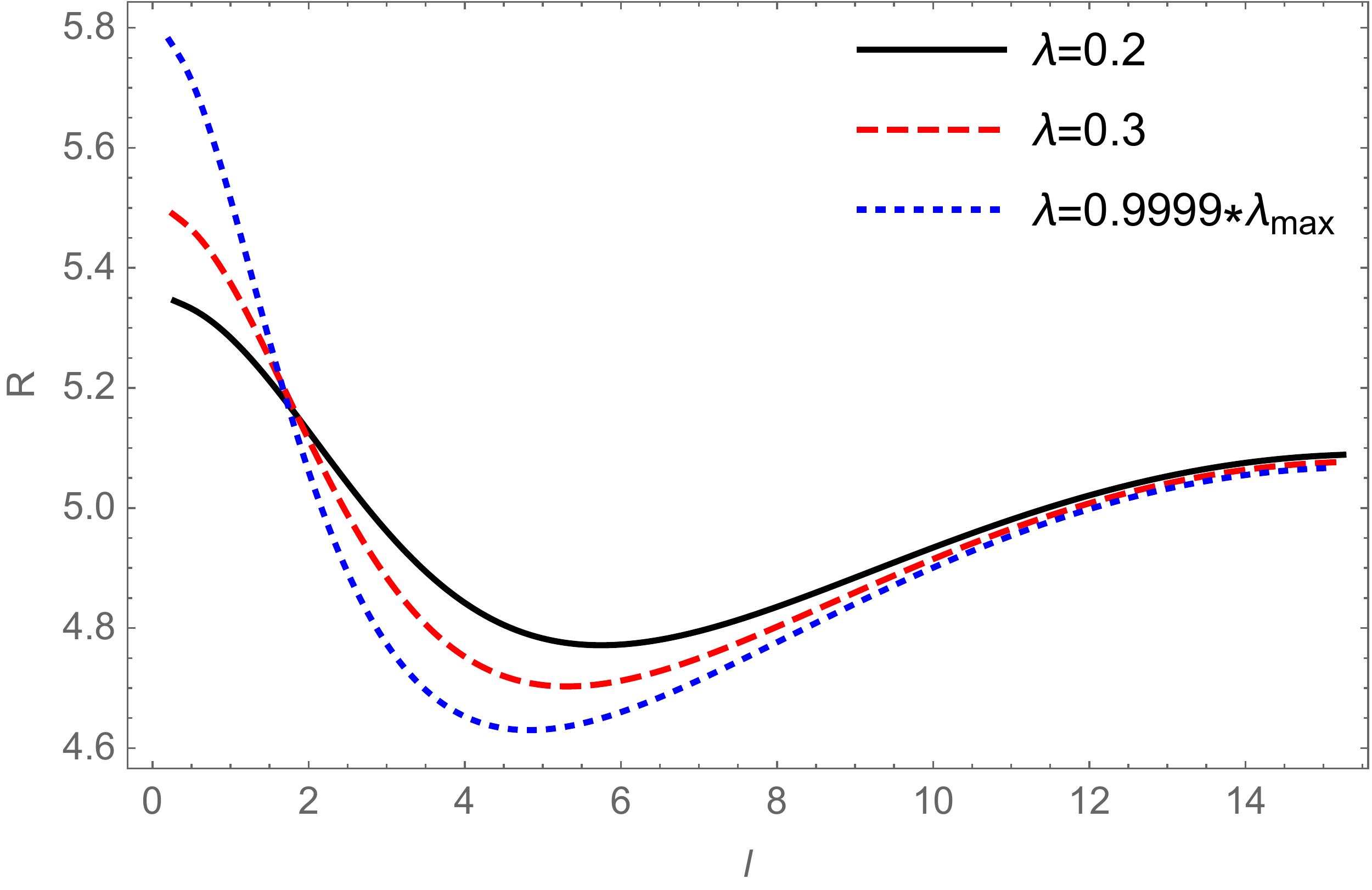}
		\caption{Curvature radius as a function of the length parameter for $a$=0.5, $Q$=0.4 and $M=1$. The parameter $\lambda$=0.2, 0.3, and $0.9999\times\lambda_{max}$ with $\lambda_{max}$=0.3752.}\label{curv}
	\end{figure}

	Since $\alpha$ and $\beta$ are parameterized by $r$, we can adopt these forms to calculate the local curvature of the shadow $R$ and $l$ in terms of $r$ as done in Refs.~\cite{Weiliu,WeiChuan}. Finally, we plot $R$ versus $l$ in Fig.~\ref{curv}. In this plot, the first point on the perimeter of the shadow has the largest curvature value $R$, corresponding, by convention, to $l=0$ (where $l$ is the shadow segment length), and the last point corresponds to $l=l_s/2$, allowing to drop the $\mathcal{Z}_2$ symmetry in the plot. We  find that the curvature $R$ first decreases with the length parameter $l$, and then increases. This result is consistent with that of the Kerr BH shadows \cite{Weiliu}. {Along} each curve, there are one maximum and one minimum. In particular, the maximum increases while the minimum decreases with increasing the parameter $\lambda$.
	
	When the shapes of the shadow are obtained from astronomical observations, we can use the curvature radius to fit the results and then obtain the values of the BH parameters. In \cite{WeiChuan}, we discuss several different ways to determine the BH spin and the inclination angle of the observer for a Kerr BH. These provide possible applications on testing the nature of a BH through the shadow.
	
	%\begin{figure}[!htb]
	%\includegraphics[width=8cm]{Permeter.pdf}
	%\caption{Behavior of the perimeter $l_s$ of shadow as a %function of $\lambda$ for $a$=0.5 and $Q$=0.4.}\label{perme}
	%\end{figure}
	
	\begin{figure}[!htb]
		\includegraphics[width=8cm]{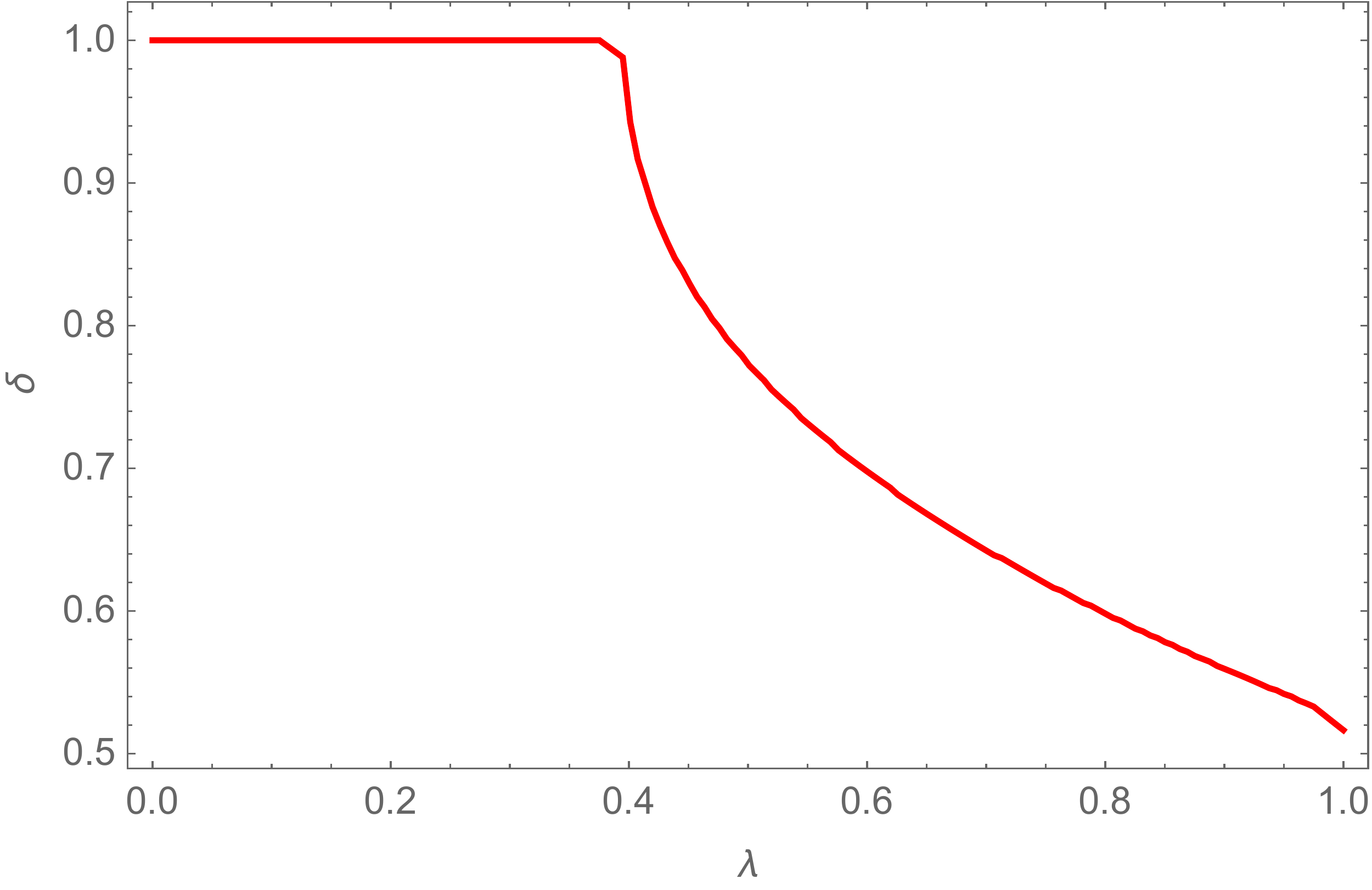}
		\caption{Topological quantity $\delta$ as a function of $\lambda$ for $a$=0.5, $Q$=0.4 and $M=1$. For small $\lambda$, $\delta$ assumes the value 1, while for $\lambda>0.3752$, $\delta$ decreases with $\lambda$.}\label{Topp}
	\end{figure}
	
	As we know, topology plays an important role in  physics. In our investigation, topology can be used to describe differences between BHs and horizonless solutions~\cite{Weiliu}. To reveal the particular topological properties of the shadow we introduce the topological covariant quantity~\cite{Weiliu},
	\begin{equation}
	\delta=\frac{1}{2\pi}\left(\int\frac{dl}{R}+\sum_i \theta_i\right).
	\end{equation}
	Since in the current case, the light ring is always unstable, the second term vanishes. Taking $a$=0.5, $Q$=0.4 and $M=1$ as an example, we numerically calculate $\delta$ in terms of $\lambda$. The result is displayed in Fig.~\ref{Topp}. From this figure we see that, for small $\lambda$, $\delta$ assumes the value 1. While for $\lambda>0.3752$, $\delta$ decreases indicating a topological change corresponding to the transition from rotating BH $\to$ rotating horizonless solution. Note that for $a=0.5,$ $Q=0.4$, $M=1$ and $\lambda<0.3752$, the rotating solution is a BH with more than one horizon; for $\lambda=0.3752$ it is an extremal BH; and for $\lambda>0.3752$ it is a horizonless solution. This indicates a possible topological phase transition from a BH to a horizonless solution, a kind of gravitational vacuum condensate stars or  gravastars without both an event horizon and a singularity at the center~\cite{mazur}. Gravastars are compact objects and may arise due to the Bose-Einstein condensation in gravitational systems resulting with an interior structure filled with vacuum energy and with an exterior effective Schwarzschild geometry if $a=0$. So, the variation of $\lambda$ indicates a change of the rotating solution from a BH to a compact object. Therefore, we  conclude that the deviation from 1 of $\delta$ is a topological phase transition. The behavior of $\delta$ can act as a topological quantity to reflect the topological information of the spacetime structure.
	
	\section{Connection between the shadow radius and QNMs\label{secrq}}
	
	It is well known that QNMs in the eikonal regime are related to the angular velocity of the last circular null geodesic, while the imaginary part was related to the Lyapunov exponent, $\Lambda$, which determines the instability time scale of the orbit \cite{cardoso}. Then,  the relation between QNMs and black hole lensing has been established by analyzing the photon sphere and light ring in a static spacetime or stationary spacetime, respectively \cite{Stefanov:2010xz,Guo,Liu,WeiLiu2}. However, it is convenient to express this connection in terms of the shadow radius and the real part of QNMs. Such a connection was  {obtained recently in \cite{Jusufi:2019ltj} (see also \cite{Liu20,Jusufi:2020agr})},
	\begin{equation}
	\omega_{\Re} = \lim_{l \gg 1} \frac{l}{R_S}.
	\end{equation}
	This result was proved to be valid  for the static spherical spacetime,   {and}  accurate in the eikonal limit   $l \gg 1$.  Very recently
the correspondence between the shadow radius and the real part of QNMs frequencies was improved to the sub-leading regime to 
half of its value  \cite{Cuadros-Melgar:2020kqn} (see also \cite{Guo:2020nci,Cai:2020kue,Jusufi:2020mmy})
\begin{equation}
	\omega_{\Re} = \lim_{l \gg 1} R_S^{-1}\left(l+\frac{1}{2}\right).
	\end{equation}
	Of course, in the large angular momentum regime, i.e., $l>>1$, we recover Eq. (50). Thus, we can write 
	\begin{equation}
	\omega_{QNM}=\lim_{l \gg 1} R_S^{-1}\left(l+\frac{1}{2}\right) -i \left(n+\frac{1}{2}\right)|\Lambda|.
	\end{equation}
	It is interesting to note that the above correspondence sometimes works well even for small values of $l$. It provides an alternative way to compute the real part of the QNMs by means of the shadow radius.
	\begin{table}[tbp]
        \begin{tabular}{|l|l|l|l|l|l|}
        \hline
    \multicolumn{1}{|c|}{ } &  \multicolumn{1}{c|}{  $l=1, n=0$ } 
    & \multicolumn{1}{c|}{  $l=2, n=0$ } &   \multicolumn{1}{c|}{  $l=3, n=0$ } 
     & \multicolumn{1}{c|}{} \\
\hline
  $Q$ & \quad $\omega_{\Re}$ & \quad $\omega_{\Re}$ & \quad $\omega_{\Re}$  
     & \quad $R_S$  \\ 
        \hline
0.1 & 0.2898919233 & 0.4831532055 & 0.6764144877    & 5.174342157 \\ 
0.2 & 0.2913902646 & 0.4856504410 & 0.6799106174   & 5.147735467 \\ 
0.3 & 0.2939658648 & 0.4899431080 & 0.6859203512    & 5.102633263\\
0.4 & 0.29774983241 & 0.4962497208 &0.6947496090   & 5.037786211 \\
0.5 & 0.3029582758 & 0.5049304598 & 0.7069026436    & 4.951176844\\
        \hline
        \end{tabular}
        \caption{ Numerical values for the shadow radius and the 
    real part of QNMs obtained via Eq. (51). Here we use a constant $\lambda=0.1$ and change $Q$.}
\end{table}
\begin{table}[tbp]
        \begin{tabular}{|l|l|l|l|l|l|}
        \hline
    \multicolumn{1}{|c|}{ } &  \multicolumn{1}{c|}{  $l=1, n=0$ } 
    & \multicolumn{1}{c|}{  $l=2, n=0$ } &   \multicolumn{1}{c|}{  $l=3, n=0$ } 
     & \multicolumn{1}{c|}{} \\
\hline
  $\lambda$ & \quad $\omega_{\Re}$ & \quad $\omega_{\Re}$ & \quad $\omega_{\Re}$  
     & \quad $R_S$  \\ 
        \hline
0.0 & 0.2931522710 & 0.4885871182 & 0.6840219656   & 5.116794746 \\ 
0.1 & 0.2939658648 & 0.4899431080 & 0.6859203512    & 5.102633264 \\
0.2 & 0.2948115346 & 0.4913525578 & 0.6878935808   & 5.087996308 \\
0.3 & 0.2956924989 & 0.4928208315 & 0.6899491641    & 5.072837511\\
0.4 & 0.2966125526 & 0.4943542542 &  0.6920959560  & 5.057102227 \\
        \hline
        \end{tabular}
        \caption{Numerical values for the shadow radius and the 
    real part of QNMs obtained via Eq. (51). Here we use a constant $Q=0.3$ and change $\lambda$.}
\end{table}
	In Table II and Table III, we present the numerical calculations for the real part of QNMs obtained by means of the shadow radius. In the following we are going to study the QNMs of scalar and electromagnetic fields in the spacetime of static EYM BH using the WKB method.
	
	\subsection{QNMs of a scalar field}
	
	{Before we consider the problem of QNMs, let us point out that in this section we are going to simplify the problem by setting the rotation of the black hole to zero, i.e. $a=0$. For the metric (\ref{static}), we introduce the tortoise coordinate,
		\begin{equation}
		dr_{\star}=\frac{dr}{f(r)},
		\end{equation}
		in order to study }  perturbations of {a massless scalar field, described by the equation}
	\begin{equation}
	\frac{1}{\sqrt{-g}}\partial_{\mu}\left(\sqrt{-g} g^{\mu \nu} \partial_{\mu}\Phi  \right)=0.
	\end{equation}
	
	{Separation of variables of the function $\Phi$ in terms of the spherical harmonics yields}
	\begin{equation}
	\Phi(t,r,\theta,\phi)=\frac{1}{r}\,e^{-i\omega t}Y_{l}(r,\theta)\Psi(r),
	\end{equation}
	with  $l=0,1,2,...$  being the multipole numbers. {Then,} one can show that the perturbations are governed  by a Schr\"odinger wave-like equation
	\begin{equation}
	\lb{eq46}
	\frac{d^2\Psi}{dr_{\star}^2}+\left(\omega^2-V_S(r)\right)\Psi=0,
	\end{equation}
	{where the function $\Psi$ satisfies} the following boundary conditions
	\begin{equation}
	\lb{eq47}
	\Psi(r_{\star})=C_{\pm}\,\exp\left( \pm i\,\omega \,r_{\star}  \right),\,\,\,\,\,r\to \pm \infty,
	\end{equation}
	where $\omega$ can be written in terms of the real and imaginary parts,  i.e., $\omega=\omega_{\Re}-i \omega_{\Im}$,  {where  the imaginary part } is proportional to the decay rate of a given mode. The
	{effective potential  $V_S(r)$ of the perturbations for the scalar field is given by,}
	\begin{eqnarray}
	\lb{eq48}
	V_S(r)&=&\left[1+\left(\frac{r^4}{r^4+2 \lambda}\right)\left(-\frac{2M}{r}+\frac{Q^2}{r^2}\right)\right]\\\notag
	&\times& \left[\frac{l(l+1)}{r^2} +\frac{2Mr^5-2Q^2 r^4-12 M\lambda r+4 Q^2\lambda}{(r^2+2\lambda)^2} \right].
	\end{eqnarray}
	
	{To solve Eqs.(\ref{eq46}) and  (\ref{eq48}) with the boundary conditions (\ref{eq47}), we use the WKB approximation} to compute the quasi-normal  frequencies. The WKB method is widely used for numerical computations of QNMs and is based on the analogy with the problem of wave scattering near the peak
	of a potential barrier in quantum mechanics, where $\omega$
	plays a role of energy \cite{Schutz,Iyer}.
	In this work we are going to use the sixth order WKB approximation for calculating QNMs developed by Konoplya  \cite{KonoplyaWKB}.
	
	In Table IV and Table V, we present the results for the scalar perturbations by varying the magnetic charge $Q$ and the parameter $\lambda$, respectively. Note that we have not presented the calculations of QNMs for the  fundamental mode $l=n=0$ in Table IV and Table V. This is simply related to the fact that the WKB method is applicable {only when $l>n$ and does not give a satisfactory precision} for this fundamental mode.
	\begin{table}[tbp]
		\begin{tabular}{|l|l|l|l|l|}
			\hline
			\multicolumn{1}{|c|}{ spin 0 } &  \multicolumn{1}{c|}{  $l=1, n=0$ } & \multicolumn{1}{c|}{  $l=2, n=0$ } & \multicolumn{1}{c|}{ $l=2, n=1$ }\\\hline
			$Q$ & $\omega \,(WKB)$ &  $\omega \,(WKB)$ &  $\omega \,(WKB)$  \\ \hline
		%	0.0 & 0.2920-0.0978 i & 0.4836-0.0968 i & 0.4638-0.2956 i \\
			0.1 & 0.2942-0.0968 i & 0.4857-0.0958 i & 0.4662-0.2924 i  \\
			0.2 & 0.2958-0.0969 i & 0.4882-0.0959 i & 0.4689-0.2927 i    \\
			0.3 & 0.2984-0.0971 i & 0.4925-0.0961 i & 0.4734-0.2932 i  \\
			0.4 & 0.3023-0.0972 i & 0.4989-0.0963 i & 0.4801-0.2937 i  \\
			0.5 & 0.3076-0.0974 i & 0.5076-0.0965 i & 0.4892-0.2941 i \\
			0.6 & 0.3148-0.0974 i & 0.5193-0.0966 i & 0.5015-0.2940 i\\
			0.7 & 0.3242-0.0969 i & 0.5350-0.0963 i & 0.5177-0.2927 i\\
			0.8 & 0.3369-0.0954 i & 0.5562-0.0950 i & 0.5392-0.2883  i \\\hline
		\end{tabular}
		\caption{The real and imaginary parts of  the quasinormal frequencies of the scalar field with {different values of $Q$. In all these cases we have set} $\lambda=0.1$. }
	\end{table}

	\begin{table}[tbp]
		\begin{tabular}{|l|l|l|l|l|}
			\hline
			\multicolumn{1}{|c|}{ spin 0 } &  \multicolumn{1}{c|}{  $l=1, n=0$ } & \multicolumn{1}{c|}{  $l=2, n=0$ } & \multicolumn{1}{c|}{ $l=2, n=1$ }\\\hline
			$\lambda$ & $\omega \,(WKB)$ &  $\omega \,(WKB)$ &  $\omega \,(WKB)$  \\ \hline
			0.0 & 0.2975-0.0982 i & 0.4912-0.0972 i & 0.4717-0.2969  i  \\
			0.1 & 0.2984-0.0971 i & 0.4925-0.0961 i & 0.4734-0.2932 i  \\
			0.2 & 0.2993-0.0958 i & 0.4939-0.0949 i & 0.4749-0.2892  i    \\
			0.3 & 0.3001-0.0944 i & 0.4953-0.0936 i & 0.4762-0.2849 i  \\
			0.4 & 0.3008-0.0929 i & 0.4967-0.0922 i & 0.4771-0.2802 i  \\
			0.5 & 0.3013-0.0913 i & 0.4980-0.0906 i & 0.4776-0.2753 i \\
			0.6 & 0.3016-0.0896 i & 0.4994-0.0890 i & 0.4774-0.2702  i\\
			0.7 & 0.3018-0.0880 i & 0.5006-0.0872 i & 0.4767-0.2652 i \\\hline
		\end{tabular}
		\caption{The real and imaginary parts of  the quasinormal frequencies of the scalar field with {different values of $\lambda$. In all these cases we have set} $Q=0.3$. }
	\end{table}
	
	\begin{figure*}[!htb]
		\includegraphics[width=8.2cm]{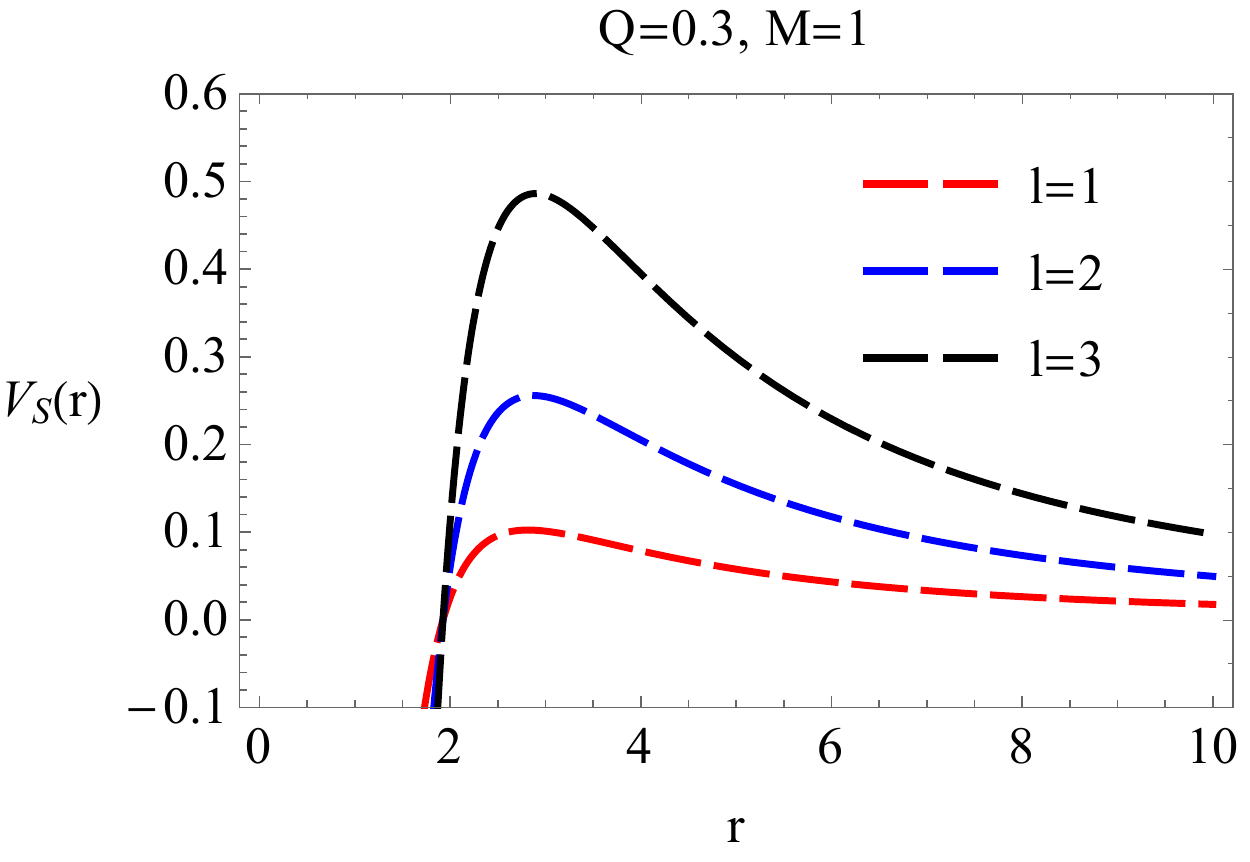}
		\includegraphics[width=8.2cm]{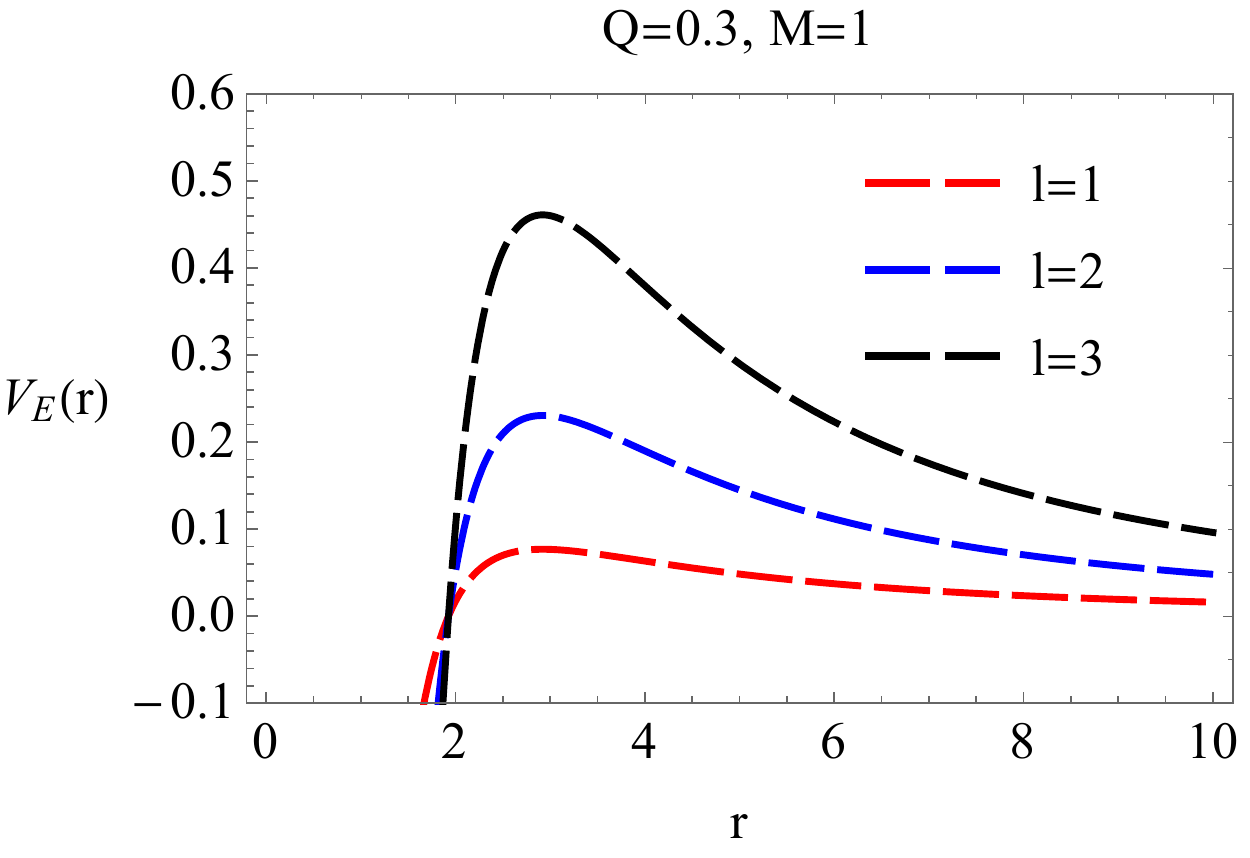}
		\caption{Left panel: {The} effective potential of {the}
			scalar field for different values of $l$. Right panel: {The} effective potential of {the}
			electromagnetic field for different values of $l$.  }
	\end{figure*}
	\begin{figure*}[!htb]
		\includegraphics[width=8.2cm]{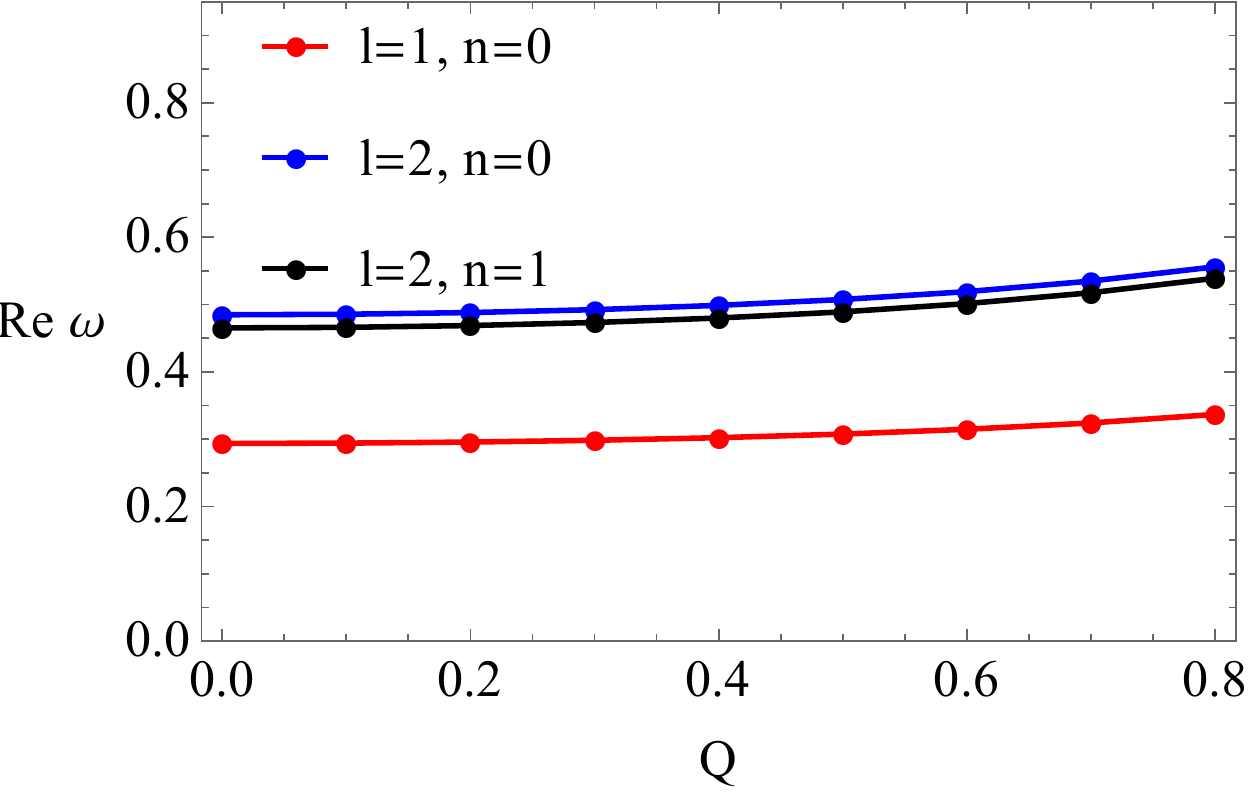}
		\includegraphics[width=8.2cm]{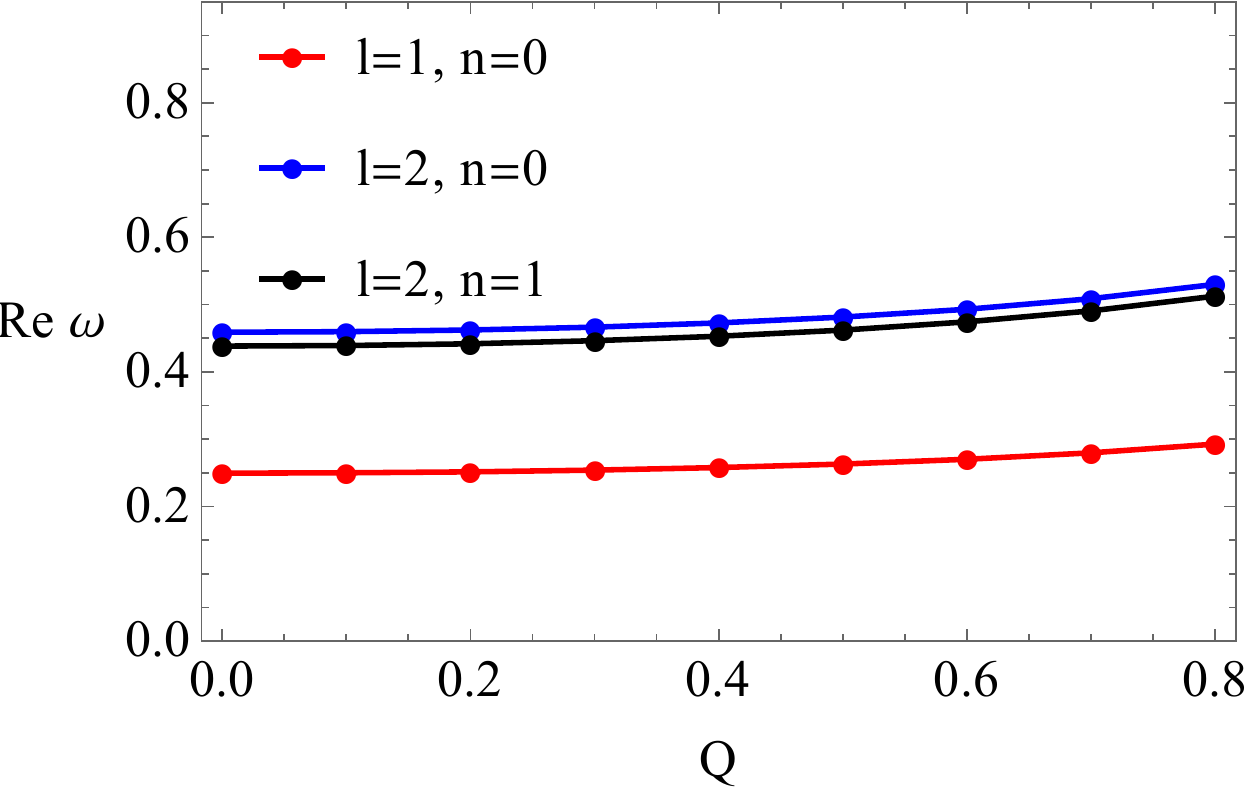}
		\caption{Left panel: The real part of QNMs for the scalar field vs the magnetic charge $Q$. In both plots we have set $\lambda=0.1$. Right panel: The real part of QNMs for the electromagnetic field vs the magnetic charge $Q$. In both plots we have set $\lambda=0.1$. }
	\end{figure*}

\begin{figure*}[!htb]
		\includegraphics[width=8.2cm]{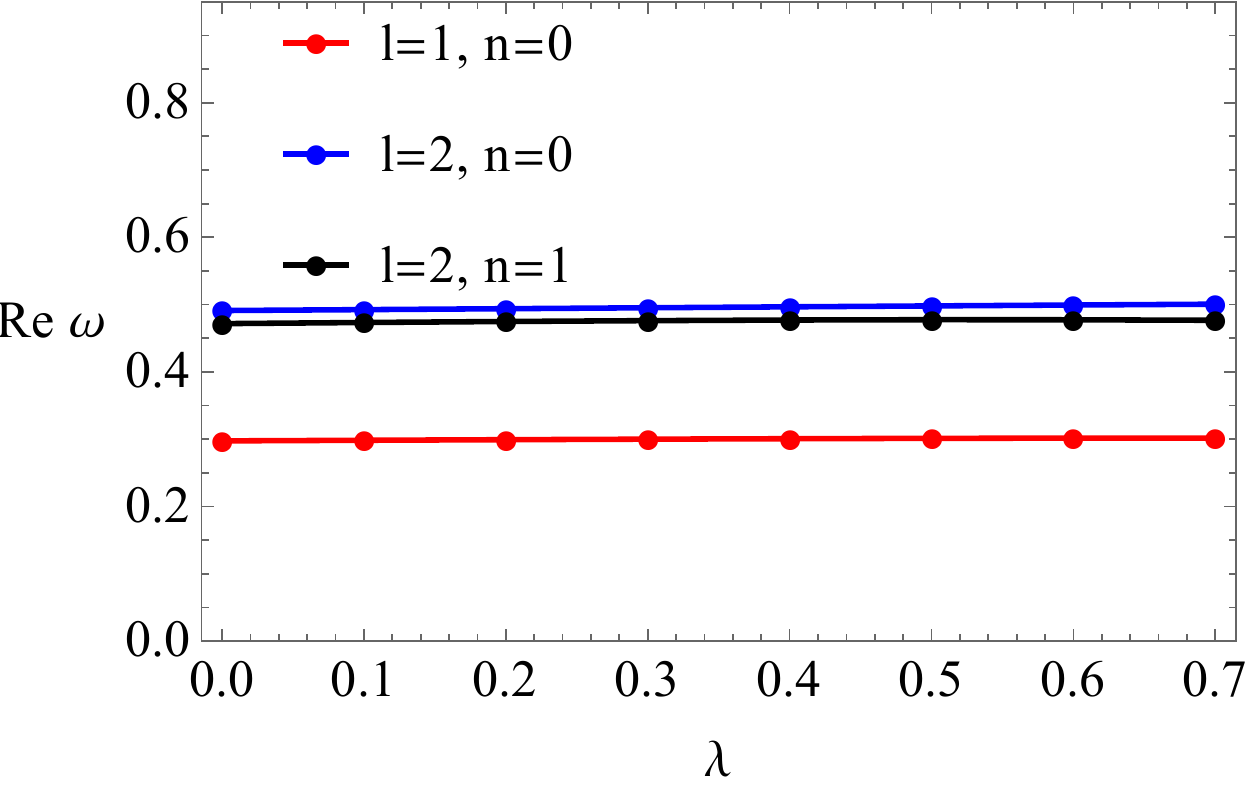}
		\includegraphics[width=8.2cm]{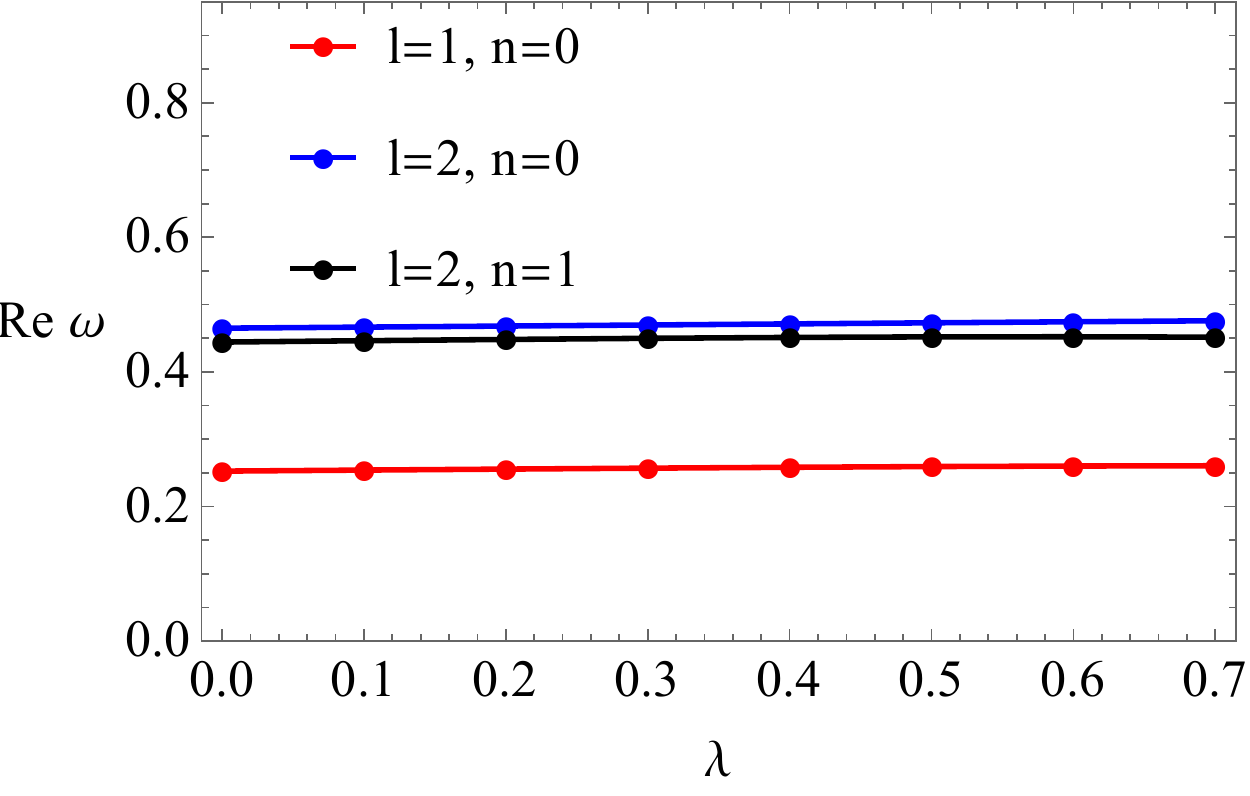}
		\caption{Left panel: The real part of QNMs for the scalar field vs the parameter  $\lambda$ and constant magnetic charge $Q=0.3$. Right panel: The real part of QNMs for the electromagnetic field vs the parameter  $\lambda$ and constant magnetic charge $Q=0.3$.  }
	\end{figure*}
	
	\begin{table}[tbp]
		\begin{tabular}{|l|l|l|l|l|}
			\hline
			\multicolumn{1}{|c|}{ spin 1 } &  \multicolumn{1}{c|}{  $l=1, n=0$ } & \multicolumn{1}{c|}{  $l=2, n=0$ } & \multicolumn{1}{c|}{ $l=2, n=1$ }\\\hline
			$Q$ & $\omega \,(WKB)$ & $\omega \,(WKB)$ & $\omega \,(WKB)$   \\ \hline
		%	0.0 & 0.2482-0.0926 i & 0.4576-0.0950 i & 0.4365-0.2907 i \\
			0.1 & 0.2501-0.0917 i & 0.4598-0.0940 i & 0.4392-0.2875 i  \\
			0.2 & 0.2515-0.0918 i & 0.4623-0.0942 i & 0.4418-0.2878 i    \\
			0.3 & 0.2541-0.0920 i & 0.4665-0.0943 i & 0.4464-0.2883 i  \\
			0.4 & 0.2579-0.0923 i & 0.4728-0.0946 i & 0.4530-0.2888 i  \\
			0.5 & 0.2631-0.0926 i & 0.4815-0.0948 i & 0.4621-0.2893 i \\
			0.6 & 0.2702-0.0928 i& 0.4931-0.0949 i & 0.4744-0.2893 i\\
			0.7 & 0.2798-0.0925 i& 0.5088-0.0947 i & 0.4907-0.2880 i\\
			0.8 & 0.2928-0.0911 i& 0.5301-0.0934 i & 0.5126-0.2835 i\\\hline
		\end{tabular}
		\caption{The real and imaginary parts of  {the} quasinormal frequencies of the electromagnetic field  with different values of $Q$. In all these cases we have set  $\lambda=0.1$.}
	\end{table}
	\begin{table}[tbp]
		\begin{tabular}{|l|l|l|l|l|}
			\hline
			\multicolumn{1}{|c|}{ spin 1 } &  \multicolumn{1}{c|}{  $l=1, n=0$ } & \multicolumn{1}{c|}{  $l=2, n=0$ } & \multicolumn{1}{c|}{ $l=2, n=1$ }\\\hline
			$\lambda$ & $\omega \,(WKB)$ & $\omega \,(WKB)$ & $\omega \,(WKB)$   \\ \hline
			0.0 & 0.2526-0.0932  i & 0.4650-0.0955  i & 0.4443-0.2921 i \\
			0.1 & 0.2541-0.0920 i & 0.4665-0.0944 i & 0.4464-0.2883 i\\
			0.2 & 0.2556-0.0908 i & 0.4681-0.0931 i & 0.4482-0.2841 i   \\
			0.3 & 0.2570-0.0894 i & 0.4697-0.0918 i & 0.4499-0.2796  i  \\
			0.4 & 0.2583-0.0878 i & 0.4713-0.0903 i & 0.4512-0.2747 i  \\
			0.5 & 0.2594-0.0861 i & 0.4729-0.0886 i & 0.4521-0.2694 i \\
			0.6 & 0.2602-0.0842  i& 0.4745-0.0869 i &0.4522-0.2639  i\\
			0.7 & 0.2606-0.0823  i& 0.4760-0.0849 i & 0.4516-0.2583 i \\\hline
		\end{tabular}
		\caption{The real and imaginary parts of  {the} quasinormal frequencies of the electromagnetic field  with different values of $\lambda$. In all these cases we have set  $Q=0.3$.}
	\end{table}
	
	\subsection{QNMs of an electromagnetic field}
	
	In this section we precede to study the effect of {the } magnetic charge on the propagation of the electromagnetic field. To do so, we recall the wave equations for a test electromagnetic field,
	\begin{equation}
	\frac{1}{\sqrt{-g}}\partial_{\nu}\left[ \sqrt{-g} g^{\alpha \mu}g^{\sigma \nu} \left(A_{\sigma,\alpha} -A_{\alpha,\sigma}\right) \right]=0.
	\end{equation}
	The four-potential $A_{\mu }$ can be expanded {in terms of the}
	4-dimensional vector spherical harmonics as,
	%\begin{widetext}
	\begin{eqnarray}\notag
	A_{\mu }\left( t,r,\theta ,\phi \right) &=&\sum_{\ell ,m}\left(\left[
	\begin{array}{c}
	0 \\
	0 \\
	\frac{a(t,r)}{\sin \left( \theta \right) }\partial _{\phi }Y_{\ell
		m}\left( \theta ,\phi \right) \\
	-a\left( t,r\right) \sin \left( \theta \right) \partial _{\theta }Y_{\ell
		m}\left( \theta ,\phi \right)%
	\end{array}%
	\right] \right. \\
	&+& \left.\left[
	\begin{array}{c}
	f(t,r)Y_{\ell m}\left( \theta ,\phi \right) \\
	h(t,r)Y_{\ell m}\left( \theta ,\phi \right) \\
	k(t,r)\partial _{\theta }Y_{\ell m}\left( \theta ,\phi \right) \\
	k(t,r)\partial _{\varphi }Y_{\ell m}\left( \theta ,\phi \right)%
	\end{array}%
	\right] \right) ,
	\end{eqnarray}%
	in which $Y_{\ell m}\left( \theta ,\phi \right) $ {denotes} the spherical
	harmonics.  Without going to details,  {we find the following}  second-order differential equation for the radial part
	\begin{equation}
	\lb{eq51}
	\frac{d^{2}\Psi \left( r_{\ast }\right) }{dr_{\ast }^{2}}+\left[ \omega
	^{2}-V_{E}\left( r_{\ast }\right) \right] \Psi \left( r_{\ast }\right) =0,
	\end{equation}%
	with {the  effective potential}
	\begin{equation}
	V_E(r)=\left[1+\left(\frac{r^4}{r^4+2 \lambda}\right)\left(-\frac{2M}{r}+\frac{Q^2}{r^2}\right)\right]\,\frac{l(l+1)}{r^2}.
	\end{equation}
	 In Table VI and Table VII we show the results for the electromagnetic perturbations by varying the magnetic charge $Q$ and the parameter $\lambda$, respectively. 
	From  Fig. (12) we see that the effective potentials for both fields are indeed affected by the magnetic charge $Q$. From Fig. (13) we see that by increasing the magnetic charge $Q$, while having a constant $\lambda$, the real part of QNMs describing scalar and electromagnetic fields increase. A similar result is obtained when we increase the parameter $\lambda$ while having a fixed value of $Q$, namely the real part of QNMs increase monotonically, as can be seen in Fig. (14). Although in this case the effect of the magnetic charge on the real part of QNMs is smaller compared to the first case.  From Tables IV-VII, it can also be seen that, in general, the absolute values of the imaginary part of QNMs decreases with the increase of the magnetic charge $Q$ and $\lambda$, respectively.  This means that the field perturbations in the spacetime of EYM black hole having $Q > 0$ or $\lambda > 0$ oscillate more rapidly compared to the vacuum Schwarzschild BH, however in terms of  damping, the field perturbations decay more slowly compared to the  Schwarzschild  BH.  In addition to that, we see that for the scalar field perturbations the values of the real part of QNMs in absolute {values} are higher than those for  the electromagnetic field perturbations (see Tables IV-VII). Thus the scalar field perturbations will oscillate more rapidly compared to the electromagnetic field perturbations, in the same time the scalar field ones damp more rapidly than electromagnetic 
field ones.  Once we  compute the real part of QNMs and find that $\omega_{\Re }$ increases with $Q$ with a constant $\lambda$, we can make use of the inverse relation between  $\omega_{\Re }$ and the shadow radius $R_S$
	\begin{equation}
	R_S(Q)=\lim_{l>>1}\frac{l+\frac{1}{2}}{\omega_{\Re }(Q)}|_{\lambda=const},
	\end{equation}
	which decreases with increasing $Q$ as can be seen from Fig. 15 (left panel). This fact is verified in Fig. 6 where we have shown that the shadow radius decreases by increasing $Q$. Similarly, having the real part of QNMs with a varying $\lambda$ and a constant $Q$, we can use
	\begin{equation}
	R_S(\lambda)=\lim_{l>>1}\frac{l+\frac{1}{2}}{\omega_{\Re }(\lambda)}|_{Q=const}.
	\end{equation}
	and show that the shadow radius monotonically decreases with increasing $\lambda$, as can be seen from Fig. 14 (right panel). This is consistent with Fig. 7 where we have shown that the shadow radius decreases by increasing $\lambda$. Finally, we can compare the numerical results for the real part of QNMs obtained from the shadow radius presented in Tables II-III, with the ones obtained via the WKB method presented in Tables IV-VII. We observe that even for the fundamental modes with small $l$, the accuracy between two methods works well for the case of the scalar field perturbations. Increasing $l$, the accuracy between the two methods increases. 
	
		\begin{figure*}[!htb]
		\includegraphics[width=8 cm]{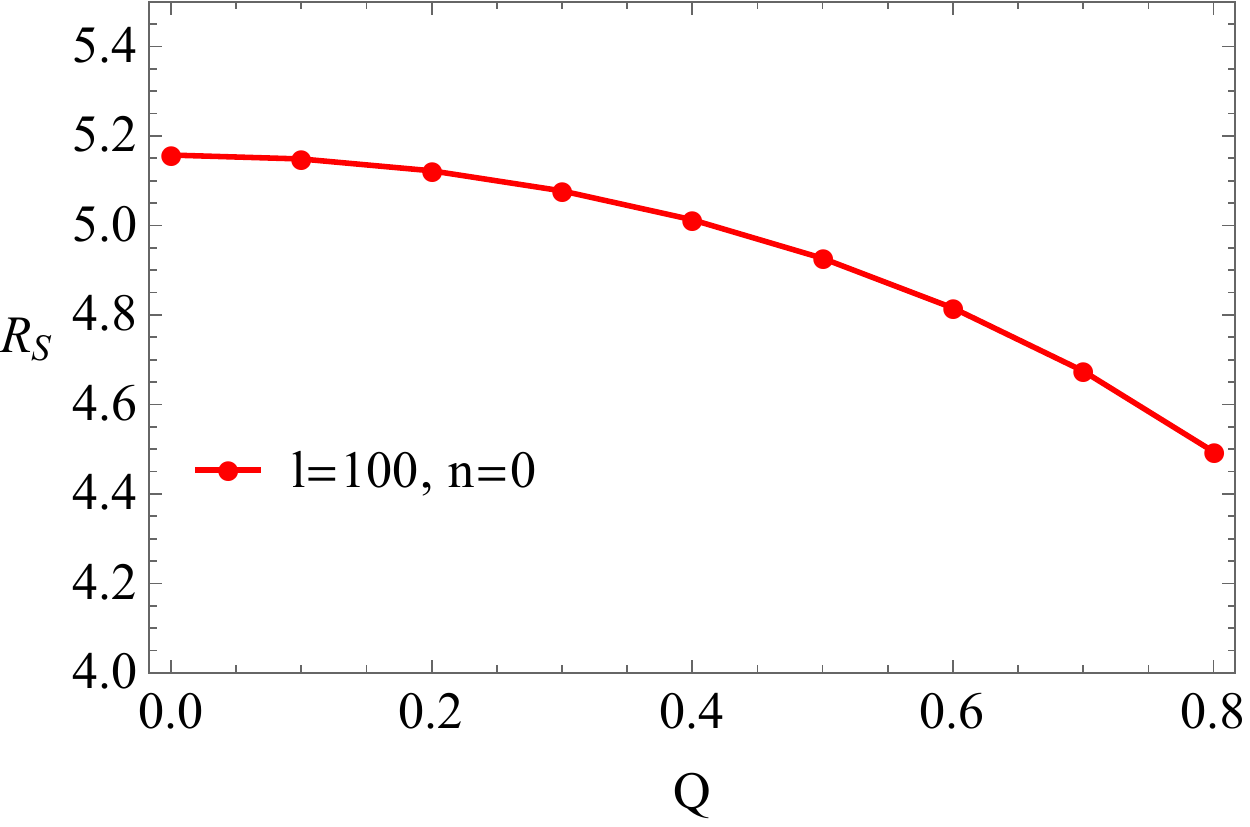}
		\includegraphics[width=8 cm]{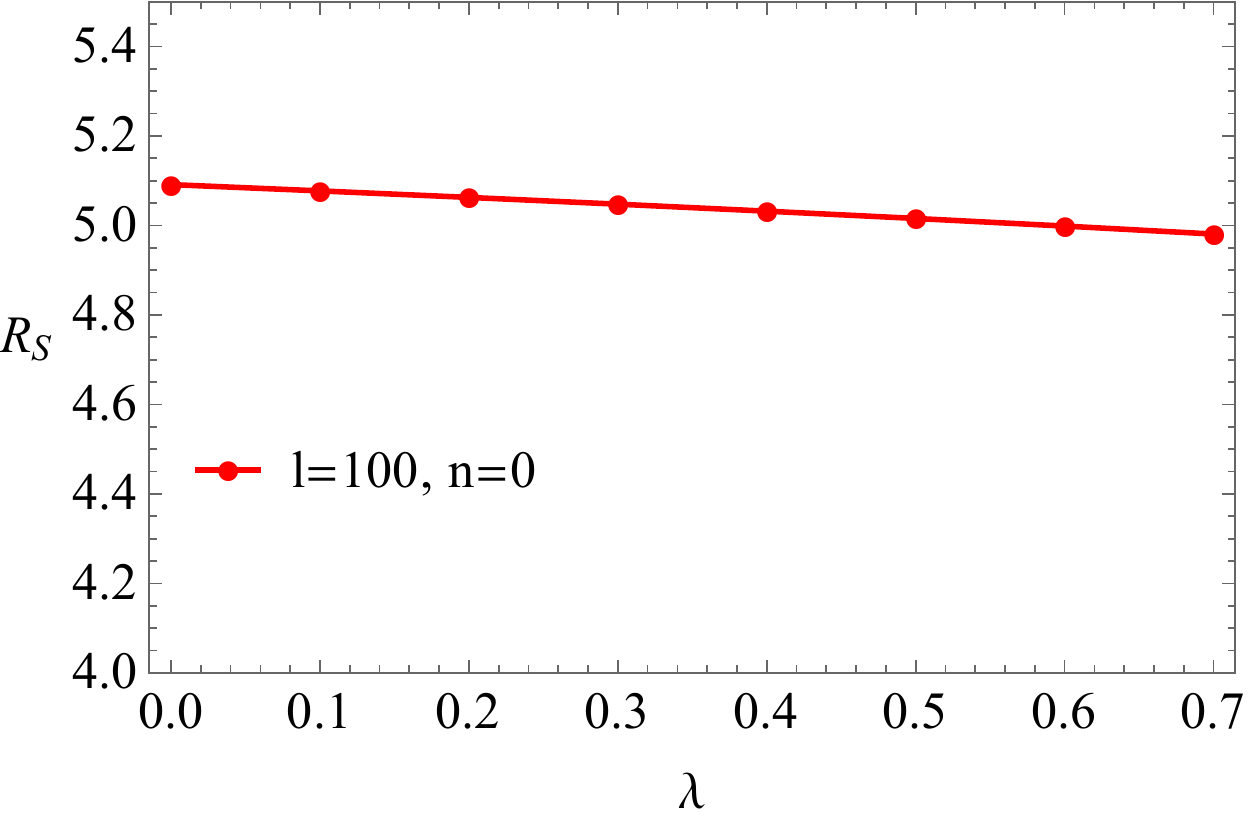}
		\caption{Left panel: The plot of  {the} shadow radius as a function of {the} magnetic charge $Q$ obtained directly from the real part of QNMs given by Eq. (58). We have set $\lambda=0.1$. Right panel: The plot of  {the} shadow radius as a function of {the} parameter $Q$ obtained from the real part of QNMs given by Eq. (59). We have set $Q=0.3$.}
	\end{figure*}
	\begin{figure*}[!htb]
		\includegraphics[width=8.2 cm]{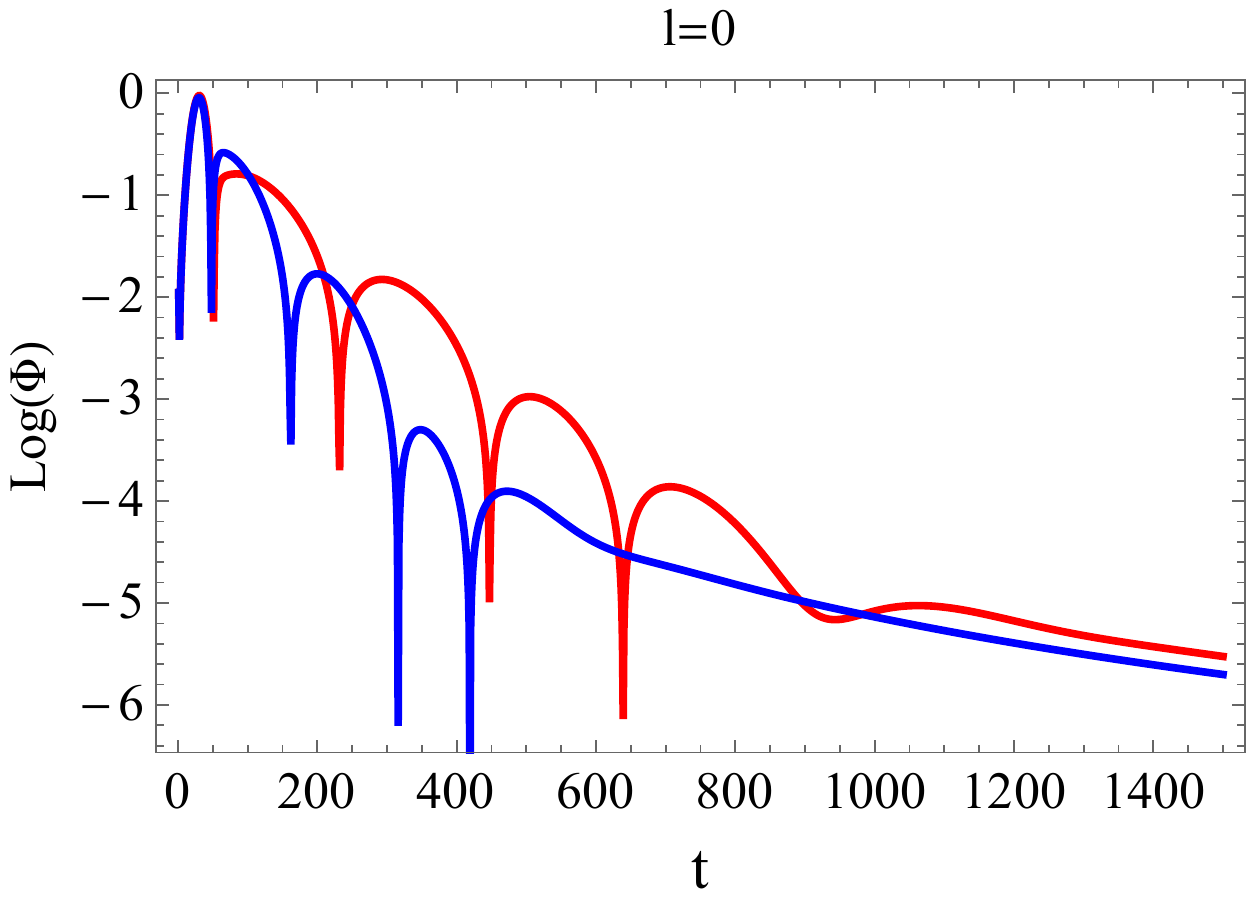}
		\includegraphics[width=8.2 cm]{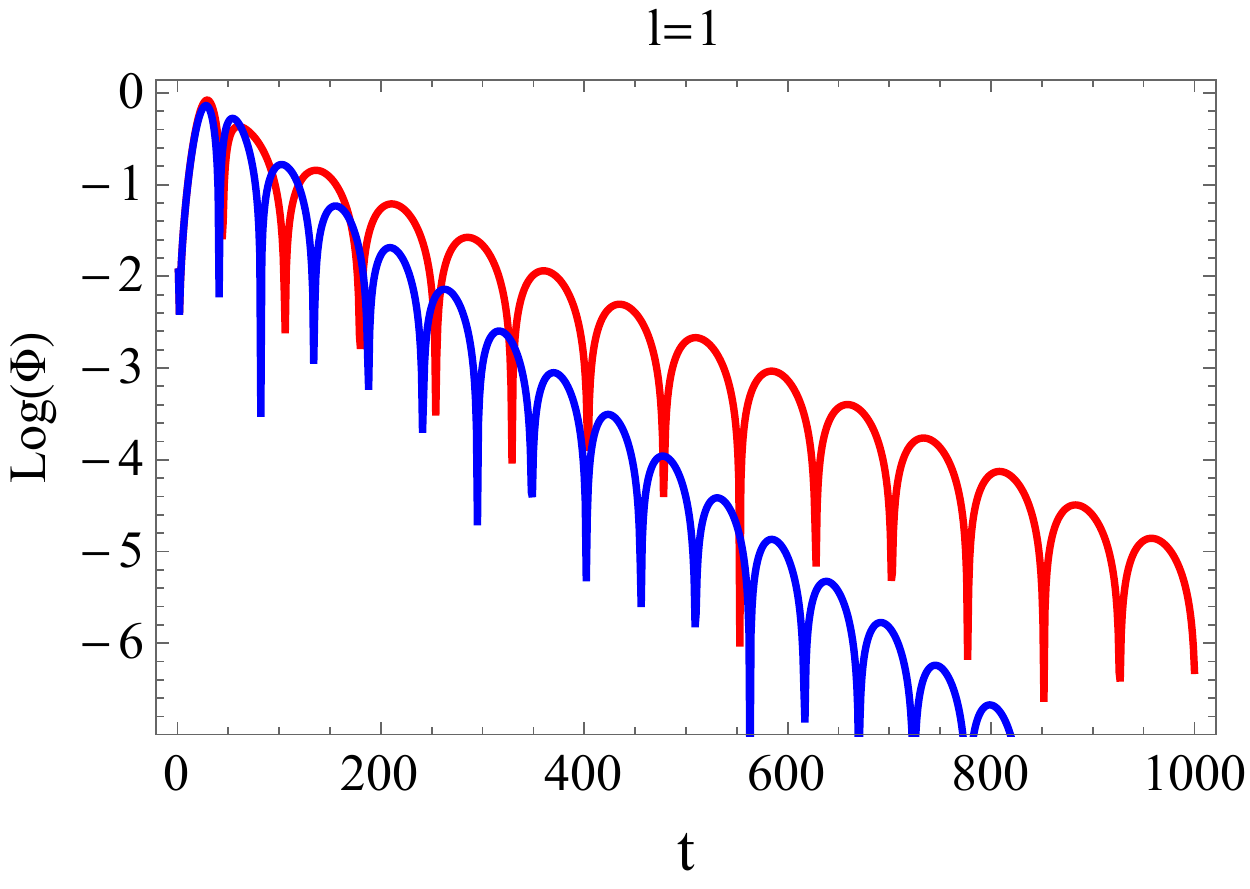}
		\caption{ The time domain profile for the scalar perturbations. Left panel: The  red curve is the time domain profile for the EYM BH and the blue curve for the Schwarzschild BH with $l=0$. Right panel: The
			red curve is the time domain profile for the EYM BH and the blue curve for the Schwarzschild BH with $l=1$.  We have set $\lambda=0.1$ and $r_h=1$ in both plots. }
	\end{figure*}	
	\begin{figure*}[!htb]
		\includegraphics[width=8.2 cm]{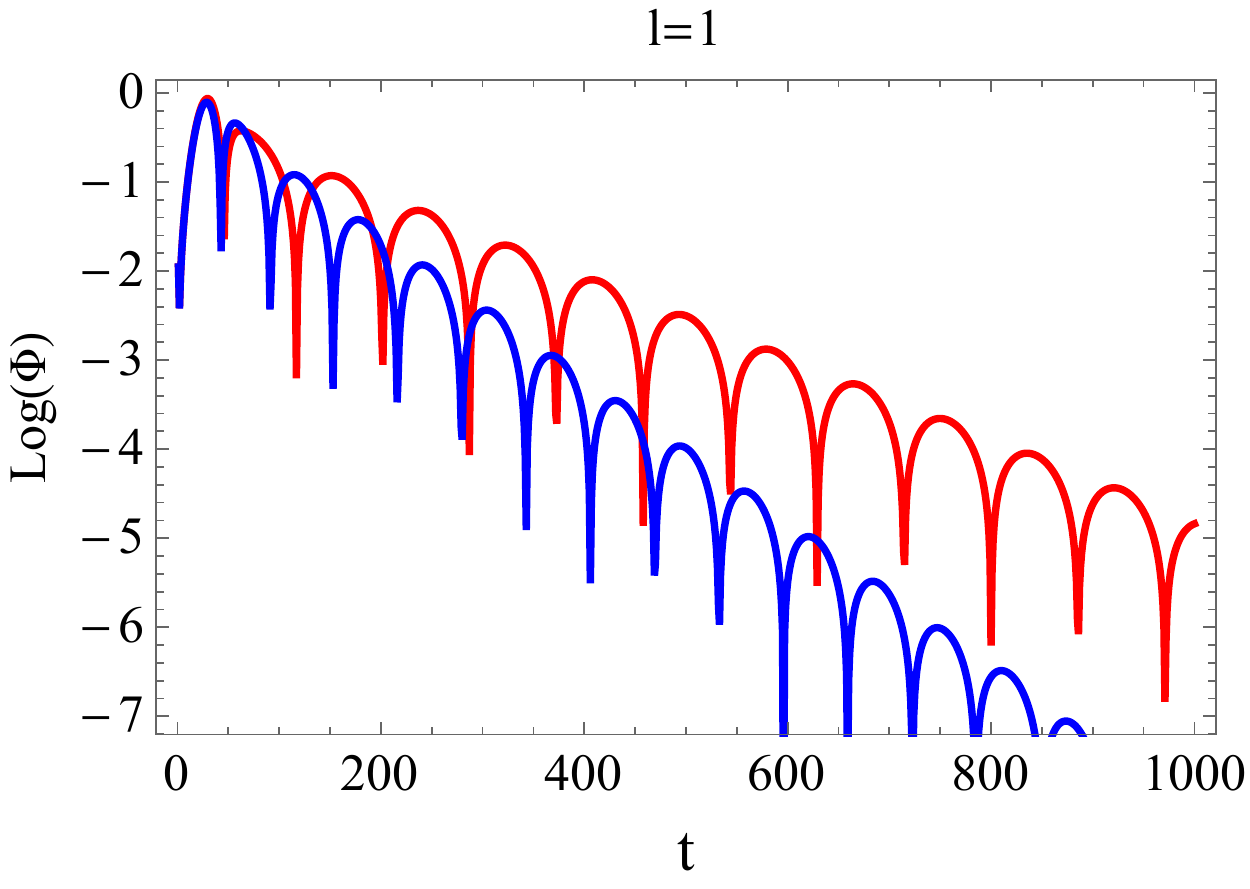}
		\includegraphics[width=8.2 cm]{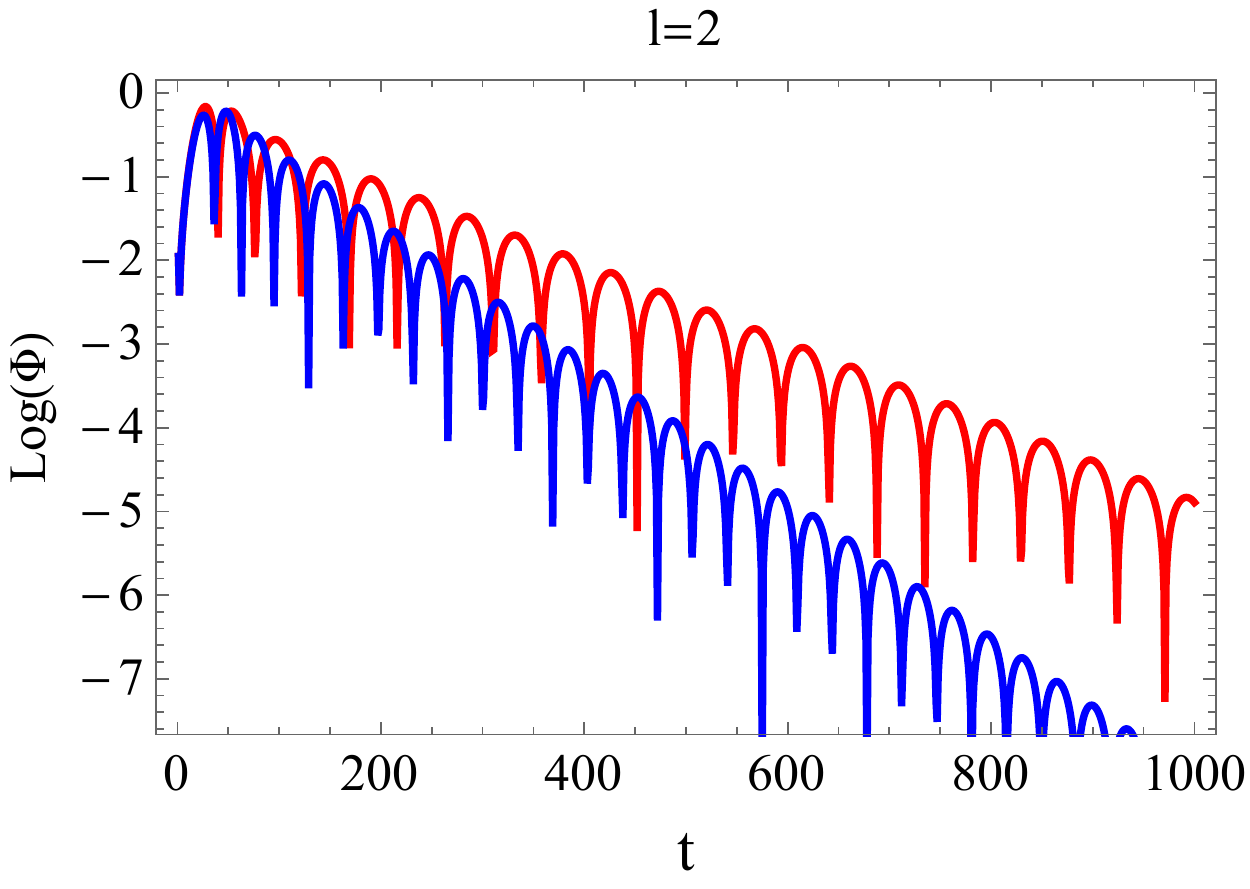}
		The time domain profile for the electromagnetic perturbations. Left panel: The red curve is the time domain profile for the EYM BH and the blue curve for the Schwarzschild BH with $l=1$. Right panel: The  red curve is the time domain profile for the EYM BH and the blue curve for the Schwarzschild BH with $l=2$.  We have set $\lambda=0.1$ and $r_h=1$ in both plots. 
	\end{figure*}
	
	\subsection{The time domain integration method}
	
	Let us now explore the the dynamical
	evolution  of the scalar and electromagnetic perturbations using the  time domain integration method. Toward this goal, first we simplify our computations by employing the following relation
	\begin{equation}
	2M=f(r_h, \lambda, Q)=r_h+\frac{Q^2}{r_h}+\frac{2\lambda}{r_h^3}.
	\end{equation}
	This means that we fix the mass of the BH and analyze when the line $y = 2M$ meets the curve $y =
	f(r_h, \lambda, Q$) at the point of minimum. We call this mass as the
	critical mass, $M = M_c$, and the horizon is denoted by $r = r_H$. In other words, the Cauchy and event horizons coincide. With that information in mind, we can rewrite Eq. (\ref{eq5}) as follows \cite{Balakin:2015gpq}
	\begin{equation}
	g(r)=\frac{(r-r_h)^2}{r^4+2\lambda}\left[r^2+\frac{2 \lambda}{r_h^2} \left(\frac{2 r}{r_h}+1  \right)  \right].
	\end{equation}
	
	Introducing the tortoise coordinate $r_{\star}=\int dr/g(r)$, we find that  it is possible to write the wave equations {(\ref{eq46}) and (\ref{eq51})} as follows
	\begin{equation}
	\frac{\partial^2 \Phi}{\partial r_{\star}^2}-\frac{\partial^2 \Phi}{\partial t^2}=V_{S/E}(r)\Phi,
	\end{equation}
	where $V_{S/E}(r)$ represents the effective potential for the scalar and electromagnetic field, respectively. One can  determine { the  oscillation shape of the QNMs, by utilizing the finite difference method to study the dynamical evolution of the field perturbations} in the time domain and examine the stability of the EYM BH.
	To do so, we first re-write the wave equation in terms of the variables $u$ and $v$,
	\begin{equation}
	\frac{\partial^2 \Phi}{\partial u \partial v}+\frac{1}{4}\,V(r)\Phi=0,
	\end{equation}
	where $u=t-r_{\star}$ and $v=t+r_{\star}$, respectively. To solve this two-dimensional wave equation we use a numerical method known as the finite difference method based on the following equation \cite{Li:2014fka}
	\bqn
	\lb{eq58}
	&& \Phi(u+\delta u,v+\delta v)=\Phi(u,v+\delta v)+\Phi(u+\delta u,v)\nb\\
	&& ~~~ -\Phi(u,v)
	-\delta u \delta v\, \Theta\, \frac{\Phi(u+\delta u,v)+\Phi(u,v+\delta u)}{8}\nb\\
	&& ~~~ +\mathcal{O}(\epsilon^4),
	\eqn
	where
	\begin{equation}
	\lb{eq59}
	\Theta=V\left( \frac{2v-2u+\delta v-\delta u}{4} \right).
	\end{equation}
	Next, we suppose the initial perturbation as a gaussian pulse, centered on $v_c$ given by
	\begin{equation}
	\Phi(u=u_0,v)=e^{-\frac{\left(v-v_c\right)^2}{2 \sigma^2}}.
	\end{equation}
	{With this initial condition, from Eqs.(\ref{eq58}) and (\ref{eq59}) we find numerically the function $\Phi$, and plot it out in Figs. (16) and (17) for different values of $l$. From these figures  we can see that the decaying rates of the scalar and electromagnetic perturbations in the EYM BH spacetime are slower than that  of  the Schwarzschild BH, and   end up in a tail. This conclusion is also supported by our numerical results obtained above by the WKB approximations. }

	\begin{figure*}[!htb]
		\centering
		\includegraphics[width=0.495\textwidth]{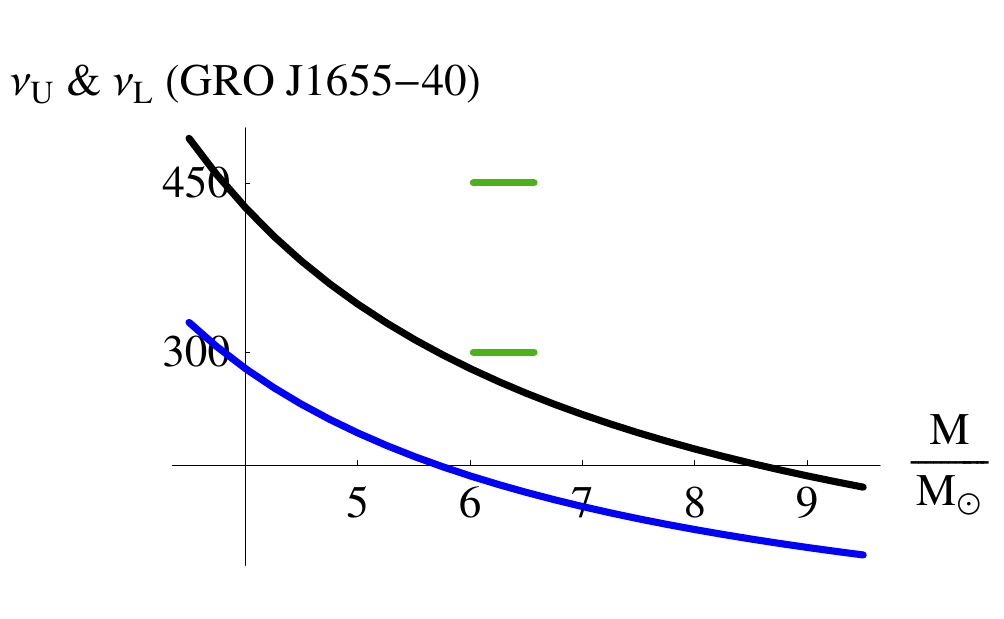}
		\includegraphics[width=0.495\textwidth]{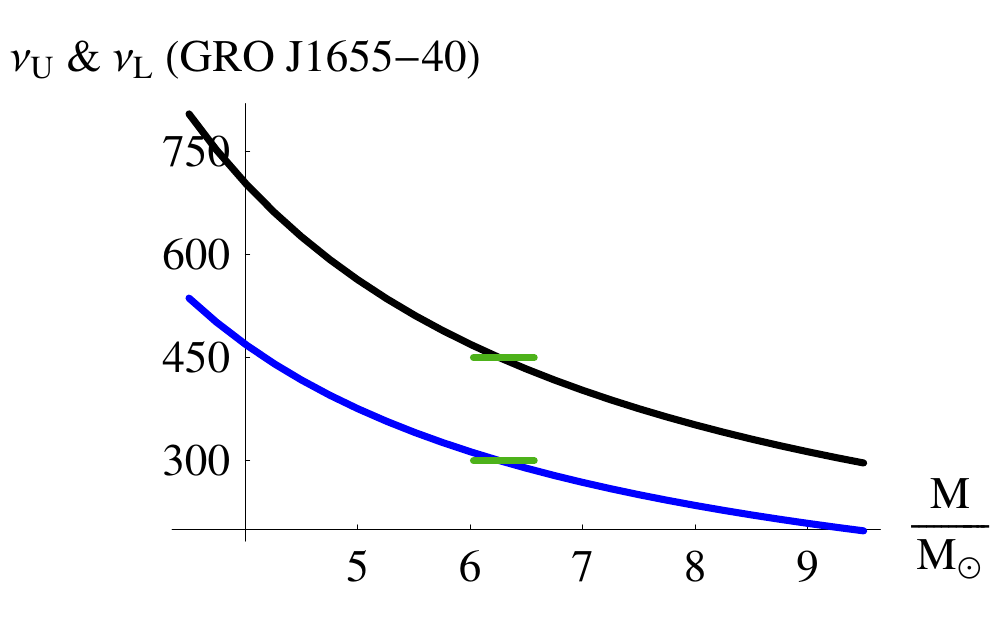}
		\includegraphics[width=0.495\textwidth]{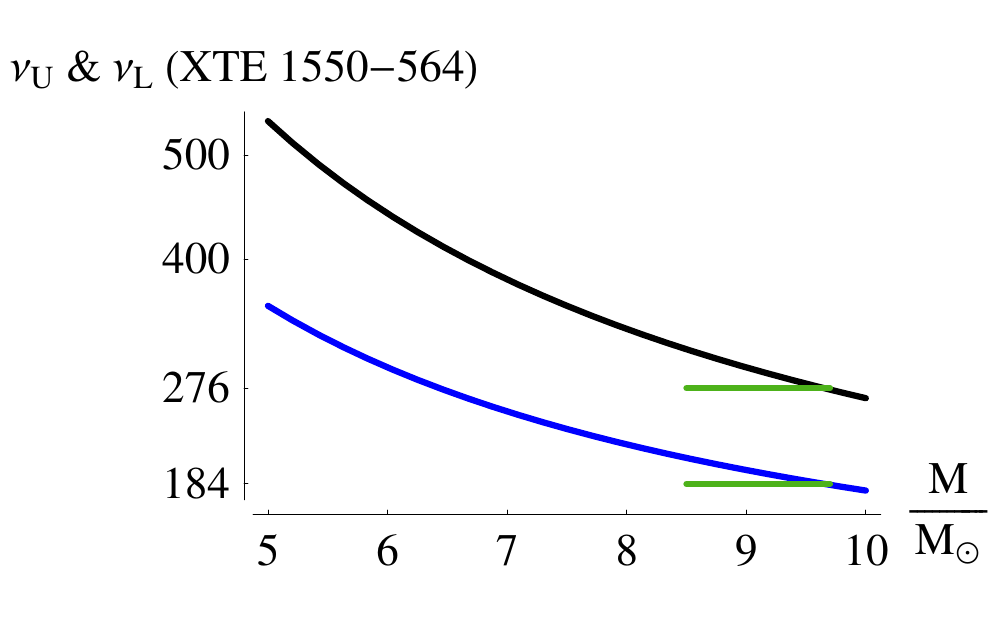}
		\includegraphics[width=0.495\textwidth]{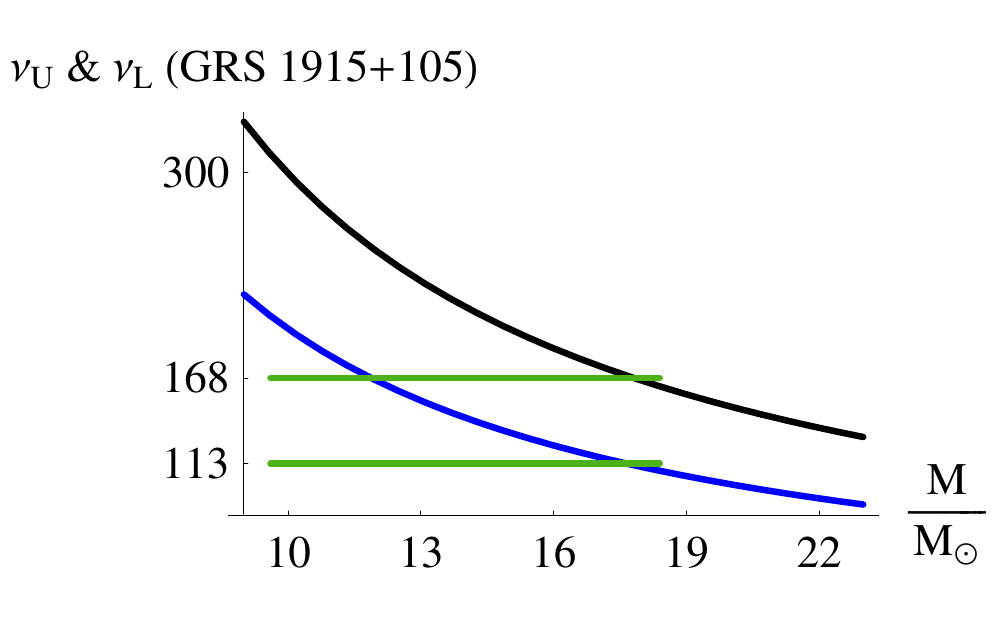}
		\caption{Fitting the uncharged particle oscillation upper and lower frequencies to the observed frequencies (in Hz) for the microquasars GRO J1655-40, XTE J1550-564 and GRS 1915+105 at the 3/2 resonance radius. In these plots each microquasar is treated as a rotating EYM BH given by Eq.\eqref{metric} with $Q=0$. The black curves represent $\nu_U=\nu_\theta$, the blue curves represent $\nu_L=\nu_r$ with $\nu_U/\nu_L=3/2$, and the green curves represent the mass limits as given in~\eqref{pr1}, \eqref{pr2} and~\eqref{pr3}. Upper Left Panel: The microquasar GRO J1655-40 treated as a Kerr BH with $M/M_\odot=6.3$ and $a/r_g=0.70$ ($r_g\equiv GM/c^2$). We see that the black (blue) curve does not cross the upper (lower) mass error band. This panel has been added for comparison. Upper Right Panel: The microquasar GRO J1655-40 treated as a rotating EYM BH with $M/M_\odot=6.3$, $a/r_g=0.70$ and $\lambda/r_g^4=2.14$. We see that the black (blue) curve crosses the upper (lower) mass error band. Lower Left Panel: The microquasar XTE J1550-564 treated as a rotating EYM BH with $M/M_\odot=9.1$, $a/r_g=0.51$ and $\lambda/r_g^4=4.30$. The black (blue) curve crosses the upper (lower) mass error band. Lower Right Panel: The microquasar GRS 1915+105 treated as a rotating EYM BH with $M/M_\odot=9.7$, $a/r_g=0.99$ and $\lambda/r_g^4=0.005$. The black (blue) curve crosses the upper (lower) mass error band.}
		\label{figqpos}
	\end{figure*}

	\section{Quasiperiodic oscillations (QPOs)\label{secqpos}}
	
For the numerical calculations to be carried out in this section, we need the numerical values of some physical constants including the solar mass  $M_\odot=1.9888\times 10^{30}$, the gravitational constant $G=6.673\times 10^{-11}$, and the speed of light in vacuum $c=299792458$, all given in SI units. These same constants will be written explicitly in some subsequent formulas of this section.
	
	In the power spectra of Fig.~3 of Ref.~\cite{res},  we clearly see two peaks  at 300 Hz and 450 Hz,  representing, respectively, the possible occurrence of the lower $\nu_L=300$ Hz quasi-periodic oscillation (QPO), and of the upper $\nu_U=450$ Hz QPO from the Galactic microquasar GRO J1655-40. Similar peaks have been obtained for the microquasars XTE J1550-564 and GRS 1915+105 obeying the remarkable
	{relation},  $\nu_U/\nu_L=3/2$~\cite{qpos1}. Some of the physical {quantities} of these three microquasars and their uncertainties are as follows~\cite{res,res2}:
	\begin{multline}\label{pr1}
	\text{GRO J1655-40 : }\frac{M}{M_\odot}=6.30\pm 0.27,\;\frac{a}{r_g}=0.70\pm 0.05\\
	\nu_U=450\pm 3 \text{ Hz},\;\nu_L=300\pm 5 \text{ Hz},
	\end{multline}
	\begin{multline}\label{pr2}
	\text{XTE J1550-564 : }\frac{M}{M_\odot}=9.1\pm 0.6,\;\frac{a}{r_g}=0.405\pm 0.115\\
	\nu_U=276\pm 3 \text{ Hz},\;\nu_L=184\pm 5 \text{ Hz},
	\end{multline}
	\begin{multline}\label{pr3}
	\text{GRS 1915+105 : }\frac{M}{M_\odot}=14.0\pm 4.4,\;\frac{a}{r_g}=0.99\pm 0.01\\
	\nu_U=168\pm 3 \text{ Hz},\;\nu_L=113\pm 5 \text{ Hz},
	\end{multline}
	where $r_g\equiv GM/c^2$.
	
	These twin values of the QPOs are most certainly due to the phenomenon of resonance which occurs in the vicinity of the ISCO, where the accreting particles perform radial and vertical oscillations around almost circular orbits. These two oscillations couple generally non-linearly to yield resonances in the power spectra~\cite{res3,res4}.
	
	So, in the first part of this section we will be concerned with stable circular orbits in the symmetry plane and their perturbations, since these orbits are mostly borrowed by in-falling matter in accretion processes.
	
	From now on we consider stable circular orbits in the $\theta=\pi/2$ plane. First of all, we need to set up the equations governing an unperturbed circular motion. Once this is done, we will derive the equations that describe a perturbed circular motion around a stable unperturbed circular motion. In a third step we will {separate out} the set of equations governing the perturbed circular motion.
	
	The unperturbed circular motion is a geodesic motion obeying the equation,
	\begin{equation}\label{p1}
	\frac{d u^{\mu}}{d \tau}+\Gamma^{\mu}_{\alpha\beta}u^{\alpha}u^{\beta}=
	0,
	\end{equation}
	where $u^{\mu}=d x^{\mu}/d \tau =\dot x^{\mu}$ is the four-velocity. Here the connection $\Gamma^{\mu}_{\alpha\beta}$ is related to the unperturbed metric~\eqref{metric}. For a circular motion in the equatorial plane ($\theta=\pi/2$), $u^{\mu}=(u^t,\,0,\,0,\,u^{\phi})=u^t(1,\,0,\,0,\,\omega)$, where $\omega=d\phi/d t$ is the angular velocity of the test particle. The only equation describing such a motion is the $r$ component of~\eqref{p1} and the normalization condition $g_{\mu\nu}u^{\mu}u^{\nu}=-c^2$, which take,  respectively, the following forms,
	\begin{align}
	\label{p2a}&\partial _r g_{tt} (u^t)^2+2 \partial _r g_{t\phi}u^t u^{\phi }+\partial _r g_{\phi \phi } (u^{\phi })^2=0,\\
	\label{p2b}&g_{tt} (u^t)^2+2 g_{t\phi} u^t u^{\phi }
	+g_{\phi \phi } (u^{\phi })^2=-c^2,
	\end{align}
	where the metric and its derivatives are evaluated at $\theta=\pi/2$. From them, we  obtain
	\begin{align}
	&\omega =\frac{-\partial _rg_{t \phi }\pm \sqrt{\left(\partial _rg_{t \phi }\right)^2-\partial _rg_{t t} \partial _rg_{\phi  \phi }}}{\partial _rg_{\phi
			\phi }},\nonumber\\
	&u^t=\frac{c}{\sqrt{-\left(g_{t t}+2 \partial _rg_{t \phi } \omega +g_{\phi  \phi } \omega ^2\right)}},\nonumber\\
	\label{p3}&u^{\phi }=\omega  u^t ,
	\end{align}
	where the upper sign corresponds to prograde circular orbits and the lower sign corresponds to retrograde orbits.
	
	%along with~\eqref{p1} and
	
	If the motion is perturbed, the actual position is now denoted by $X^{\mu}=x^{\mu}+\eta^{\mu}$ and the 4-velocity by $U^{\mu}=u^{\mu}+\dot{\eta}^{\mu}$ (where $~\dot{}\equiv d /d\tau$) with $u^{\mu}$ being the unperturbed values given in~\eqref{p3}. {First substituting it  to}
	\begin{equation}
	\frac{d U^{\mu}}{d \tau}+\Gamma^{\mu}_{\alpha\beta}(X^{\sigma})U^{\alpha}U^{\beta}=
	0,
	\end{equation}
	where $\Gamma^{\mu}_{\alpha\beta}(X^{\sigma})$ is the perturbed connection, { and then keeping only linear terms in $\eta^\mu$ and its derivatives
		(and also considering  \eqref{p1}), we finally arrive at~\cite{Kerr1,qposknb}
		\begin{equation}\label{p4}
		\ddot{\eta}^{\mu}+2\Gamma^{\mu}_{\alpha\beta}u^{\alpha}\dot{\eta}^{\beta}
		+\partial_{\nu}\Gamma^{\mu}_{\alpha\beta}u^{\alpha}u^{\beta}\eta^{\nu}
		=0,
		\end{equation}
		where the background connection $\Gamma^{\mu}_{\alpha\beta}$ and its derivatives are  all evaluated at $\theta=\pi/2$.} As shown in~\cite{qposknb}, Eqs.~\eqref{p4} decouple and take the form of oscillating radial (in the $\theta=\pi/2$ plane) and vertical (perpendicular to the $\theta=\pi/2$ plane) motions obeying the following harmonic equations:
	\begin{align}
	&\ddot{\eta}^{r}+\Omega_{r}^2\eta^{r}=0,
	&\ddot{\eta}^{\theta}+\Omega_{\theta}^2\eta^{\theta}=0.
	\end{align}
	The locally measured frequencies ($\Omega_{r},\,\Omega_{\theta}$) are related to the spatially-remote observer's frequencies ($\nu_{r},\,\nu_{\theta}$) by
	\begin{align}\label{pr4}
	&\nu_r=\frac{1}{2\pi}~\frac{1}{u^t}~\Omega_{r},
	&\nu_\theta=\frac{1}{2\pi}~\frac{1}{u^t}~\Omega_{\theta},
	\end{align}
	where $u^t$ is given in~\eqref{p3} and~\cite{qposknb}
	\begin{align}
	&\Omega_{\theta}^2\equiv (\partial_{\theta}\Gamma^{\theta}_{ij})u^iu^j,\qquad (i,\,j=t,\,\phi),\\
	\label{p5b}&\Omega_{r}^2\equiv (\partial_{r}\Gamma^{r}_{ij}-4\Gamma^{r}_{ik}\Gamma^{k}_{rj})u^iu^j,\qquad (i,\,j,\,k=t,\,\phi).
	\end{align}
	In these expressions the summations extend over ($t,\,\phi$). It is understood that all the functions appearing in~\eqref{p3}, \eqref{pr4} and~\eqref{p5b} are evaluated at $\theta=\pi/2$.
	
	In therms of
	\begin{equation}\label{rg}
	x\equiv\frac{r}{r_g},\quad a_0\equiv\frac{a}{r_g},\quad\lambda_0\equiv\frac{\lambda}{r_g^4},\quad r_g\equiv\frac{GM}{c^2},
	\end{equation}
	the expressions of ($\nu_{r},\,\nu_{\theta}$) measured in Hz take the form
	\begin{equation}\label{vrvt}
	\nu _r=\frac{c^3}{2\pi  G M } \sqrt{N} \sqrt{\frac{N_r}{D_r}},\;\ \nu _{\theta }=\frac{c^3}{2\pi G M} \sqrt{N} \sqrt{\frac{N_\theta}{D_\theta}},
	\end{equation}
	where ($N,\,N_r,\,D_r,\,N_\theta,\,D_\theta$) are given in Appendix B.
	
	As we {mentioned} earlier, the twin values of the QPOs observed in the microquasars are most certainly due to the phenomenon of resonance resulting from the coupling of the vertical and radial oscillatory motions~\cite{res3,res4}. The most common models for resonances are parametric resonance, forced resonance and Keplerian resonance. {It is the general belief} that the resonance observed in the three microquasars~\eqref{pr1}, \eqref{pr2} and~\eqref{pr3} is  {of the nature of the parametric resonance and is given by}
	\begin{equation}\label{as1}
	\nu_U = \nu_\theta,\qquad \nu_L =\nu_r \ ,
	\end{equation}
	with
	\begin{equation}\label{as2}
	\frac{\nu_\theta}{\nu_r}=\frac{n}{2} ,\qquad n\in \mathbb{N}^+ .
	\end{equation}
	{In most of the applications of the} parametric resonance one considers the case $n=1$~\cite{b1,b2,b3,b4}, where in this case $\nu _{r}$ is the natural frequency of the system and $\nu_{\theta }$ is the parametric excitation ($T_{\theta }=2T_{r}$, the corresponding periods), that is, the vertical oscillations supply energy to the radial oscillations causing resonance~\cite{b4}. However, since $\nu_\theta>\nu_r$ in the vicinity of ISCO, where accretion occurs and QPO resonance effects take place, the lower possible value of $n$ is 3 and in this case $\nu_{r}$ becomes the parametric excitation that supplies energy to the vertical oscillations.
	
	Thus, the observed ratio $\nu_U/\nu_L=3/2$ is theoretically justified by making the assumptions~\eqref{as1} and~\eqref{as2} with $n=3$. Numerically we have to show that the plot of $\nu_\theta$ ($\nu_r$) versus $M/M_\odot$ crosses the upper (lower) mass band error, given in~\eqref{pr1}, \eqref{pr2} and~\eqref{pr3}, as $a_0$ assumes values in its defined band error and $\lambda_0>0$ runs within some interval {to be defined  later  as the interval of its constrained values.}
	
	{Curves that fit the  upper and lower oscillation frequencies  of the uncharged test particles to the observed frequencies (in Hz) of the microquasars GRO J1655-40, XTE J1550-564 and GRS 1915+105 at the 3/2 resonance radius are presented in Fig.~\ref{figqpos}. In these plots each microquasar is treated as a rotating EYM BH~\eqref{metric} with $Q=0$. The black curves represent $\nu_U=\nu_\theta$ versus $M/M_\odot$, the blue curves represent $\nu_L=\nu_r$ versus $M/M_\odot$ with $\nu_U/\nu_L=3/2$, and the green curves represent the mass limits as given in~\eqref{pr1}, \eqref{pr2} and~\eqref{pr3}. For comparison we start with the upper left panel where the microquasar GRO J1655-40 is treated as a Kerr BH ($\lambda_0=0$) with $M/M_\odot=6.3$ and $a/r_g=0.70$. We see that the black (blue) curve does not cross the upper (lower) mass error band. In the other remaining three panels, where each microquasar is treated as a rotating EYM BH~\eqref{metric}, we see how the black (blue) curve crosses the upper (lower) mass error band for each of the microquasars. The curve fittings allow} us to fix the following limits for $\lambda_0$:
	\begin{equation}\label{li}
	0<\lambda_0\lesssim 4.3.
	\end{equation}
	
	It is worth noting that the {ratio} $\nu_U/\nu_L=3/2$ may sometimes admit two $x$-roots. In our plot we have chosen the root that is closer to $x_{\text{ISCO}}$ where the events of accretion and QPOs occur.
	
	For completeness and comparison, two other plots (not shown in this paper) similar to those in the upper left panel of Fig.~\ref{figqpos} have been sketched for the microquasars XTE J1550-564 and GRS 1915+105 treating them as the Kerr BH ($\lambda_0=0$). For the microquasar XTE J1550-564 the plots show no intersections of the curves ($\nu_U,\,\nu_L$) with the mass error bands, and for the microquasar GRS 1915+105 intersections exist but these are certainly due to the large mass band error for this microquasar~\eqref{pr3}.
	
	Other {curves that fit  the data of the three microquasars~\eqref{pr1}, \eqref{pr2} and~\eqref{pr3} have been given either} via the immersion of a Schwarzschild BH into a test magnetic filed~\cite{fit} or via the consideration of generalized theories of gravity~\cite{fit2}.

	\section{Conclusion\label{seccon}}
	
	In this paper we have obtained a rotating regular magnetic BH solution {of the EYM theory,  by applying   the NJAAA  to  a spherical symmetric solution}. We have then investigated the ergosurface and the BH shadow. We {have} found that  { the magnetic charge $Q$ causes deformations to both of} the size and shape of the BH shadow. For a given value of  {the angular momentum $a$ and the} inclination angle $\theta_0$, the presence of the magnetic charge  {$Q$} shrinks the shadow and  {enhances} its deformation with respect to the shadow {of}  the Kerr spacetime. In other words, the shadow radius decreases due to  {the presence of the} magnetic charge $Q$. Among other things, we have constructed the embedding diagram for the rotating EYM BH and examined the energy conditions. {In particular,  it has been found that the strong energy condition} in general is not satisfied. The particular topological property of the shadow has been revealed upon studying the behavior of the topological quantity $\delta$ as a function of $\lambda$. At some critical value $\lambda_c$ we have found that there is a possible topological phase transition. In this transition the rotating EYM BH first becomes  extremal and then turns to a horizonless compact object without spacetime singularities at the center. 
	
In addition,  we have studied the connection between the real part of QNMs in the eikonal limit and {the} shadow radius. First, using the WKB approximation to the sixth order we have shown that the quasi-normal frequencies in the spacetime of the EYM BH  deviate from those of {the} Schwarzschild BH,  {that is,} $\omega_{\Re }$ increases with increasing $Q$. We have shown that the same result is obtained if we fix the magnetic charge $Q$ and increase the parameter $\lambda$, although the effect is very small. This suggest that the shadow radius $R_S$ decreases due to the inverse relation given by Eq.(58) and Eq. (59), respectively. We have verified this result by means of the geodesic approach with the shadow images given in Figs. 6-7. Despite the fact the effect of $\lambda$ is small, we have used the M87 black hole parameters and shown that the rotating EYM black hole can be distinguished from the Kerr-Newman black hole with a magnetic charge. The difference between the angular diameters of their shadows is given by the interval $\Delta \theta_s \in (0.11-0.61)\mu$as with $\lambda \in (0.1-0.5)$.  In addition, we studied observational constraints on the EYM parameter $\lambda$ via frequency analysis of QPOs and the EHT data of shadow cast by the M87 central black hole. It is interesting to note that EHT data offers more tighter constraints on the parameter $\lambda$ as compared to QPO's associated with microquasars.
	
We have also examined the dynamical evolution  of the scalar and electromagnetic perturbations using the  time domain integration. We have shown that the decaying rates of the scalar and electromagnetic perturbations in  the rotating EYM BH  are  slower  than that of  the Schwarzschild BH, and  end up in a tail. 
	
Finally, we have considered the QPOs and their resonances generated by a test particle undergoing a circular motion in the symmetric plane of the rotating EYM BH. We have employed the usually put-forward assumptions: $\nu_U=\nu_\theta$, $\nu_L=\nu_r$ with $\nu_U/\nu_L=3/2$. With these assumptions, we have explored in details the effects of the parameter $\lambda_0$ on the frequencies of QPOs. For the uncharged rotating EYM BHs,  we have shown that the value of $\lambda_0$ lies in an interval bounded below by $0$ and above by $4.3$. This has allowed us to obtain
good and complete curve fittings  for the three microquasars GRO J1655-40, XTE J1550-564 and GRS 1915+105, all treated as rotating neutral EYM BHs.

	\section*{Appendix A: Einstein field equations\label{AA}}
	
	{The non-vanishing  components of the Einstein tensor} are given by~\cite{Azreg-Ainou:2014pra},
	
	\begin{eqnarray}\nonumber
	\emph{G}_{tt}&=&\frac{2 \Upsilon' (a^4 \cos^4\theta-a^4\cos^2\theta+a^2 r^2+r^4-2 \Upsilon r^3)}{\Sigma^3}\\\notag
	&& - \frac{a^2 r \sin^2\theta \Upsilon''}{\Sigma^3},\\\nonumber
	\emph{G}_{rr}&=&-\frac{2 \Upsilon' r^2}{\Delta \Sigma},\\\notag\label{einstein}
	\emph{G}_{\theta\theta}&=& -\frac{\Upsilon'' a^2 r^2 \cos^2\theta+2\Upsilon' a^2 \cos^2\theta+\Upsilon'' r^3}{\Sigma},\\\nonumber
	\emph{G}_{t\phi}&=&\frac{a \sin^2\theta}{\Sigma^3} \Big[r(a^2+r^2)\Sigma \Upsilon''\\\notag
	& &+ 2 \Upsilon' \left((a^2+r^2)a^2\cos^2\theta-a^2 r^2-r^3(r-2 \Upsilon)\right)  \Big],\\\nonumber
	\emph{G}_{\phi\phi}&=&-\frac{\sin^2\theta}{\Sigma^3}\Big[r(a^2+r^2)^2\Sigma\Upsilon'' \\\notag
	& & + 2 a^2 \Upsilon'(\cos^2\theta \Big(a^4+3a^2r^2+2 r^4-2 \Upsilon r^3)\\
	&& - a^2r^2-r^4+2\Upsilon r^3 \Big)   \Big].
	\end{eqnarray}
	
	\section*{Appendix B: QPOs' expressions\label{AB}}
	
	%To complete the expressions of ($\nu_{r},\,\nu_{\theta }$) presented in~\eqref{vrvt} we provide those of
	{In terms of $x$, $a_0$, and $\lambda_0$,  the quantities  $N$, $N_r$, $D_r$, $N_\theta$ and $D_\theta$ appearing in  ~\eqref{vrvt}  are given by}
	%~\eqref{rg} taking $Q=0$ in~\eqref{metric}.
	\begin{widetext}
		\bqn
		%	\begin{align*}
		N&=& \frac{1}{[(x^4+2 \lambda _0)^2-a_0^2 (x^5-6 x \lambda _0)]^2}\Bigg[2 x^{3/2} a_0^3 \sqrt{x^4-6 \lambda _0}
		\left(x^8-4 x^4 \lambda _0-12 \lambda _0^2\right)\nb\\
		&& +2 x^{7/2} a_0 \sqrt{x^4-6 \lambda _0} \left(3 x^8+4 x^4 \lambda _0-4 \lambda _0^2\right)
		+\left(x^4+2\lambda _0\right)^2 \left(-3 x^7+x^8+2 x^3 \lambda _0+4 x^4 \lambda _0+4 \lambda _0^2\right)\nb\\
		&& +a_0^2 \left(-3 x^{12}-3 x^{13}+20 x^8 \lambda _0+6 x^9
		\lambda _0-12 x^4 \lambda _0^2+60 x^5 \lambda _0^2+72 x \lambda _0^3\right)\Bigg],\\
		N_r&=&\left(x^5+2 x \lambda _0\right)^2 \left(-6 x^7+x^8-12 x^3 \lambda _0+36 x^4 \lambda _0-60 \lambda _0^2\right)+a_0^4 \left(-3 x^{13}+70
		x^9 \lambda _0-324 x^5 \lambda _0^2+72 x \lambda _0^3\right)\nb\\
		&& +2 x^{5/2} a_0 \sqrt{x^4-6 \lambda
			_0} \left(6 x^{11}+3 x^{12}+24 x^7 \lambda _0-46 x^8 \lambda _0+24 x^3 \lambda _0^2-92 x^4 \lambda _0^2+24 \lambda _0^3\right)\nb\\
		&& +a_0^2 \Big(-6 x^{14}-15 x^{15}-3 x^{16}+24 x^{10} \lambda _0+190 x^{11} \lambda _0+40
		x^{12} \lambda _0+72 x^6 \lambda _0^2-468 x^7 \lambda _0^2+184 x^8 \lambda _0^2-792 x^3 \lambda _0^3\nb\\
		&& +160 x^4 \lambda _0^3-48 \lambda _0^4\Big)+2
		a_0^3 \Big(4 x^{23/2} \sqrt{x^4-6 \lambda _0}+3 x^{25/2} \sqrt{x^4-6 \lambda _0}-48 x^{15/2} \sqrt{x^4-6 \lambda _0} \lambda _0\nb\\
		&& -46 x^{17/2} \sqrt{x^4-6
			\lambda _0} \lambda _0+144 x^{7/2} \sqrt{x^4-6 \lambda _0} \lambda _0^2-92 x^{9/2} \sqrt{x^4-6 \lambda _0} \lambda _0^2
		+24 \lambda _0^3 \sqrt{x^5-6
			x \lambda _0}\Big),\\
		D_r&=& x (x^4+2 \lambda _0) \Bigg[2 x^{3/2} a_0^3 \sqrt{x^4-6 \lambda _0} (x^8-4 x^4 \lambda _0-12 \lambda _0^2)+2 x^{7/2} a_0
		\sqrt{x^4-6 \lambda _0} \left(3 x^8+4 x^4 \lambda _0-4 \lambda _0^2\right)\nb\\
		&&+\left(x^4+2 \lambda _0\right)^2 \left(-3 x^7+x^8+2 x^3 \lambda _0+4 x^4
		\lambda _0+4 \lambda _0^2\right)+a_0^2 \Big(-3 x^{12}-3 x^{13}+20 x^8 \lambda _0+6 x^9 \lambda _0-12 x^4 \lambda _0^2\nb\\
		&& +60 x^5 \lambda _0^2+72 x \lambda_0^3\Big)\Bigg],\\
		N_\theta&=&(x^4-6 \lambda _0) (x^5+2 x \lambda _0)^2-2 x^{5/2} a_0 \sqrt{x^4-6 \lambda _0} (3 x^8+4 x^4 \lambda _0-4 \lambda
		_0^2)+a_0^4 (3 x^9-20 x^5 \lambda _0+12 x \lambda _0^2)\nb\\
		&&+a_0^2 (9 x^{11}+3 x^{12}-44 x^7 \lambda _0+10 x^8 \lambda _0-60 x^3
		\lambda _0^2+4 x^4 \lambda _0^2-8 \lambda _0^3)+a_0^3 \Big(-4 x^{15/2} \sqrt{x^4 -6 \lambda _0}\nb\\
		&& -6 x^{17/2} \sqrt{x^4-6 \lambda _0}+24 x^{7/2}
		\sqrt{x^4-6 \lambda _0} \lambda _0-8 x^{9/2} \sqrt{x^4-6 \lambda _0} \lambda _0+8 \lambda _0^2 \sqrt{x^5-6 x \lambda _0}\Big),\\
		D_\theta&=& x \Bigg[2 x^{3/2} a_0^3 \sqrt{x^4-6 \lambda _0} \left(x^8-4 x^4 \lambda _0-12 \lambda _0^2\right)+2 x^{7/2} a_0 \sqrt{x^4-6 \lambda _0}
		\left(3x^8+4 x^4 \lambda _0-4 \lambda _0^2\right)\\
		&&+\left(x^4+2 \lambda _0\right)^2 \left(-3 x^7+x^8+2 x^3 \lambda _0+4 x^4 \lambda _0+4 \lambda _0^2\right)+a_0^2
		\Big(-3 x^{12}-3 x^{13}+20 x^8 \lambda _0+6 x^9 \lambda _0-12 x^4 \lambda _0^2\nb\\
		&& +60 x^5 \lambda _0^2+72 x \lambda _0^3\Big)\Bigg].
		%	\end{align*}
		\eqn
	\end{widetext}

	\section*{Acknowledgements} MJ would like to thank Kai Lin for supporting us with numerical codes for the study of the evolution of the scalar field perturbations. S.-W. Wei is supported by National Natural Science Foundation of China (NNSFC) with the Grant No. 11675064, while AW is supported in part by NNSFC with the Grant Nos.  11675145 and 11975203. The authors gratefully acknowledge the anonymous referee for numerous insightful remarks that helped in improvement of the manuscript.

\end{document}